\documentclass{jfm}
\usepackage{graphicx,color}
\usepackage{epstopdf, epsfig,comment,mathtools,amsmath,amssymb,amsfonts}
\usepackage{appendix}
\usepackage[normalem]{ulem}

\def\avkrev#1{\textcolor{black}{#1}}
\def\NEW#1{\textcolor{black}{#1}}

\def\ub{\mathbf{u}}

\shorttitle{Vortex crystal formation in instability-driven two-dimensional turbulence}
\shortauthor{A. van Kan, B. Favier,  K. Julien and E. Knobloch}

\title{From a vortex gas to a vortex crystal in instability-driven two-dimensional turbulence}

\author{Adrian van Kan\aff{1}
  \corresp{\email{avankan@berkeley.edu}},
  Benjamin Favier\aff{2}, Keith Julien\aff{3}, Edgar Knobloch\aff{1}}

\affiliation{\aff{1}Department of Physics, University of California at Berkeley, Berkeley, California 94720, USA
\aff{2}Aix Marseille Univ., CNRS, Centrale Marseille, IRPHE, Marseille, France
\aff{3}Department of Applied Mathematics, University of Colorado, Boulder, CO 80309, USA}
\begin{document}

\maketitle

\begin{abstract}
\end{abstract}
We study structure formation in two-dimensional turbulence driven by an external force, interpolating between linear instability forcing and random stirring, subject to nonlinear damping. Using extensive direct numerical simulations, we uncover a rich parameter space featuring four distinct branches of stationary solutions: large-scale vortices, hybrid states with embedded shielded vortices (SVs) of either sign, and two states composed of many similar SVs. Of the latter, the first is a dense vortex gas where all SVs have the same sign and diffuse across the domain. The second is a hexagonal vortex crystal forming from this gas when the linear instability is sufficiently weak. These solutions coexist stably over a wide parameter range. The late-time evolution of the system from small-amplitude initial conditions is nearly self-similar, involving three phases: initial inverse cascade, random nucleation of SVs from turbulence and, once a critical number of vortices is reached, a phase of explosive nucleation of SVs, leading to a statistically stationary state. The vortex gas is continued in the forcing parameter, revealing a sharp transition towards the crystal state as the forcing strength decreases. This transition is analysed based on the diffusion of individual vortices and using statistical physics. The crystal can also decay via an inverse cascade resulting from the breakdown of shielding or insufficient nonlinear damping acting on SVs. Our study highlights the importance of the forcing details in two-dimensional turbulence and reveals the presence of nontrivial SV states in this system, specifically the emergence and melting of a vortex crystal.

\section{Introduction}

The study of two-dimensional (2D) and quasi-2D turbulence has a long history \citep{boffetta2012two}, from the discovery of the dual inverse energy, forward enstrophy cascade phenomenology  \citep{fjortoft1953changes,kraichnan1967inertial} to early numerical simulations \citep{lilly1969numerical} and laboratory experiments \citep{sommeria1986experimental}. \NEW{Such problems are of interest as idealized models of geophysical fluid dynamics \citep{pedlosky1987geophysical} and, more recently, active fluid flows where energy-consuming microswimmers can drive vortices and jets \citep{dombrowski2004self}.} In a finite system, the nonlinear transfer of kinetic energy from small to large scales in an inverse cascade generates large-scale coherent structures, typically vortices or jets, called \textit{condensates} \citep{smith1993bose}. Inverse energy cascades are also observed in simulations of highly anisotropic 3D flows within thin layers \citep{smith1996crossover,celani2010turbulence}, rapidly rotating turbulence \citep{deusebio2014dimensional}, strongly stratified flows \citep{sozza2015dimensional} \avkrev{and can arise in magnetohydrodynamic systems as well \citep{seshasayanan2014edge,dallas2015self,pouquet2019helicity}}. Such quasi-2D inverse energy cascades can also lead to large-scale condensation if large-scale damping is weak \citep{seshasayanan2018condensates,van2019condensates,musacchio2019condensate}. Moreover, condensation is known to occur in rapidly rotating convection \citep{julien2012statistical,rubio2014upscale,favier2014inverse,guervilly2014large} and convection driven by an imposed heat flux \citep{vieweg2021supergranule}, and has been reported in active fluid flows \citep{linkmann2019phase,linkmann2020condensate,puggioni2022giant}. \NEW{There is also an extensive literature on experimental studies of quasi-2D large-scale condensation \citep{sommeria1986experimental,paret1997experimental,xia2011upscale,xia2017two,fang2021spectral}.} Recent reviews of quasi-2D turbulence are provided by \cite{alexakis2018cascades} and \cite{alexakis2023quasi}.

In addition to structure formation at the largest scales, another type of self-organisation widely observed in fluid flows is the \textit{vortex crystal}, a regular array of smaller-scale vortices. For instance, such structures are observed in rotating convection \citep{boubnov1986experimental,boubnov1990temperature,zhong1991asymmetric,vorobieff2002turbulent,boubnov2012convection}, in experiments on magnetised electron columns \citep{driscoll1999relaxation,schecter1999vortex} and quantum fluids \citep{tosi2012geometrically}, as well as in active fluids, including dense suspensions of microswimmers such as sperm cells \citep{riedel2005self}. The polar vortices on Jupiter \citep{aA2018,siegelman2022polar} provide another particularly compelling example. \NEW{Vortex crystals have also been found in 2D turbulence subject to spectral truncation at large scales \citep{smithr1994finite}} \textcolor{black}{or forced with a mixture of random and deterministic forcing \citep{jimenez2007spontaneous}.} The emergence and melting of active fluid vortex crystals has already been investigated \citep{james2021emergence}, with a focus on the 2D case. More generally, 2D chiral lattices\avkrev{, which also arise in the study of active solids \citep{baconnier2022selective},} are currently of great interest in physics because they support topologically protected edge states \citep{nash2015topological,mitchell2018geometric,mitchell2018tunable} and fluid dynamical vortex crystals provide, in principle, a simple laboratory realisation of such systems. 

Sustaining any fluid flow in a stationary state against dissipation requires the injection of energy by a forcing mechanism. To facilitate a detailed analysis of the complexities of turbulence, highly idealized forcing functions are often considered. For instance, one can choose a time-independent forcing as in the case of Kolmogorov flow \citep{arnol1960kolmogorov,meshalkin1961investigation,borue1996numerical,gallet2013two}, or a stochastic forcing with a constant energy injection rate \citep{novikov1965functionals}. The latter choice in particular has been widely adopted in numerous studies of forced 2D turbulence, see e.g. \citet{smith1993bose}, \citet{boffetta2007energy}, \citet{chan2012dynamics}, \citet{laurie2014universal} and \citet{frishman2018turbulence}. \NEW{In both cases, the forcing is specified independently of the flow state, a property} that makes the problem more amenable to a comprehensive analysis. However, many real fluid flows result from instabilities, for instance of convective, shear or baroclinic type \citep{chandrasekhar2013hydrodynamic,salmon1980baroclinic,vallis2017atmospheric}, which are explicitly flow-state dependent.
Similarly, models of active fluid flows feature scale-dependent viscosities which can be
negative at certain scales \citep{slomka2017geometry}, a fact consistent with the measured rheology of such flows \citep{lopez2015turning}. Such scale-dependent viscosities also arise in eddy viscosity modelling, where the molecular viscosity $\nu$ is modified by terms involving small-scale velocities to represent the effect of smaller-scale motions on larger-scale motions. Eddy viscosity, including with a negative sign, has been studied in a variety of 2D and 3D flows \citep{kraichnan1976eddy,sivashinsky1985negative,bayly1986positive,yakhot1987negative,dubrulle1991eddy,gama1994negative,alexakis2018three}. Negative eddy viscosities are also encountered in applications within the context of backscatter parameterisations \citep{prugger2022geophysical,prugger2023rotating}. Schemes of this type are used in ocean modelling \citep{jansen2014parameterizing,juricke2020ocean}. In addition, negative-viscosity forcing has been considered in a study of axisymmetric turbulence \citep{qin2020transition}, while linearly forced isotropic turbulence at moderate Reynolds numbers has been studied by \cite{bos2020linearly}.

For flows driven by instabilities, the driving explicitly depends on the velocity field and the injection rate of kinetic energy is proportional to the squared velocity amplitude of the forcing-scale modes. Flows resulting from instabilities can differ starkly from flows driven by random stirring. For instance, it is known that the transition to two-dimensional turbulence is non-universal and depends qualitatively on the choice of the forcing function \citep{linkmann2020non}. Moreover, instability-driven turbulence can deviate significantly from Kraichnan’s picture of the inverse cascade and condensation. For instance, active flows typically do not display an inverse cascade, but form mesoscale vortices \citep{wensink2012meso}. Such coherent vortices are observed to form spontaneously in 2D turbulence driven by a negative eddy viscosity forcing \citep{gama1991two} and are often associated with screening \citep{jimenez2021collective,grooms2010model}. \avkrev{In nearly inviscid, inertial fluid flows, the resulting shielded vortices typically evolve into tripoles \citep{carton1989generation} consisting of a central vortex and two satellite vortices of opposite sign $180^\circ$ apart, as seen in both laboratory experiments \citep{van1991laboratory} and DNS \citep{orlandi1992numerical}.} In instability-driven 2D turbulence, the formation of such tripolar shielded vortices has been found to facilitate the spontaneous suppression of the inverse cascade \citep{van2022spontaneous}.

A particular challenge for flows driven by spectrally localised negative viscosities is that the resulting linear instability may grow without bound despite the presence of an advective nonlinearity. This fact was remarked upon in the context of early direct numerical simulations of 2D turbulence \citep{gama1991two,sukoriansky1996large}, and continues to be discussed in the context of geophysical fluid models with backscatter \citep{prugger2022geophysical}. In the presence of a nonlinear damping term, this unphysical unbounded growth is readily saturated. Different physical considerations may lead to such nonlinear damping terms depending on the application. In the context of active matter, a cubic damping term appears in the classical Toner-Tu model of flocking \citep{toner1998flocks}, derived from symmetry considerations and renormalization arguments. \NEW{The Toner-Tu model, which continues to be the subject of theoretical and numerical investigations \citep{gibbon2023analytical}}, was later adapted to the study of active fluid flows by the addition of a fourth-order spatial derivative term reminiscent of the Swift-Hohenberg equation \citep{dunkel2013minimal,dunkel2013fluid}. The resulting Swift-Hohenberg-Toner-Tu model describes active stresses in terms of a scale-localised negative viscosity and has been of great interest \citep{james2021emergence,puggioni2022giant,kiran2023irreversibility}. A review of recent progress based on these and other models of active turbulence is given by \cite{alert2022active}. In the geophysical context, many studies of nearly two-dimensional turbulence assume a linear Rayleigh drag law to model the effect of bottom friction \citep{boffetta2012two}. However, there is also a large body of work which considers a quadratic (turbulent) bottom drag law, see e.g. \cite{jansen2015parameterization} and \cite{gallet2020vortex}. Such a quadratic drag law may be obtained from dimensional considerations and is widely used in theoretical and numerical numerical ocean models \citep{gill1982atmosphere,willebrand2001circulation,egbert2004numerical,couto2020mixing}. 

\textcolor{black}{Following earlier work of \cite{jimenez2007spontaneous}}, a recent study \citep{van2022spontaneous} investigated 2D turbulence driven by a hybrid forcing that interpolates between a spectrally localised negative viscosity forcing and a random driving force acting on the same length scales while injecting energy at a constant rate. \avkrev{This combination of two well-established forcing mechanisms, each of which has separately led to fundamental insights into turbulence, allows for an exploration of new aspects of non-universality in 2D turbulence.} With a cubic nonlinear damping term to saturate the linear instability, extensive direct numerical simulations (DNS) by \cite{van2022spontaneous} revealed a number of transitions as the forcing function varies from stochastic to instability-like, from a large-scale condensate to a hybrid state consisting of large-scale circulation patterns with embedded mesoscale shielded vortices, and finally to a gas of shielded vortices characterised by a spontaneously broken symmetry, with all vorticity extrema in the core of the same sign at late times. \NEW{Here and in the following, we use the term \textit{mesoscale} to indicate a scale intermediate between the small forcing scales and the system size. This usage differs somewhat from the established definition in the geophysical literature. For instance, in the context of the Earth's atmosphere, the scales most unstable to baroclinic instability (comparable to the Rossby radius of deformation), are typically on the order of $1000$ kilometers (the \textit{synoptic scale} of weather systems), while the atmospheric mesoscales are substantially smaller (tens to hundreds of kilometers). In contrast, the oceanic \textit{mesoscale} is the analogue of the atmospheric synoptic scale  \citep{cushman2011introduction}.}

In the shielded vortex gas, the inverse energy cascade was found to be suppressed at large scales, while the number of shielded vortices in the domain slowly increases via a random nucleation process \citep{van2022spontaneous}. Owing to the significant numerical effort required to investigate the ultimate saturation of this process, the late-time limit of this slow evolution was not studied and remains an open problem. Here, we employ extensive DNS of this system with very long integration times to advance significantly beyond the results presented by \cite{van2022spontaneous} and investigate in detail\avkrev{, for the first time,} the late-time evolution of the broken symmetry shielded vortex gas state, revealing an approximately self-similar evolution towards a dense, statistically stationary state. This state is shown to persist over a wide range of forcing strengths, and to undergo a crystallisation transition at a critical parameter threshold. 

The remainder of this paper is structured as follows: in Sec.~\ref{sec:setup} we describe the numerical setup for our simulations, followed in Sec.~\ref{sec:late_time_near_gam1} by a discussion of the late-time evolution of the shielded vortex gas and the convergence to a statistically stationary state. Next, in Sec.~\ref{sec:crystallization} we describe the crystallisation transition observed at weak instability growth rates and quantify it using tools from statistical physics and crystallography. In Sec.~\ref{sec:vorticity_profile} we describe the dependence of the vorticity profile of the shielded vortices and the number of such vortices on the forcing parameters. In Sec.~\ref{sec:conditions_suppression_inverse_cascade} we discuss the conditions required to suppress the expected inverse cascade, thereby leading to the vortex crystal state, while in Sec.~\ref{sec:overview_state_space} we provide an overview of the state space including all stable stationary solutions we have identified in the system. The paper concludes in Sec.~\ref{sec:conclusions} with a discussion of our results in the context of existing and future research on instability-driven turbulence.

\section{Setup}
\label{sec:setup}
We study the 2D Navier-Stokes equation governing the evolution of an incompressible velocity field $\mathbf{u}=(u,v)$ on the flat torus $[0,2\pi]^2$ with nonlinear damping, hyperviscosity and a hybrid forcing function $\mathbf{f}_\gamma$, namely
\begin{align}
    \partial_t  \ub + \ub \cdot \nabla \ub =& - \nabla p - \nu_n (-\nabla^2)^n \ub - \beta |\ub|^m \ub + \mathbf{f}_\gamma, \label{eq:nse}\\
    \nabla\cdot \ub =& ~ 0 \label{eq:inc},
\end{align}
where the integers $n$ and $m$ control the order of the hyperdiffusion and damping operators, respectively, with $n,m\geq1$, and
\begin{equation}
    \mathbf{f}_\gamma=   \gamma \mathcal{L} [\mathbf{u}] + (1-\gamma)\mathbf{f}_\epsilon.\label{eq:forcing}
\end{equation}
Here, the forcing control parameter $\gamma\in[0,1]$, and $\mathcal{L}[\mathbf{u}]$ is a linear operator whose Fourier transform is given by
\begin{eqnarray}
    \widehat{\mathcal{L}[\mathbf{u}]}(\mathbf{k}) = \nu_* k^2 \hat{\mathbf{u}}(\mathbf{k}),\quad \nu_*>0, \label{eq:lin_force}
\end{eqnarray}
for wavenumbers $\mathbf{k}$ in the annulus $k=|\mathbf{k}| \in [k_1,k_2]$, and $ \widehat{\mathcal{L}[\mathbf{u}] }(\mathbf{k})=0$ otherwise. We denote the largest length scale in the forcing range by $\ell_1\equiv 2\pi/k_1$. This linear forcing term is associated with a maximum growth rate $\sigma\equiv\nu_* k_2^2$. \NEW{An important nondimensional number characterising this system is the ratio
\begin{equation}
r(\gamma) = \frac{\gamma\sigma}{\nu_n k_2^{2n}}
\end{equation}
between the maximum instability growth rate, which occurs at the wavenumber $k=k_2$, and the rate of \avkrev{hyperviscous energy} dissipation rate at that wavenumber. We choose $\nu_*$ such that $r(\gamma)$ varies from $r(0)=0$ to $r\gg 1$ as $\gamma$ increases from $0$ to $1$ (in all the runs described below, we take $\nu_*=0.002$, $k_2=40$, $n=4$, $\nu_4=10^{-14}$, such that $r(\gamma=1) \approx 48.8$, see Table \ref{tab:overview_all_runs} for details of the parameters used). We note that the ratio $r$ has also been identified as a key control parameter in models of active turbulence \citep{linkmann2019phase,linkmann2020condensate}, where the case $n=1$ (regular viscosity) was considered.} The second term in (\ref{eq:forcing}) involves the solenoidal zero-mean white-in-time stochastic force $\mathbf{f}_\epsilon(\mathbf{x},t)$ with random phases acting within a thin shell of wavenumbers centered on the most linearly unstable wavenumber $k=k_2$, injecting kinetic energy at a rate $\epsilon$.

We record the domain-averaged kinetic energy (density) $E\equiv\langle \mathbf{u}^2\rangle$ and the enstrophy (density) $\Omega\equiv\langle \omega^2 \rangle$, where $\langle \cdot \rangle$ denotes the domain average, as well as the vorticity $\omega\equiv\partial_x v - \partial_y u$. A further important quantity used to characterise the structure of fluid flows is the energy spectrum $E(k)$, defined by 
\begin{equation}
    E(k) = \sum\limits_{\mathbf{q}: k-1/2\leq |\mathbf{q}|<k+1/2} |\hat{\mathbf{u}}(\mathbf{q})|^2
\end{equation}
which characterises the distribution of energy across scales in terms of the Fourier transform of the velocity field $\hat{\mathbf{u}}(\mathbf{q})$. The system defined above is further characterised by two dissipation-related nondimensional parameters
\begin{align}
    Re_n = U_{\rm rms}L_I^{2n-1}/\nu_n,  \hspace{1cm} Re_{\beta,m} = \frac{1}{\beta (U_{\rm rms})^{m-1} L_I},
\end{align}
with the r.m.s. velocity $U_{\rm rms}=\sqrt{\langle \mathbf{u}^2\rangle}$ and the integral length scale $L_I$, defined spectrally as $L_I = \sum_{k} \frac{2\pi}{k} E(k)/E$. Note that the Reynolds numbers thus defined can only be evaluated {\it a posteriori}. In addition, the problem depends on the {\it a priori} parameter $\gamma$, which controls the relative amplitude of the random and deterministic forcing terms. 
An alternative nondimensional but \textit{a posteriori} parameter $\Gamma$ can be defined by the ratio of the energy injection rates $\gamma \sigma U_{\mathrm{rms}}^2$ and $(1-\gamma)^{2}\epsilon$ associated with the deterministic and stochastic forces, respectively:
\begin{align}
    {\Gamma} = \frac{\gamma\sigma U_{\mathrm{rms}}^2}{(1-\gamma)^{2}\epsilon}.
    \label{eq:Gamma}
\end{align}
When this parameter is large, the instability forcing provides the dominant contribution to the energy injection.

\begin{table}
    \centering
    \footnotesize
    \begin{tabular}{|c|c|c|c|c|c|c|c|c|}
         Set & \#runs & $\gamma$&$Re_n$ & $Re_{\beta,m}$ & Initial condition & Setup\\ \hline 
         A & 49  & $0-1.0$ & $3.2\times $$10^{11}$--$1.2\times$$ 10^{20}$ & $216$--$1360$ & random small-amp. & standard \\  \hline
         B & 66  & $0-1.0$ & $3.2\times $$10^{11}$--$1.2\times$$10^{20}$ & $216$--$3600$ &  vortex gas/crystal & standard\\ \hline
         C & 4 &  $0.05-1.0$ & $8.2\times$$10^{15}$--$2.7$$\times 10^{20}$ & $1.7$--$8.5$ & small-amp. & $m=1$ \\ \hline 
         D & 10 & $0.05$ & $1.2\times$$ 10^{11}$--$4.5\times$$10^{11}$& -- -- &vortex crystal & filtered damping, $m=2$ \\ \hline
         E & 10  & $0.95$& $2.2\times10^{11}$ & $1260$ &random small-amp. & $\epsilon=0$ \\ \hline
         F & 5  & $0.005-0.035$& $O(10^{15})-O(10^{17})$ & $\lesssim 0.2$ &vortex crystal & $\epsilon=0$ \\ \hline
    \end{tabular}
    \caption{Summary of the runs performed in this study. All runs were done at a moderate resolution, $n_x=n_y=512$, to facilitate long-time integration. The parameters for the standard setup read $\nu_* = 0.002$, $k_1=33$, $k_2=40$, $\epsilon=1$, $\beta=10^{-4}$, $n=4$, $\nu_4 = 10^{-14}$. Runs in set A are initialised from small-amplitude, random initial conditions, while runs in set B are initialised in the vortex gas state obtained in set A or vortex crystal states obtained from that by continuation in $\gamma$. In set C, the cubic damping term is replaced by a quadratic term {$\beta |{\bf u}|{\bf u}$ with $\beta=0.1$, all other parameters remaining the same}. In set D, the cubic damping is spectrally filtered in order to systematically assess its importance for the maintenance of the vortex crystal. The runs in set E are identical to runs in set A, but with the random forcing set to zero. Set F similarly repeats runs from set B with the random forcing term set to zero. Reynolds numbers given refer to the stationary state (except for set $F$) and indicate the range observed within each set.}
    \label{tab:overview_all_runs}
\end{table}

Equations (\ref{eq:nse})--(\ref{eq:forcing}) were solved using the MPI-parallelised, pseudospectral code GHOST (Geophysical High-Order Suite for Turbulence), cf. \cite{mininni2011hybrid}. The 2/3 rule was used for dealiasing. \avkrev{We ensure that all simulations are well resolved by checking that the enstrophy dissipation rate $D_\Omega(k) = \nu_n k^{2n+2}E(k)$ decays towards the grid scale.} A total of $144$ distinct simulations were performed, requiring around 1 million CPU hours in total. The runs are organised in six sets as described in Table \ref{tab:overview_all_runs}. Cubic damping ($m=2$) is considered in all runs except those in set C, which is shown in Appendix \ref{sec:app_quad_drag} to yield states that are qualitatively similar to those discussed here.  All runs were done at a moderate resolution $n_x=n_y=512$ to facilitate the long-time integration required to observe the phenomena of interest here. The longest runs performed for this work (in set B) lasted about $40,000$ time units measured in terms of the time scale $\sigma^{-1}$ associated with the linear instability growth rate, corresponding to a walltime of about 90 days. In sets E and F, the random forcing amplitude is set to zero to isolate the impact of the random force on the observed solutions.

\section{Late-time evolution near $\gamma=1$}
\label{sec:late_time_near_gam1}

\begin{figure}
    \centering
    \includegraphics[width=0.49\textwidth]{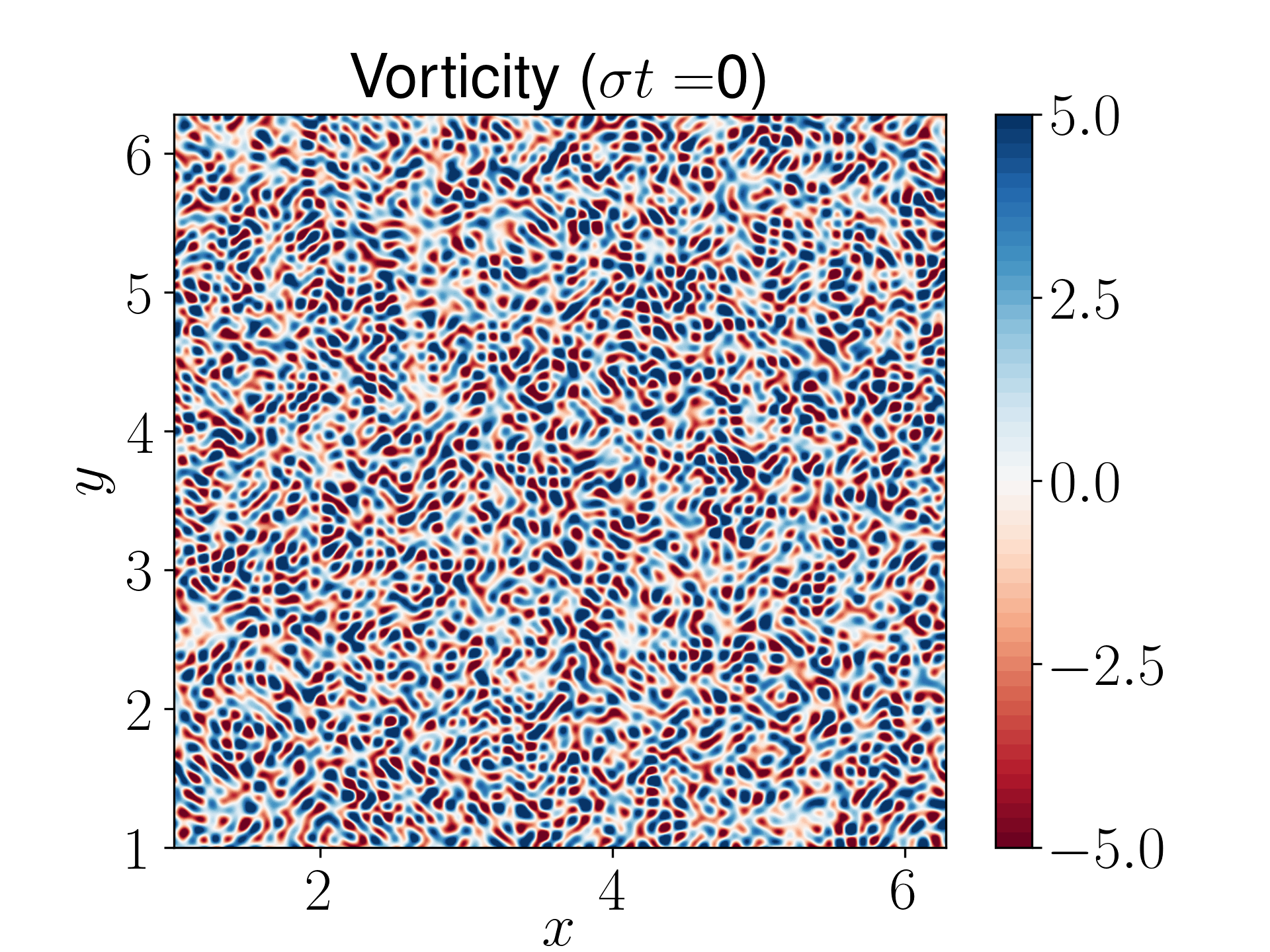}
    \includegraphics[width=0.49\textwidth]{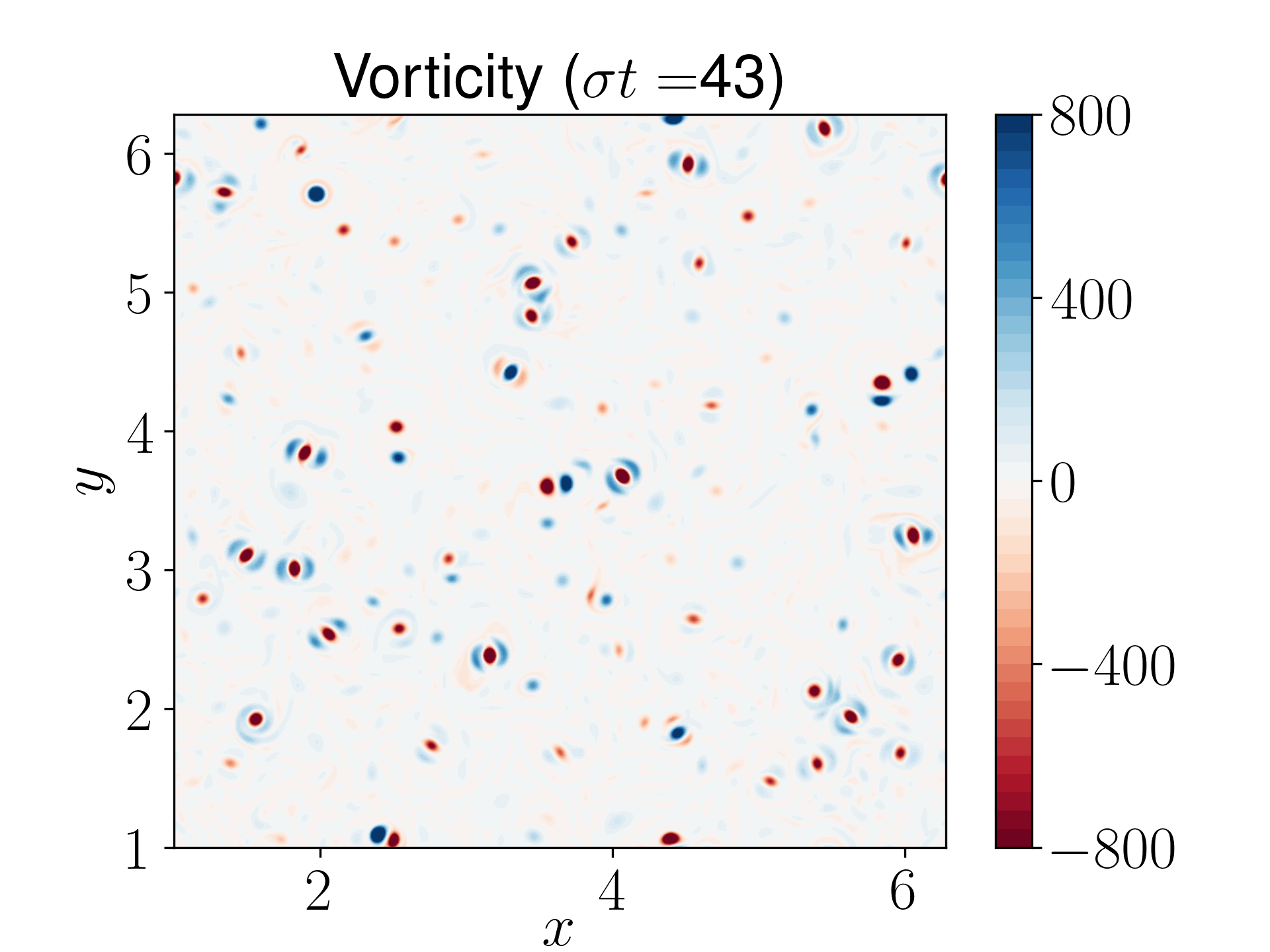}
    \includegraphics[width=0.49\textwidth]{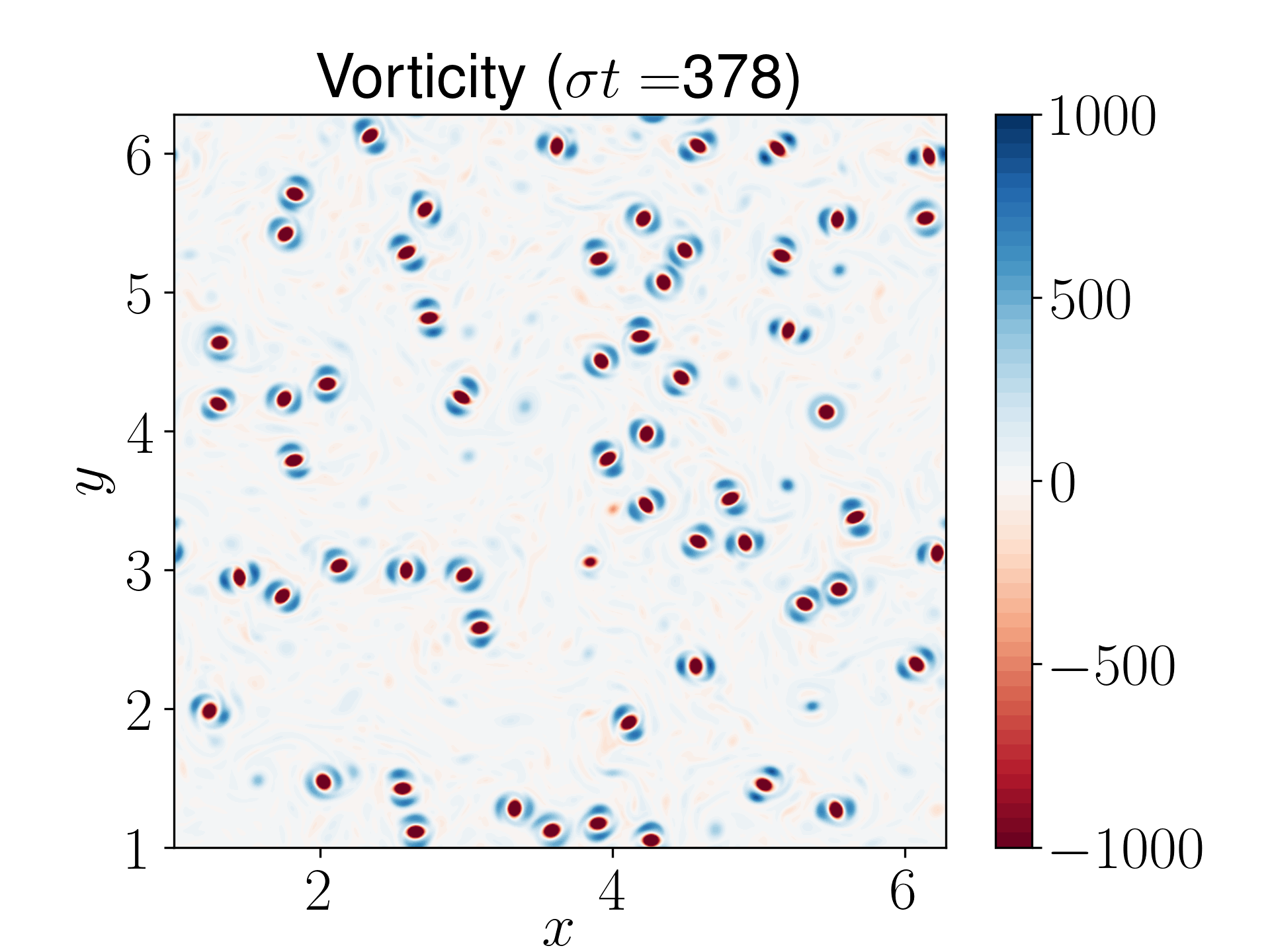}
    \includegraphics[width=0.49\textwidth]{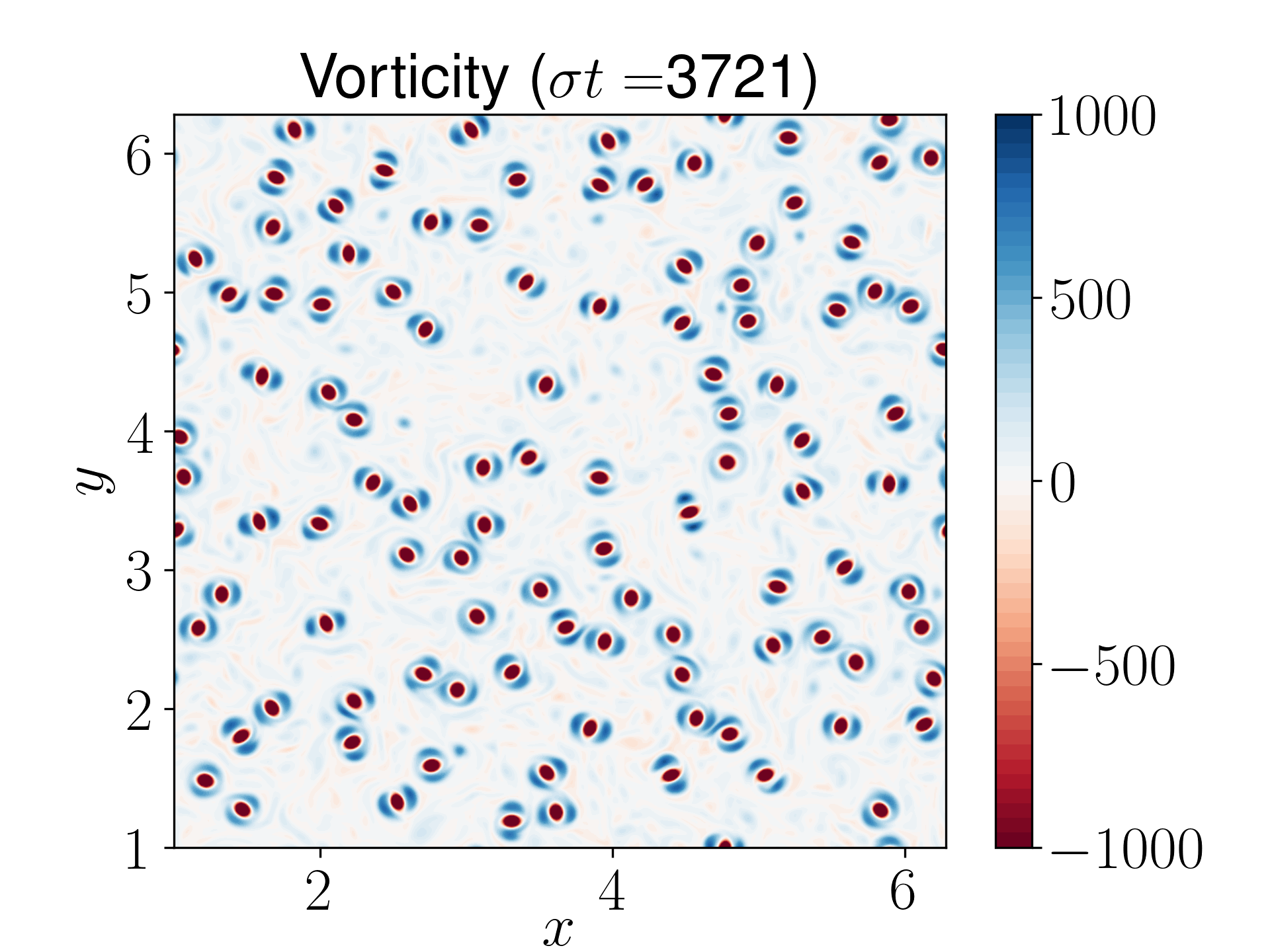}
    \includegraphics[width=0.49\textwidth]{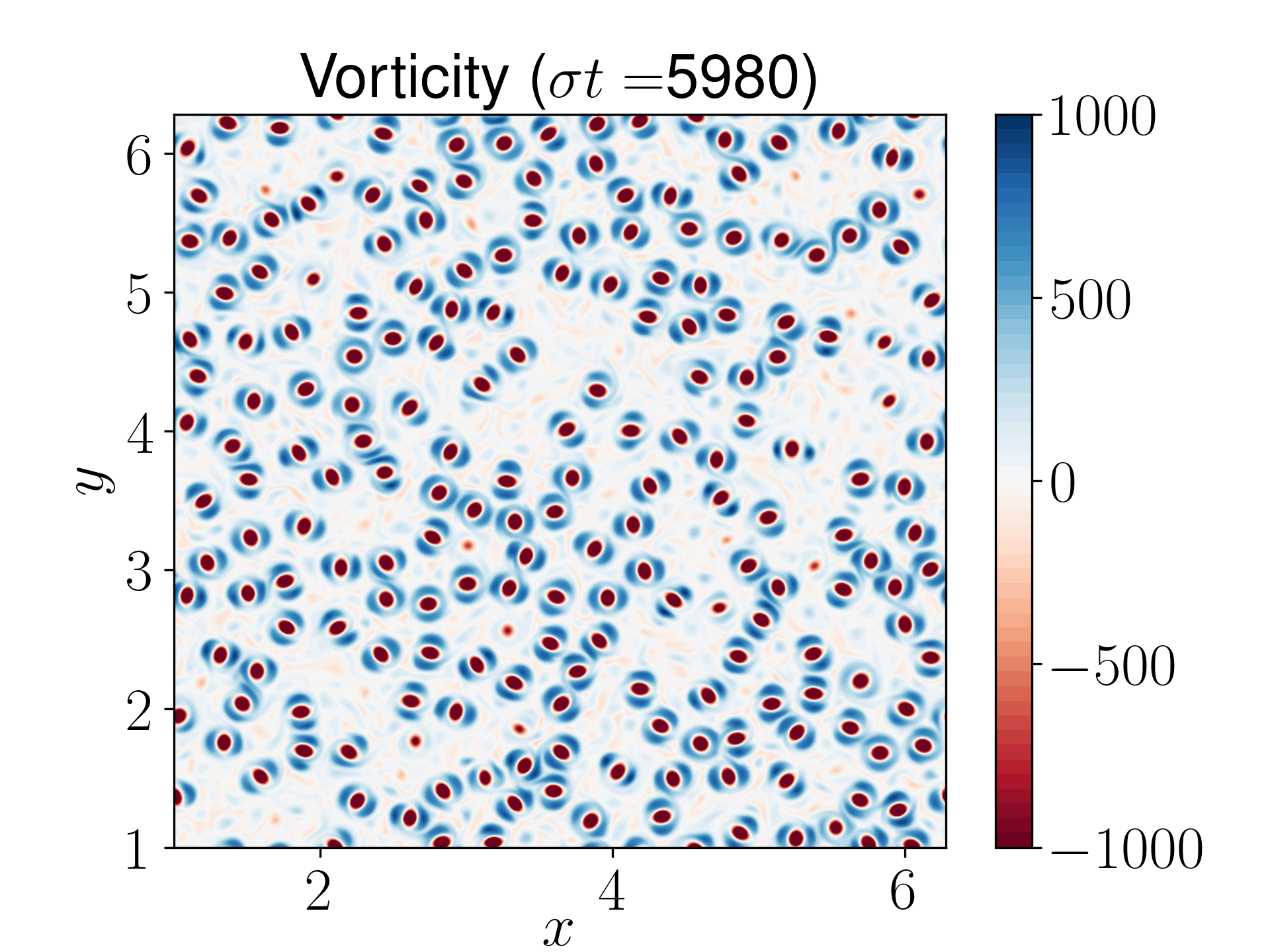}
    \includegraphics[width=0.4875\textwidth]{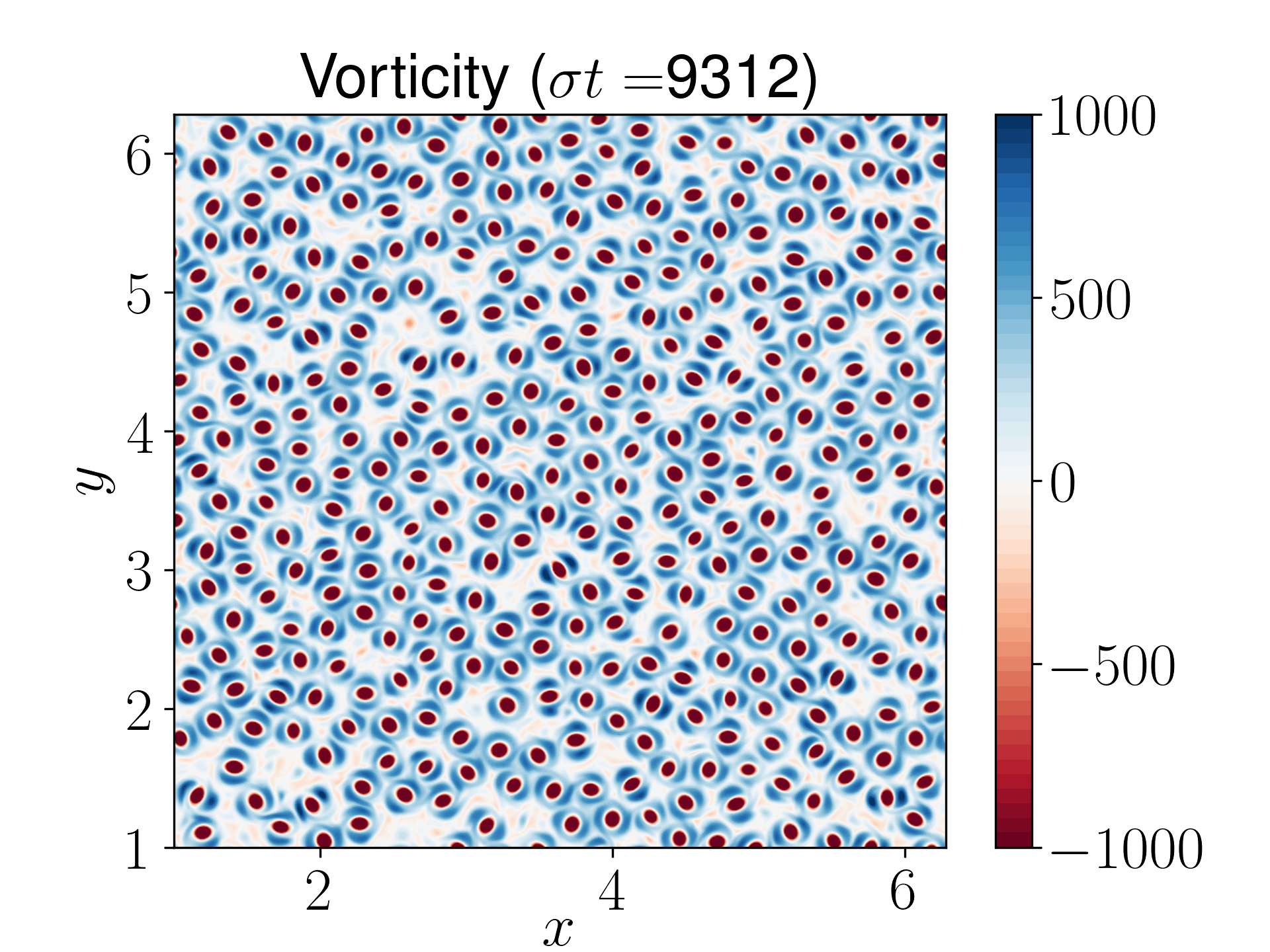}
    \caption{Visualization of the vorticity field at different times for $\gamma=1$, showing the evolution from small-amplitude, random initial conditions through a dilute to a dense vortex gas.}
    \label{fig:states_encountered}
\end{figure}

\begin{figure}
 \hspace{0.2\textwidth} (a) \hspace{0.45\textwidth} (b)\\
    \centering
    \includegraphics[width=0.49\textwidth]{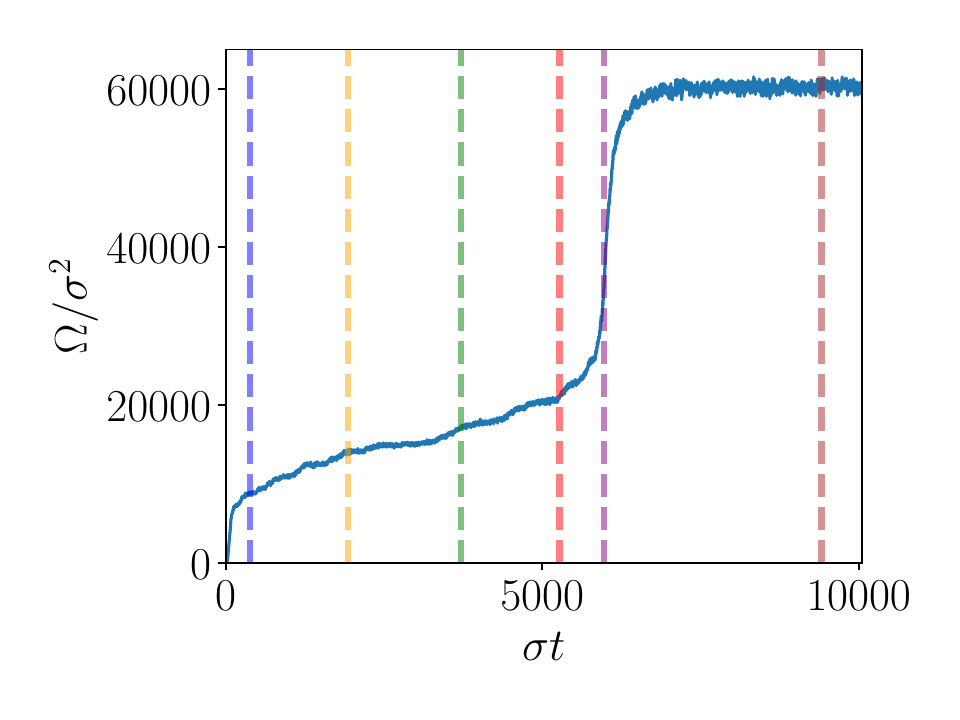}
    \includegraphics[width=0.49\textwidth]{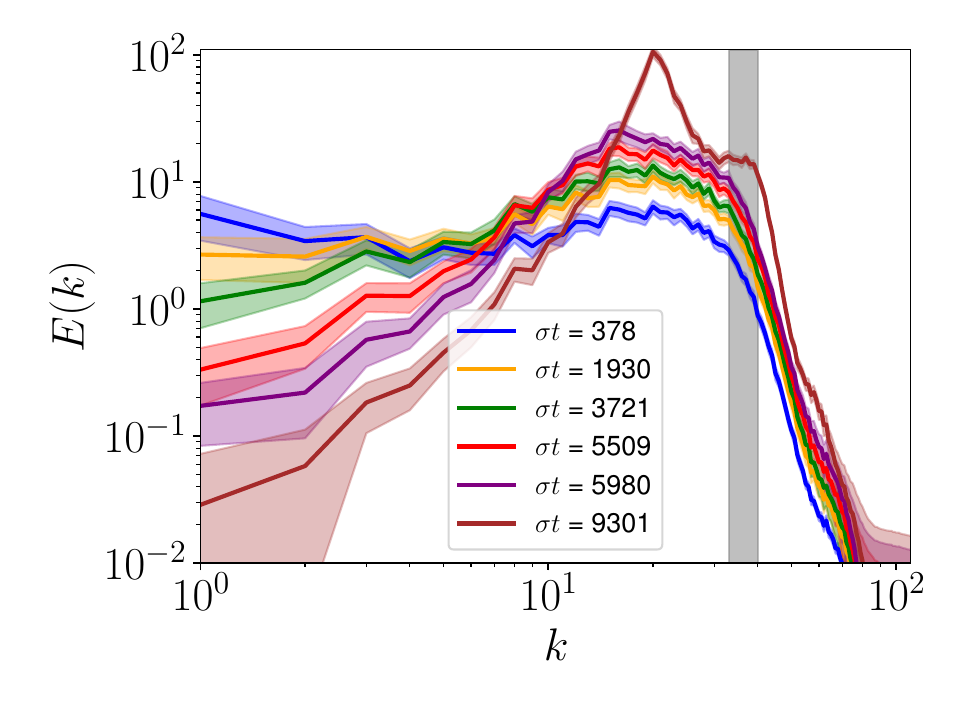}
    \caption{Panel (a): Time evolution of the enstrophy from small-amplitude random initial conditions when $\gamma=1$ (case A). Panel (b): Color-coded log-log plots of the energy spectrum versus wavenumber at the times indicated by dashed vertical lines in the left panel. \avkrev{The shaded envelopes indicate one standard deviation of the spectrum about the mean, computed over 50 snapshots.} The grey shaded region indicates the forcing range.}
    \label{fig:spectra}
\end{figure}

\begin{figure}
 \hspace{0.2\textwidth} (a) \hspace{0.45\textwidth} (b)\\
    \centering
    \includegraphics[width=0.49\textwidth]{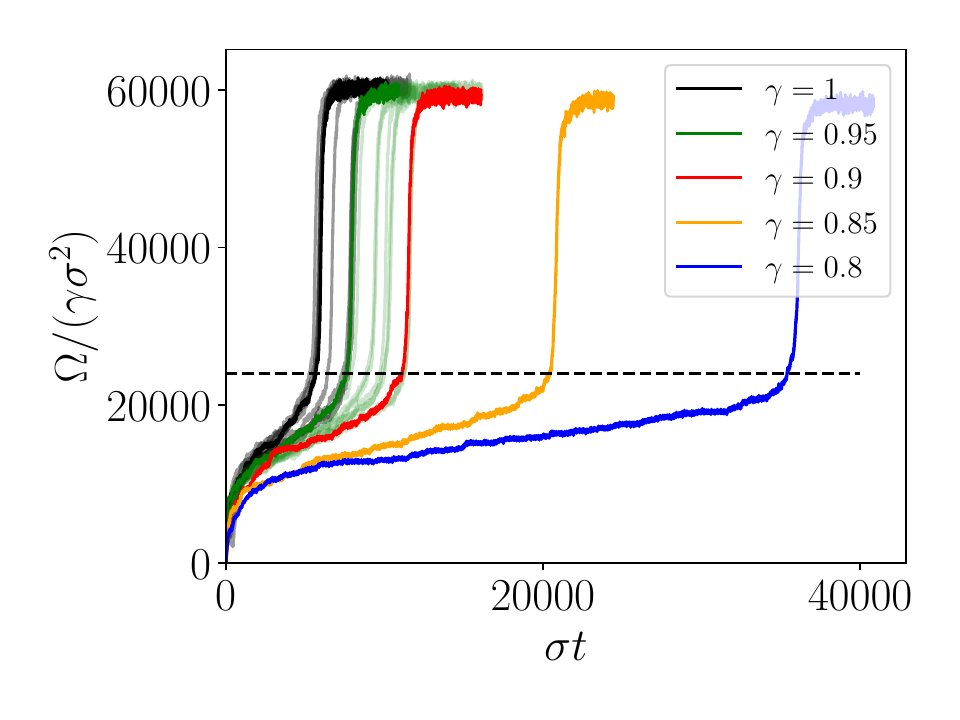}
    \includegraphics[width=0.49\textwidth]{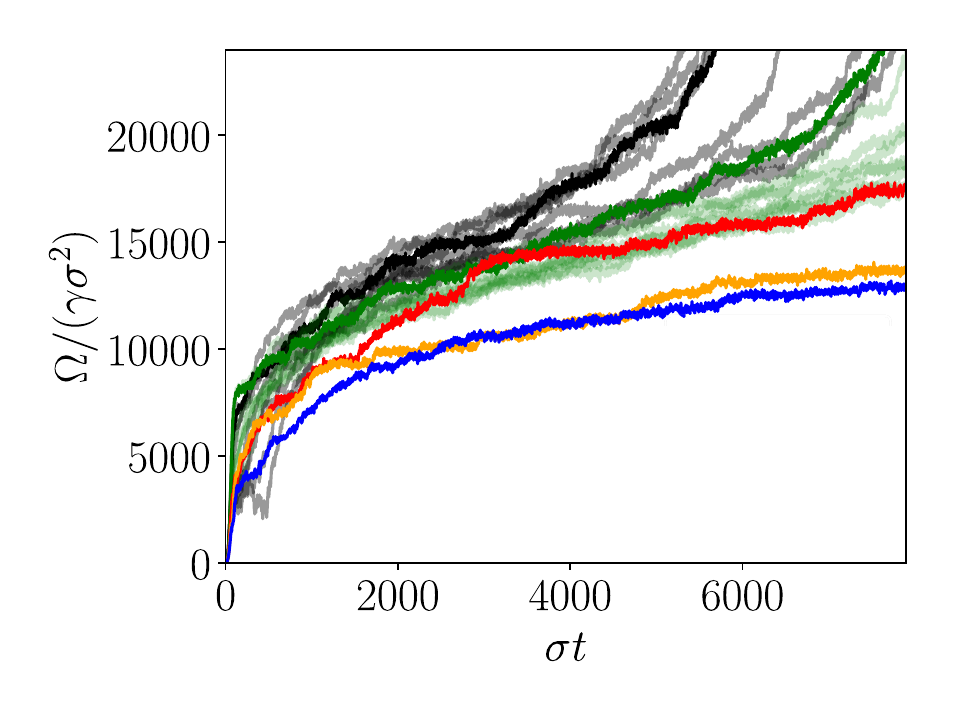}
     \hspace{0.2\textwidth} (c) \hspace{0.45\textwidth} (d)\\
    \includegraphics[width=0.49\textwidth]{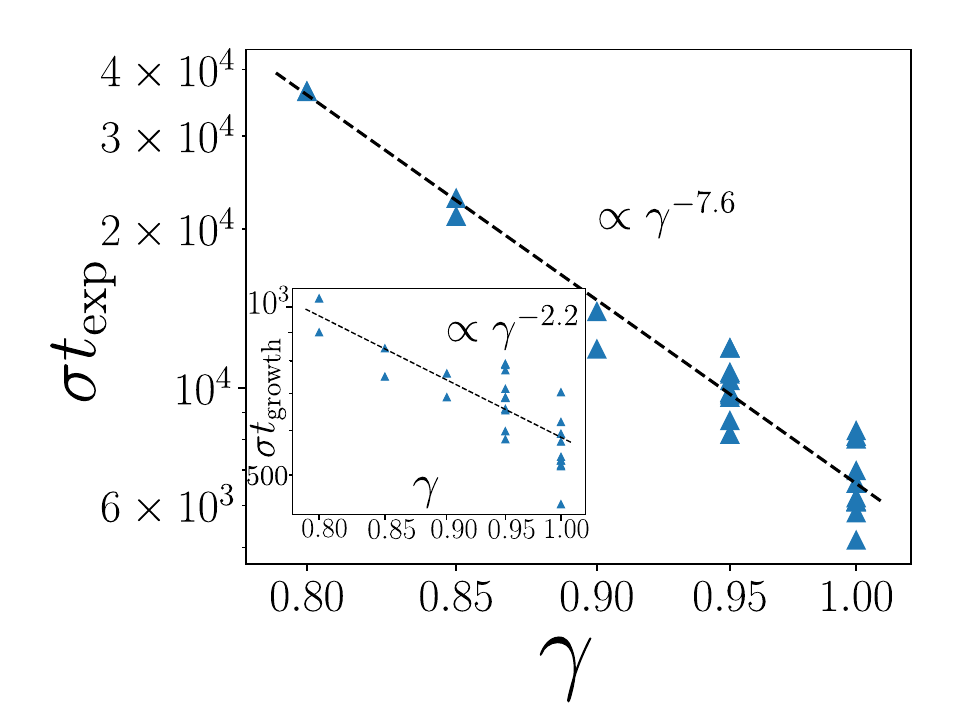}
    \includegraphics[width=0.49\textwidth]{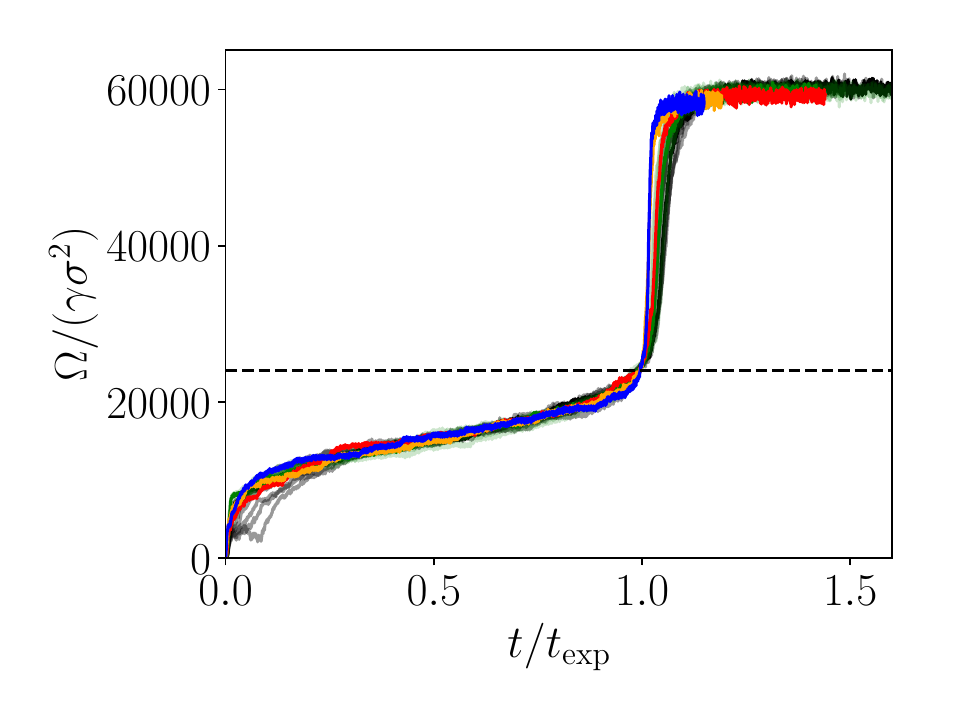}
    \caption{Panel (a): Nearly self-similar evolution of the enstrophy from small-amplitude random initial conditions to the dense vortex gas. The final enstrophy value in the stationary state scales approximately linearly with $\gamma$, while the time scale depends on $\gamma$ highly nonlinearly. Low-opacity curves at $\gamma=1$ and $\gamma=0.95$ indicate an ensemble of 10 runs performed at each of these values. \NEW{The horizontal dashed line indicates the enstrophy threshold for the onset of the rapid growth phase.} Panel (b): Zoom on early times highlighting the stochastic nature of the evolution and the deviations between different ensemble members. Panel (c): \NEW{Nondimensional time $\sigma t_{\rm exp}$ at which the explosive growth phase begins}, versus $\gamma$. An empirical power law with an exponent between $-7$ and $-8$ is observed. \NEW{Inset shows nondimensional duration $\sigma t_{\rm growth}$ of the explosive growth phase versus $\gamma$, where $t_{\rm growth}$ is defined as the difference between $t_{\rm exp}$ and the time required for the enstrophy to reach $90\%$ of its maximum value at a given $\gamma$. The results show a significantly weaker dependence of $t_{\rm growth}$ on $\gamma$ compared to $t_{\rm exp}$.} Panel (d): Near self-similarity is verified by replotting the data from the top left panel \NEW{against a time rescaled by $t_{\rm exp}$, with colors being consistent between the two panels. Since $t_{\rm exp}$ increases with $\gamma$ significantly faster than $t_{\rm growth}$, the rapid growth phase appears to sharpen as $\gamma$ decreases under this rescaling.}}
    \label{fig:late_time_enstrophy}
\end{figure}

As stated in the introduction, \NEW{the shielded vortex gas state described by \cite{van2022spontaneous} was only followed into a dilute but transient state in which the number of vortices slowly grew owing to random vortex nucleation.} In this section, we utilise much increased computational resources to study the long-time evolution of the system, based on runs from set A, as it converges to a statistically stationary state. Figure~\ref{fig:states_encountered} shows snapshots of the vortex gas in the pure instability-driven case ($\gamma=1$) at different times, nondimensionalised by the instability growth rate $\sigma$. Starting from random, small-amplitude initial conditions, a short-lived inverse cascade is followed by the emergence of shielded vortices of both parities ($\sigma t = 6.3$ in Fig.~\ref{fig:states_encountered}). In a stochastic competition between the two species, one is eventually eliminated leading to spontaneous symmetry breaking, as discussed by \cite{van2022spontaneous}. This is clearly seen at $\sigma t=378$ in Fig.~\ref{fig:states_encountered}. As time increases further, the number of vortices increases. The last snapshot, at $\sigma t=9312$, corresponds to the statistically stationary state. Upon inspection of Fig.~\ref{fig:states_encountered}, the coherent vortices are seen to be tripolar, with an elliptical core and two satellites $180^\circ$ apart. As shown by \cite{van2022spontaneous}, these tripolar vortices are shielded, meaning that the circulation generated by any given vortex becomes small beyond a finite radius, located close to the edge of its satellites and comparable to the largest forcing scale.

The corresponding time evolution of the enstrophy (defined in Sec.~\ref{sec:setup}) is shown in Fig.~\ref{fig:spectra}(a). \NEW{This quantity is closely related to the number of vortices in this system, as shown by \cite{van2022spontaneous}.} Four distinct phases can be identified: an initial, rapid increase of the enstrophy from small-amplitude initial conditions, associated with a short-lived inverse cascade, followed by a phase of slower, approximately linear, growth of enstrophy with time. The latter corresponds to random nucleation of new vortices in the  background turbulence, depicted in Fig.~\ref{fig:states_encountered}. When the enstrophy reaches around \textcolor{black}{$\Omega/\sigma^2\approx 2.5\times10^4$ (corresponding to approximately 140 vortices in a domain of area $(2\pi)^2$)}, a phase of explosive growth sets in, where the number density of vortices increases rapidly. Finally, a statistically stationary state is reached whose enstrophy is larger by a factor of about $2.5$ than the enstrophy threshold at which the rapid growth was seen to set in. It should be emphasised that the observed increase in enstrophy is due to an increasing number of vortices since the core of any given vortex remains at constant vorticity \NEW{due to a local balance between forcing and nonlinear damping}. A similar transient evolution towards a vortex crystal in the Swift-Hohenberg-Toner-Tu model of active fluids is described by \cite{PhysRevFluids.3.061101}. \avkrev{However, in this system the enstrophy of the final state is much smaller as a consequence of stronger nonlinear damping relative to the linear forcing strength and the time scale separation between the slow nucleation and the rapid explosive growth is therefore much less pronounced.}

Figure~\ref{fig:spectra}(b) shows a log-log plot of the energy spectrum $E(k)$ versus the wavenumber $k$ at the times highlighted in panel (a) by vertical dashed lines. \avkrev{The spectrum shown is averaged over 50 consecutive snapshots, with the shaded envelope indicating one standard deviation around the mean.} At the earliest time illustrated, $\sigma t =378$, the energy spectrum has a local maximum at the largest scale, a remnant of the short-lived early-time inverse cascade.
As time passes, the kinetic energy in the large scales continuously decreases and a sharp local maximum appears at an intermediate scale, \textcolor{black}{approximately twice the forcing scale,} and corresponding to the scale of the individual vortices.

Figure~\ref{fig:late_time_enstrophy}(a) shows the nondimensional enstrophy (compensated by $\gamma$) versus the nondimensional time $\sigma t$ for $\gamma=0.8$, 0.85, 0.9, 0.95, and 1. In addition to the run at $\gamma=1$ already shown in Fig.~\ref{fig:spectra}, we generated an ensemble of $10$ runs which differed only in the phases of their random, small-amplitude initial conditions. Similarly, at $\gamma=0.95$, we performed $10$ additional runs which also differed in the phases of their random, small-amplitude initial conditions and in the realisation of the stochastic forcing (by construction, no stochastic forcing is present at $\gamma=1$). Several things can be gleaned from Fig.~\ref{fig:late_time_enstrophy}(a). 
First, \textcolor{black}{the explosive nucleation of new vortices seen in Fig.~\ref{fig:spectra}(a) is triggered when the enstrophy reaches $\Omega\sim 0.4\Omega_{\rm max}$, in terms of the enstrophy $\Omega_{\rm max}$ attained in the stationary state, regardless of the value of $\gamma$ (horizontal dashed line in Fig.~\ref{fig:late_time_enstrophy}(a)). In fact, random vortex nucleation was observed for $\gamma\gtrsim 0.6$ \citep{van2022spontaneous}, but the late-time dynamics for $0.6<\gamma<0.8$ remained numerically inaccessible owing to the excessively long simulation time required to reach the final statistically stationary state at these parameter values.} Second, the enstrophy in the statistically stationary state scales linearly with $\gamma$ to a good approximation between $\gamma=0.8$ and $\gamma=1$. This is consistent with the dominant balance between the instability forcing term, which is linear in the velocity, and the cubic \NEW{damping} term in Eq.~(\ref{eq:nse}). \textcolor{black}{In contrast, the time scale $t_{\rm exp}$ required for the system to reach the explosive phase exhibits a nontrivial, sensitive dependence on $\gamma$ whose origin remains unclear.}
For the parameters considered here, the threshold configuration (shown in Fig.~\ref{fig:states_encountered} at $\sigma t=5980$) contains approximately $140$ vortices in a domain of area $(2\pi)^2$, see also Fig.~\ref{fig:number_vortices}. We note that the mean distance $d_{NN}$ between the vortex centers of nearest neighbours (computed by finding the nearest neighbour of any given vortex, and averaging over the population and over time) close to this threshold is approximately $d_{NN}\approx 2\ell_1$, which is somewhat larger than the integral scale (defined in the introduction), which is given by $L_I\approx 1.4\ell_1$ (with $\ell_1=2\pi/k_1$).


Figure~\ref{fig:late_time_enstrophy}(b) shows a zoom on the early phase of the evolution, highlighting the stochasticity of the nucleation process. Panel (c) shows the \NEW{time $t_{\rm exp}$ required to reach $40\%$ of the statistically stationary state enstrophy, where the phase of explosive growth is triggered, for all the simulations shown in panel (a), with the dashed line indicating a power-law fit giving an empirical exponent approximately equal to $-7.6$.} To date, no theoretical argument for such a power-law dependence has been identified. \NEW{The inset in panel (c) shows the duration $t_{\rm growth}$ of the explosive growth phase as a function of $\gamma$
for the same runs. This time depends on $\gamma$ less strongly than $t_{\rm exp}$,
with a power-law fit with an approximate exponent of $-2.2$ although the data shows a significant spread within ensembles.} Figure~\ref{fig:late_time_enstrophy}(d) shows the same data as panel (a) but with time rescaled by $t_{\rm exp}$, confirming the near self-similarity of the nucleation process. \NEW{Since the duration $t_{\rm growth}$ of the explosive growth phase is not proportional to $t_{\rm exp}$, the collapse during the latter phase is imperfect.} Moreover, the scaling of the stationary-state enstrophy with $\gamma$ is seen to be satisfied only approximately.

The increasingly slow nucleation of new shielded vortices as $\gamma$ decreases is a reflection of time scale competition. As $\gamma$ decreases, the time scale for the generation of a new vortex by the linear instability increases. The background turbulence, taking place in the interstitial space between the vortices already present, generates shear which disrupts the formation of new vortices and hence it may be expected that vortex nucleation slows down as $\gamma$ decreases. \NEW{For different values of $\gamma$, we measured the average strain rate $\|\mathbf{D}\| \equiv \sqrt{\mathrm{tr}(\mathbf{D} \mathbf{D}^T)}$, where $\mathbf{D} \equiv \frac{1}{2}(\nabla \mathbf{u} + (\nabla \mathbf{u})^T)$ is the rate of strain tensor, and the average mean-square vorticity of the turbulence in the interstitial space of the dilute vortex gas (not shown). We found that both of these quantities are much larger than the maximum instability growth rate, indicating that nucleation of new shielded vortices from this turbulent background is indeed a rare event.}
In principle, one may hope to deduce the dependence of $t_{\rm exp}$ on $\gamma$ from these considerations. However, the fact that the interstitial turbulence is no longer homogeneous owing to the embedded coherent, shielded vortices complicates the picture.

\begin{figure}
 \hspace{0.2\textwidth} (a) \hspace{0.45\textwidth} (b)\\
    \centering
    \includegraphics[width=0.49\textwidth]{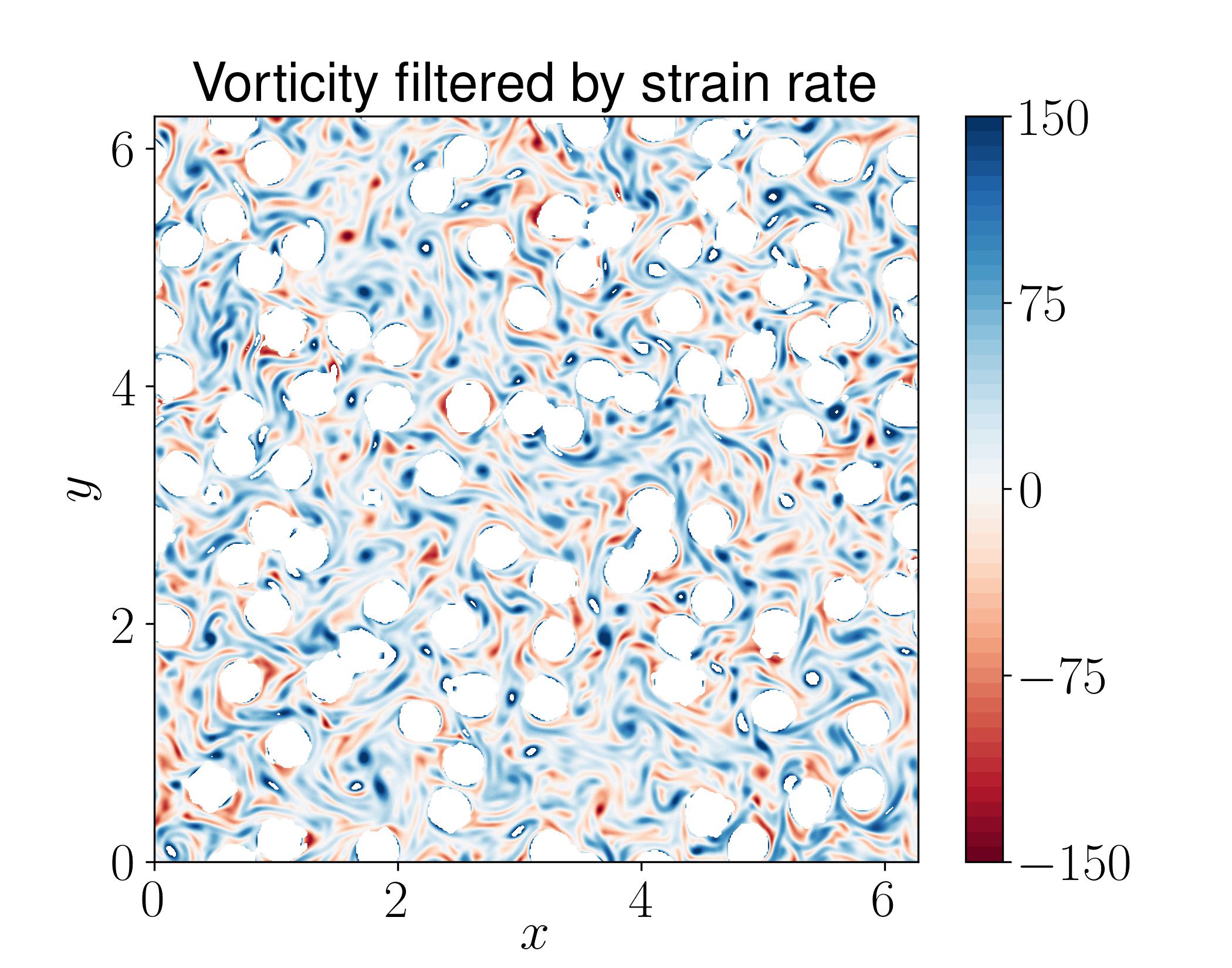}    \includegraphics[width=0.49\textwidth]{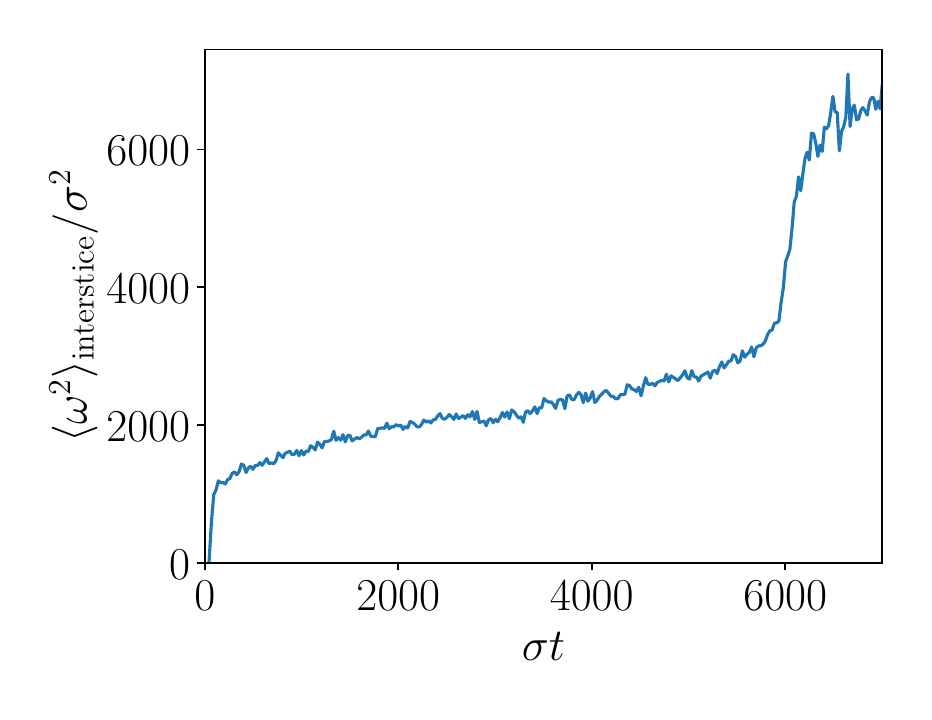}
    \caption{Panel (a): Snapshot of the vorticity field at $\gamma=1$ during the random nucleation process, highlighting the interstitial vorticity field by filtering out regions where the strain amplitude $\|\mathbf{D}\|$ (defined in the main text) is larger than $5\%$ of its maximum value (\NEW{we have also tested other threshold values and found qualitatively the same results}). High-amplitude vorticity fluctuations are preferentially found in the vicinity of shielded vortices. Panel (b): Time series of the mean squared vorticity at $\gamma=1$, averaged over the interstitial space between the coherent vortices, computed from the vorticity field shown in panel (a). The mean squared interstitial vorticity increases in time in a manner reminiscent of the full enstrophy shown in Fig.~\ref{fig:late_time_enstrophy}, with the observed interstitial values significantly smaller than the total enstrophy that is dominated by the coherent vortices. The growth in interstitial vorticity indicates that within the shrinking gaps between the coherent vortices, vorticity fluctuations increase in strength over time.  }
    \label{fig:filtered_vorticity_field}
\end{figure}

\NEW{As illustrated by Fig.~\ref{fig:filtered_vorticity_field}(a), there is a tendency for large-amplitude vorticity fluctuations to occur in the vicinity of coherent vortices. This is likely a manifestation of the coherent vortices imparting vorticity to their vicinity through filamentation or diffusion. Therefore, during the random nucleation process, as the coherent vortices increase in number, and the interstitial space is reduced, the amplitude of vorticity fluctuations in the interstices increases. This is confirmed by the time evolution of the mean-squared interstitial vorticity, shown in Fig.~\ref{fig:filtered_vorticity_field}(b). The mean-squared vorticity in the interstices is indeed seen to increase over time during the random nucleation process. At the threshold of around two fifths of the final enstrophy, a value that appears to be universal across $\gamma$, the interstices are sufficiently reduced in size that high-amplitude vorticity fluctuations can rapidly develop, serving as seeds for new shielded vortices, allowing rapid nucleation of the remaining three fifths of the final number of vortices {(note that only vortices of the same sign can mature in the vortex gas past the initial stage of spontaneous symmetry breaking, since vortices of the opposite sign undergo destructive interactions as discussed by \cite{van2022spontaneous}).} 
This type of explosive nucleation resembles behavior observed in systems undergoing crowd synchronization via quorum sensing when a large number of dynamical elements communicate with each other via a common information pool \citep{strogatz,raj}, here the interstitial vorticity. 
\avkrev{Although the initial spontaneous symmetry breaking occurs rapidly compared to the slow nucleation process in the runs discussed here, at smaller values of $\gamma$ there can be significant transients during which vortices of both signs coexist in the domain \citep{van2022spontaneous}. This is compatible with the findings of \cite{james2021emergence} in active turbulence at moderate Reynolds numbers, who report similar transients with subdomains of locally aligned vortices, whose duration grows with the size of the domain. In the present work we focus instead on the behaviour at substantially larger Reynolds numbers in a large domain of a fixed size, leaving the domain dependence to future work.}}
\textit{}

\begin{figure}
    \centering
    \includegraphics[width=0.5\textwidth]{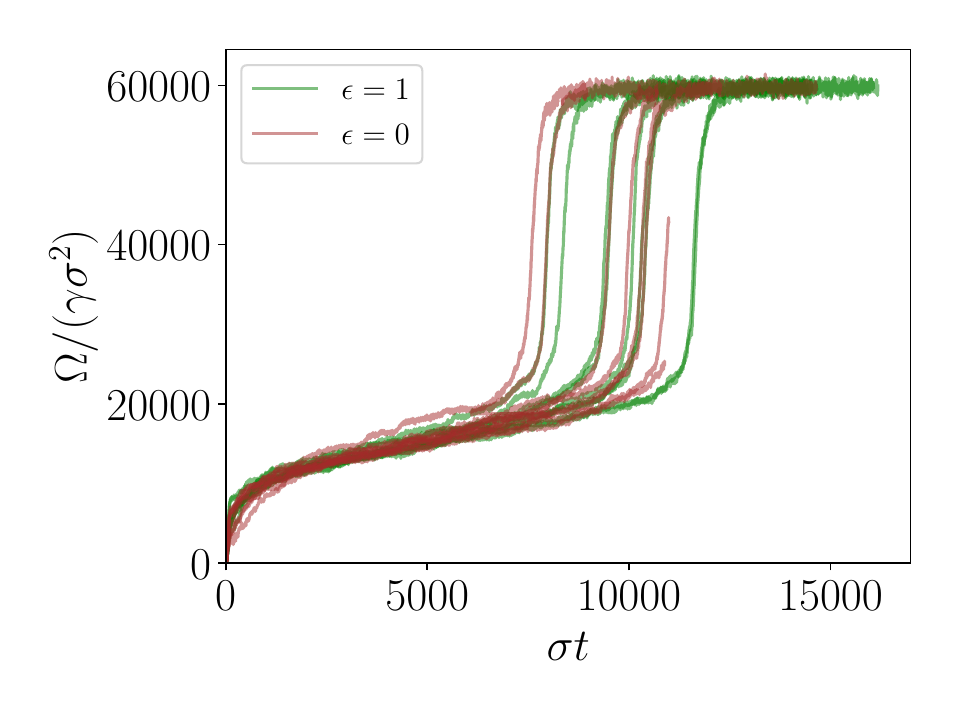}
    \caption{Time series of enstrophy from two ensembles of runs at $\gamma=0.95$. In the first ensemble, shown by the green curves, the random forcing is switched on with $\epsilon=1$ (same data as in Fig.~\ref{fig:late_time_enstrophy}) while in the second ensemble, shown by the brown curves, the random forcing amplitude is set to zero, $\epsilon=0$. \NEW{The difference between the average of $t_{\rm exp}$ over each of the two ensembles is not statistically significant compared to the standard deviation.}}
    \label{fig:enstrophy_gam0.95_eps0_eps1}
\end{figure}

As $\gamma$ changes, the relative importance of the two terms in the forcing function also changes. To determine which term is responsible for setting the observed, \NEW{increasingly long time scales of the approach to the stationary state}, we compare two ensembles of runs at $\gamma=0.95$ from sets A and E defined by the presence or absence of the random forcing term (cf. Sec.~\ref{sec:setup}). The late-time evolution of enstrophy in these two ensembles is shown in  Fig.~\ref{fig:enstrophy_gam0.95_eps0_eps1}. The deviation between the average $t_{\rm exp}$ in the two ensembles is not statistically significant compared to the standard deviation. In addition to the simulations at $\gamma=0.95$, we have also performed a run without random forcing at $\gamma=0.8$ (not shown), and found that in this case, the evolution of the enstrophy with and without the stochastic forcing is also nearly identical. These findings indicate that, at least for $\gamma$ close to unity, the stochastic forcing plays only a minor role in setting the transition to the dense state. \NEW{This is consistent with the observation that the ratio $\Gamma$ (defined in Eq.~(\ref{eq:Gamma})) between the energy injection rate associated with the instability forcing term and the random force is much greater than one. Specifically, we find that, in the stationary state, $\Gamma(\gamma=0.8)\approx 4.4\times 10^4 \gg 1$.} 

\section{Crystallisation transition at small $\gamma$}
\label{sec:crystallization}
\begin{figure}
    \includegraphics[width=0.32\textwidth,height=0.25\textwidth]{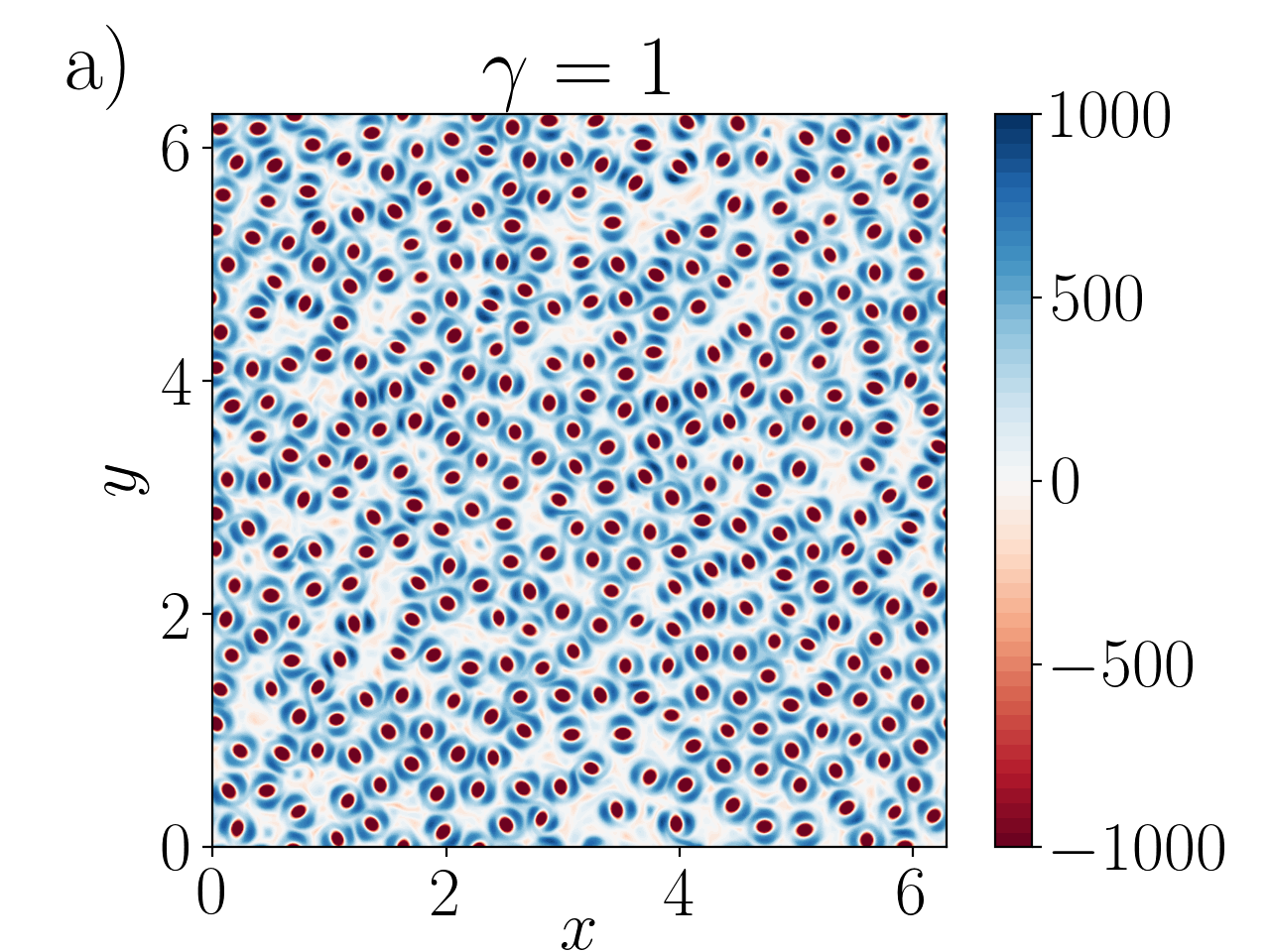}
    \includegraphics[width=0.32\textwidth,height=0.25\textwidth]{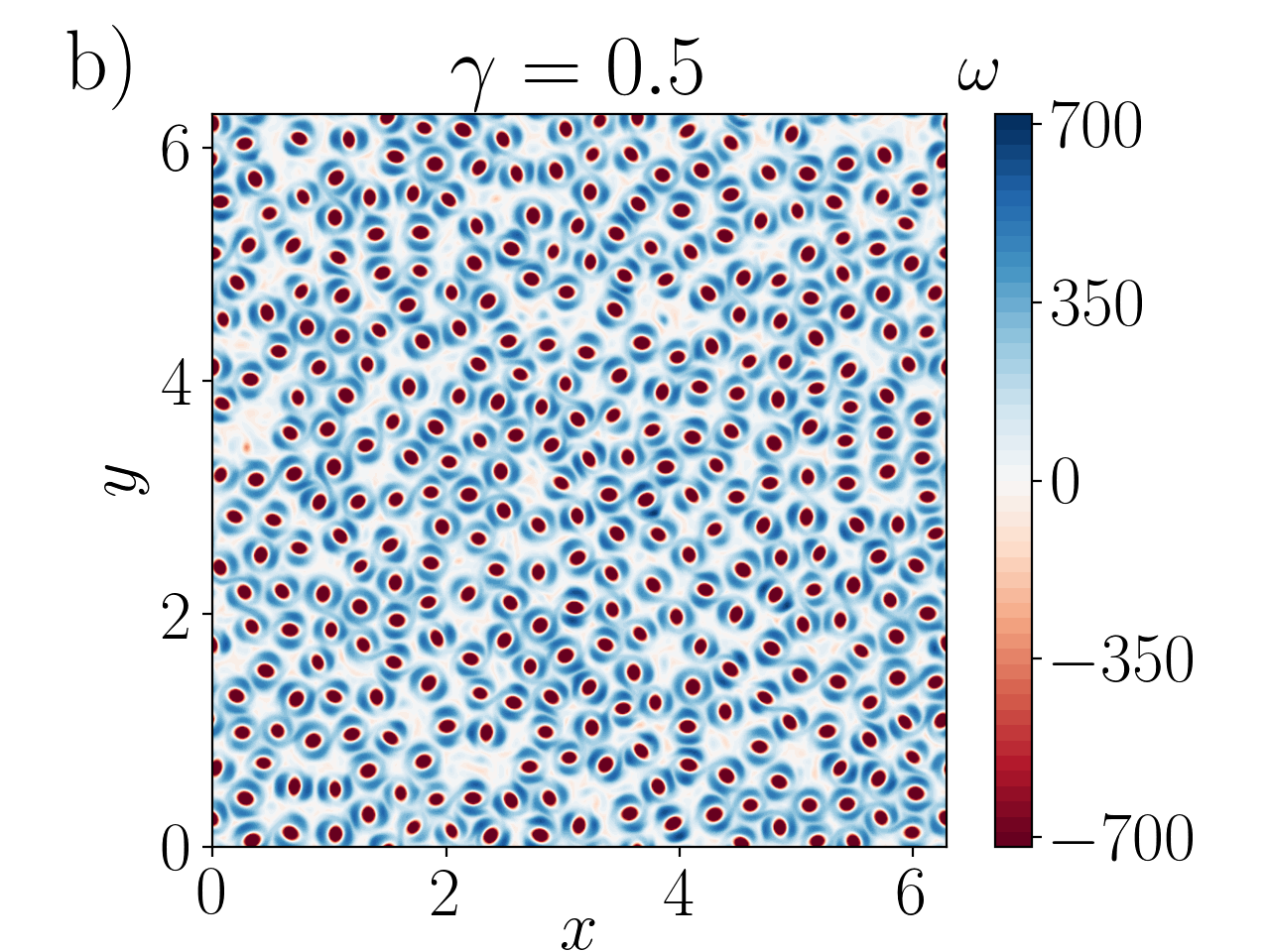}
    \includegraphics[width=0.32\textwidth,height=0.25\textwidth]{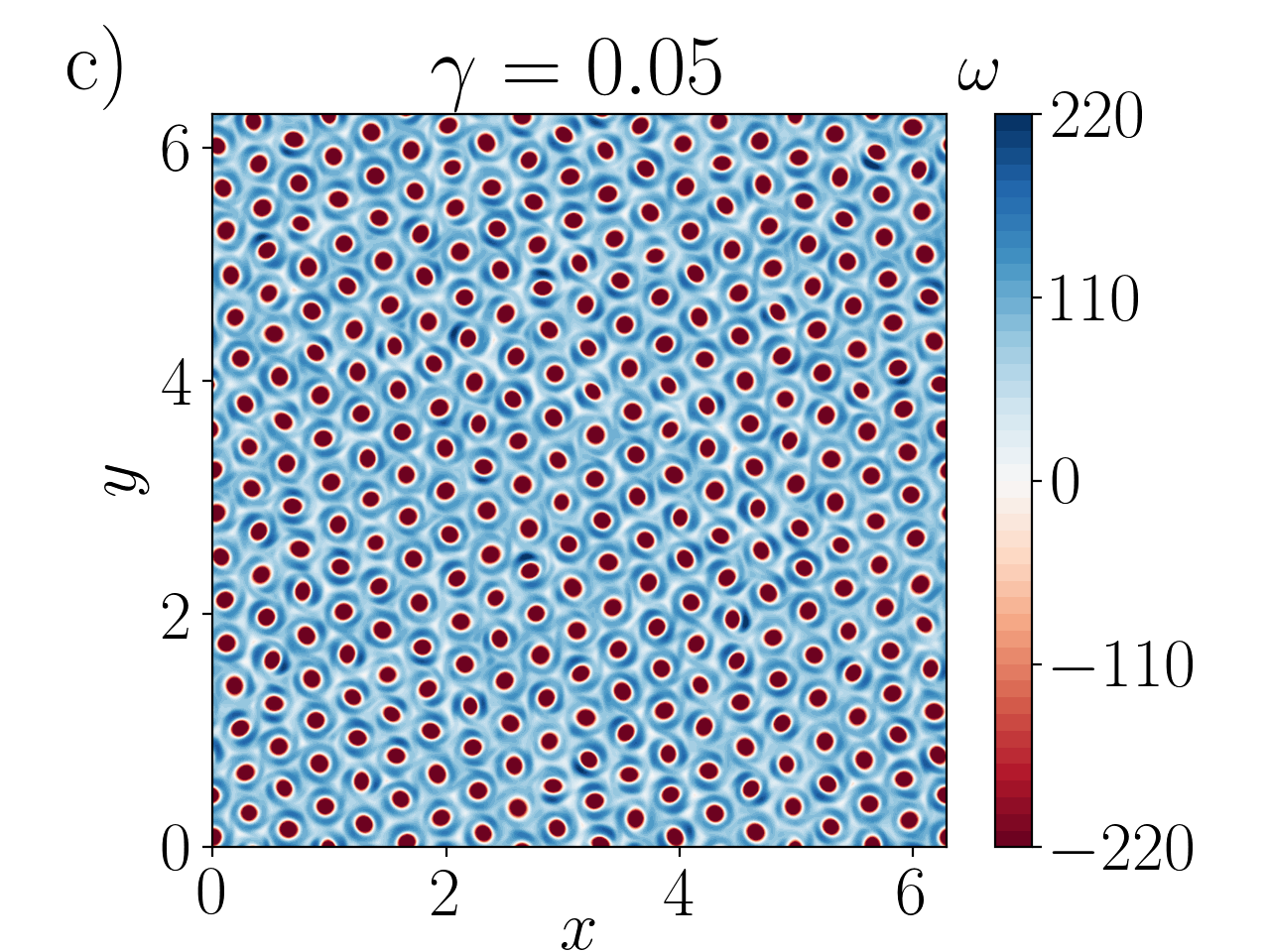}
    \caption{Snapshots of the vorticity field in the dense, statistically stationary state at $\gamma = 1$ (panel (a)), $\gamma=0.5$ (panel (b)) and $\gamma=0.05$ (panel (c)). }
    \label{fig:overview_dense_states}
\end{figure}

The dense vortex gas states found near $\gamma= 1$, whose emergence was described in the previous section, can be continued to smaller values of $\gamma$, where the observed states become much more regular.
This behaviour may appear unexpected, given that reducing $\gamma$ implies stronger stochastic forcing relative to instability forcing.
\NEW{However, for all the results described below, the instability forcing term remains dominant in the sense that the ratio $\Gamma$ remains large.} 
The dependence of the late-time flow state on $\gamma$ is illustrated in Fig.~\ref{fig:overview_dense_states} where snapshots of the vorticity field are shown for runs in set B at $\gamma=1$, $\gamma=0.5$ and $\gamma=0.05$.
In the latter case, a spontaneously formed hexagonal vortex crystal is observed.
\textcolor{black}{For identical point vortices this is the only stable vortex lattice in a periodic domain \citep{tkachenko1966stability}.}
In all three panels the vortices are of one sign only, i.e., all three panels represent symmetry-broken chiral states. \NEW{However, for every state shown in Fig.~\ref{fig:overview_dense_states}, there exists a corresponding state with the sign of the vorticity reversed.} \avkrev{The vortex state shown in Fig.~\ref{fig:overview_dense_states}(a,b) at high Reynolds numbers differs substantially from the disorganised, active turbulence state described in \cite{james2021emergence} for moderate Reynolds numbers, since we observe only vortices of a single sign, all with a pronounced tripolar structure and therefore shielded. The vortex crystal shown in Fig.~\ref{fig:overview_dense_states}(c) bears some resemblance to the active vortex lattice in \cite{james2021emergence}, although in the latter, vortices are not tripolar, but rather embedded in a background of uniform but opposite vorticity.}

In the following, the transition from the disordered vortex gas to the crystal is analysed quantitatively using different methods from statistical physics and crystallography.
\subsection{Vortex diffusion}
Supplementary movies SM1, SM2, SM3 \NEW{show the evolution of the vorticity field over time for \avkrev{$\gamma=0.5,~0.25$ and $0.05$}. As $\gamma$ is reduced, the root-mean-square speed at which individual vortices traverse the domain decreases significantly.} In particular, one observes that the individual vortices in the 
gas phase move chaotically across the domain but are trapped in the crystalline state.
Therefore, the speed at which individual vortices propagate through the domain is a natural order parameter for quantifying the transition to the crystalline state. \NEW{A particle-image-velocimetry (PIV) algorithm \citep{adrian2011particle} is suitable for this purpose.}

\begin{figure}
    \centering
    \includegraphics[width=0.32\textwidth]{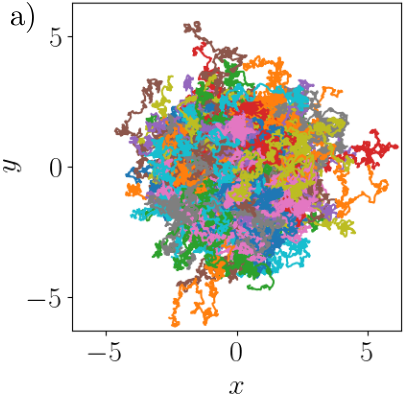}
    \includegraphics[width=0.32\textwidth]{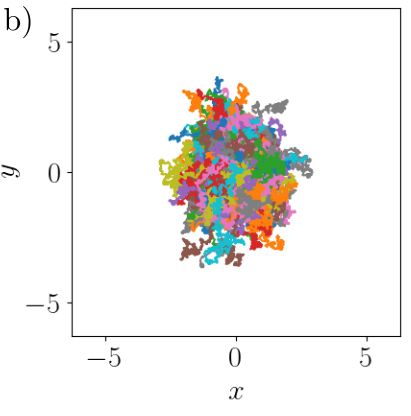}
    \includegraphics[width=0.32\textwidth]{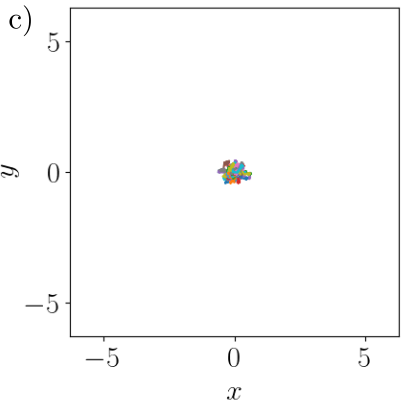}
    \caption{Trajectories of all individual vortices in the system from $t=0$ to $\sigma t = 300$, shifted to start at the origin, for $\gamma=0.5$ (panel (a)), $\gamma=0.25$ (panel (b)) and $\gamma=0.05$ (panel (c)). Each color indicates the trajectory of a particular vortex. The vortices are trapped at $\gamma=0.05$ (crystalline state), and diffuse above the melting transition ($\gamma_c\approx 0.13$), at a rate that increases with $\gamma$. }
    \label{fig:trajs}
\end{figure}

We have implemented such an algorithm in Python and applied it to the dense vortex states, taking into account the complicating feature of the periodic boundaries, which can lead to spurious doppelgängers that must be eliminated to obtain the correct trajectories. This procedure yields trajectories such as those shown in Fig.~\ref{fig:trajs}, where each color represents the trajectory of a single vortex in the system, shifted to start at the origin and extending from $t=0$ to $\sigma t = 300$. \avkrev{At $\gamma=0.5$ and $\gamma=0.25$ (Fig.~\ref{fig:trajs}a,b), the vortices diffuse with a mean squared displacement that increases with $\gamma$. By contrast, at $\gamma=0.05$ (Fig.~\ref{fig:trajs}c), i.e. in the crystalline state, the vortices remain trapped close to the origin.}

To quantify this impression we compute the mean squared displacement, a classical measure in the study of diffusion processes, over the vortex population as a function of time. Regular diffusion, which can be microscopically realised by Brownian motion, is characterised by a linear increase of the mean squared displacement with time  \citep{einstein1905molekularkinetischen}. The forced-dissipative system we are considering here is out of equilibrium, even if only weakly so, owing to the absence of an inverse energy cascade. Random motions observed in out-of-equilibrium systems are often characterised by anomalous diffusion \citep{metzler2000random,sokolov2005diffusion}, defined by a nonlinear scaling of the mean squared displacement with time.

\begin{figure}
    \centering
    \includegraphics[width=0.49\textwidth]{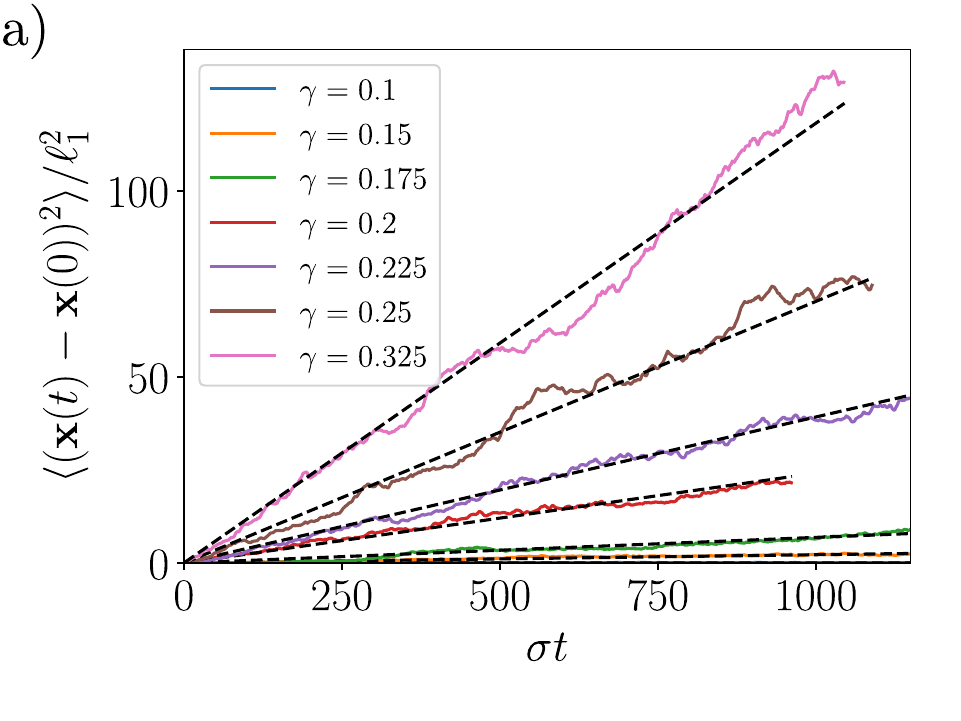}
    \\
    \includegraphics[width=0.49\textwidth]{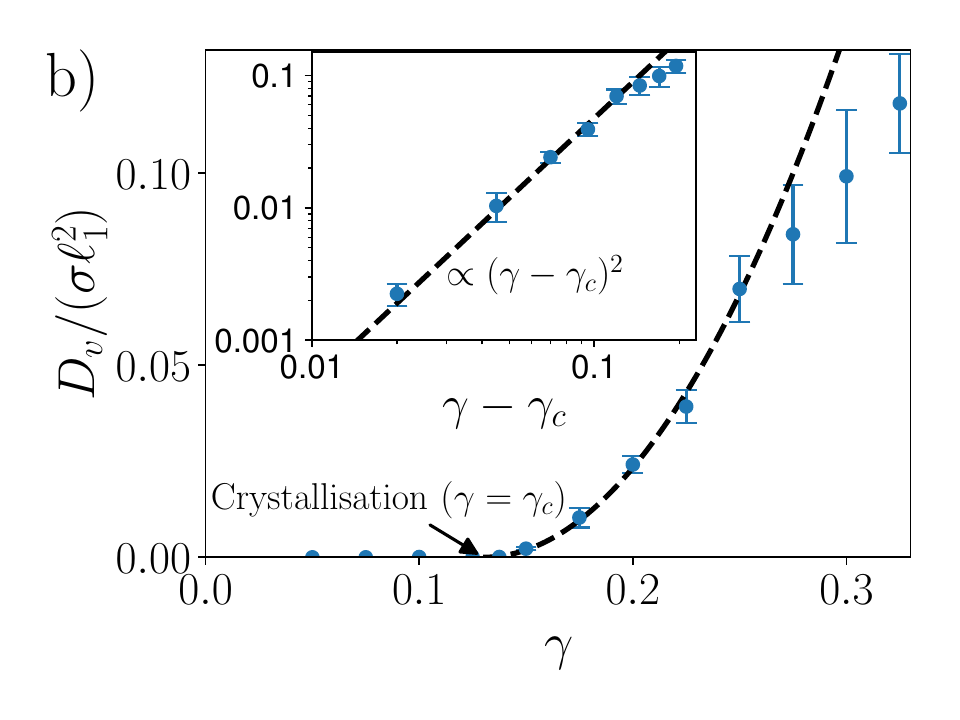}
    \includegraphics[width=0.49\textwidth]{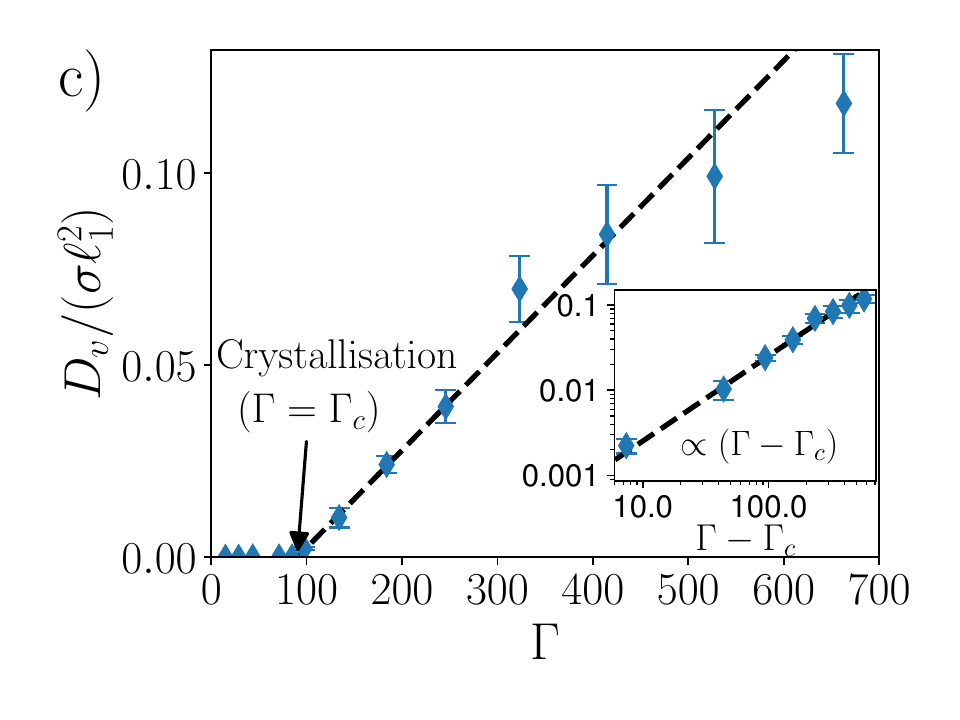}
    \caption{Panel (a): Mean squared displacement of vortices at different $\gamma$ versus time. The observed increase is approximately linear. Panel (b):  Blue triangles represent the slope $D_v$ measured from the observed mean squared displacements versus $\gamma$. Error bars indicate uncertainty in slope estimation. A clear threshold for crystallisation can be discerned at $\gamma=\gamma_c\approx 0.13$. \NEW{Black dashed line shows a quadratic fit in $\gamma-\gamma_c$ which is consistent with the data near onset.} \textcolor{black}{Inset shows a log-log plot of $D_v$ versus $\gamma-\gamma_c$, validating the approximate agreement between the data and the quadratic fit.} \avkrev{Panel (c): Same data as in panel (b), shown as a function of the ratio $\Gamma$ (defined in Eq.~(\ref{eq:Gamma})) of energy injection rates due to instability and random forcing. The slope $D_v$ scales approximately linearly with $\Gamma-\Gamma_c$, where $\Gamma_c\approx 90$.}}
    \label{fig:diffusivity_measurements}
\end{figure}

Against this backdrop, the results shown in 
Fig.~\ref{fig:diffusivity_measurements}(a) can be considered surprising: the complex mutual advection of individual vortices leads to a mean squared displacement that increases approximately \textit{linearly} with time, i.e. the vortices perform regular diffusion. The (nondimensional) slope $D_v$ of the mean square displacement over time is shown in \avkrev{panel (a)} as a function of the forcing parameter $\gamma$. Below a critical threshold, $\gamma=\gamma_c\approx 0.13$, individual vortices are trapped in the vortex crystal. Above this threshold, the diffusivity increases monotonically with $\gamma$. \NEW{The dashed line indicates a quadratic fit near onset of diffusion, which is accurate from $\gamma=\gamma_c\approx0.13$ to $\gamma\approx 0.25$}. The transition is seen to be \avkrev{continuous (or supercritical)}. \avkrev{Figure~\ref{fig:diffusivity_measurements}(c) shows the same data as panel (b), but as a function of $\Gamma$, defined in Eq.~(\ref{eq:Gamma}). The crystallisation transition occurs at $\Gamma=\Gamma_c\approx90$ with a critical exponent close to one.} Similar behaviour is found in other systems exhibiting hexagonal symmetry \citep{ammelt1998hexagon,bortolozzo2009solitary,ophaus2021two} but appears unrelated to any of the classical instabilities of a hexagonal pattern such as Eckhaus, zigzag or varicose instabilities \citep{sushchik1994eckhaus,echebarria2000instabilities}.
In the following, we refer to the transition from trapped to diffusive vortex motion observed at $\gamma=\gamma_c$ interchangeably as a \textit{crystallisation transition} or a \textit{melting transition}.
\avkrev{The observation that the melting transition discussed here is continuous is particularly interesting in view of the large literature on the search for similar continuous melting transitions in particle systems at equilibrium \citep{dash1999history}.}

In Appendix~\ref{app:phase_transition_with_different_forcings}, the location of this transition is shown to be insensitive to the width of the wavenumber band on which the random forcing acts. It is further highlighted there that in the absence of the random forcing ($\epsilon=0$), the vortex crystal develops defects which lead to residual diffusion that makes the sharp transition of Fig.~\ref{fig:diffusivity_measurements} imperfect. Thus the role of the noise associated with the stochastic forcing term in this system is once again counterintuitive, although it is well known that in simpler situations noise can indeed promote synchronisation, both in chaotic systems \citep{kurths} and in systems of nonidentical units including phase-coupled oscillators \citep{kori}.

\subsection{Radial distribution functions}
Another well-established measure of structured particle systems in statistical physics is the radial distribution function, usually denoted by $g(r)$ \citep{chandler1987introduction}. This quantity, also referred to as the \textit{pair correlation function} or \textit{pair distribution function}, measures the average density of particles near some location $\mathbf{r}$ with $|\mathbf{r}|=r$, given that a tagged particle is located at the origin. An equivalent definition of $g(r)$ is as the probability density of the quantity $N(r)/r$, where $N(r)$ is the number of vortices found within a radius $[r,r+dr]$ of a given vortex. The radial distribution function allows one to quantify the state of matter in particle systems and has been used to characterise active vortex crystal states \citep{riedel2005self}.

\begin{figure}
 \hspace{0.25\textwidth} (a) \hspace{0.45\textwidth} (b)\\

    \centering
    \includegraphics[width=0.49\textwidth]{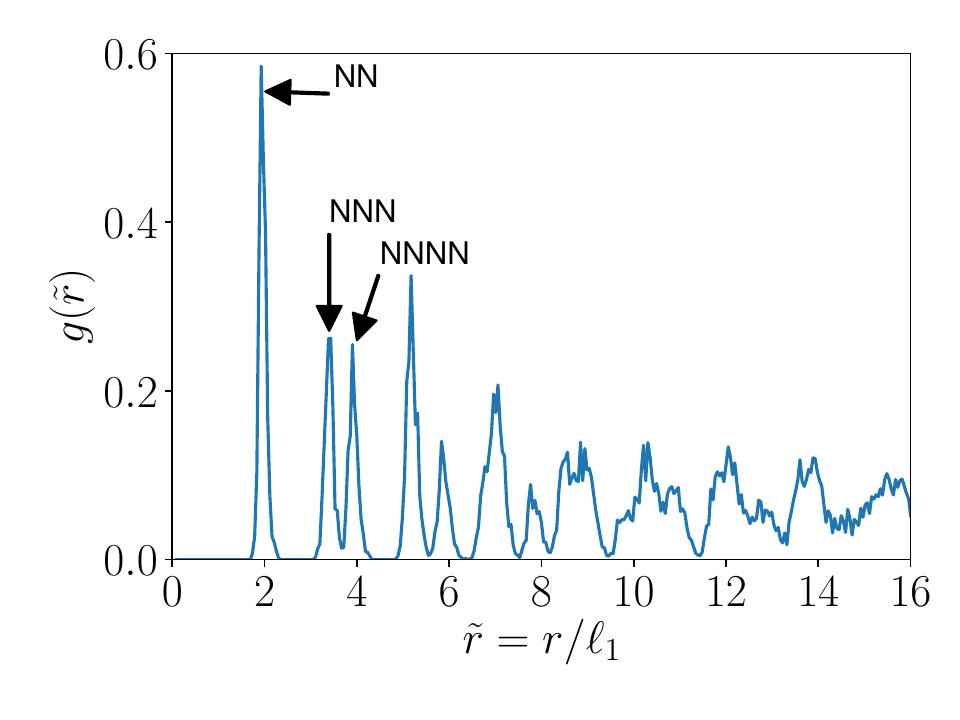}
    \includegraphics[width=0.49\textwidth]{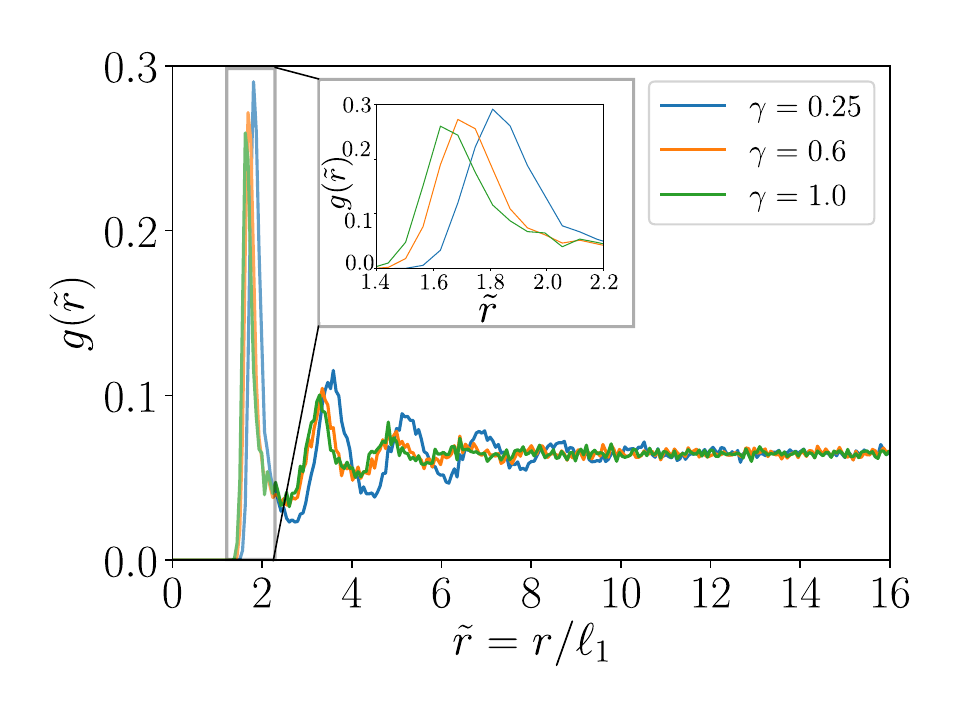}
    \caption{Radial distribution function $g(r)$ for different values of $\gamma$ in the vortex crystal state at $\gamma=0.05$ (panel (a)) and for values of $\gamma$ above the melting transition (panel (b)). In the crystalline state, there are pronounced peaks in $g(r)$ at nearest-neighbor, next-nearest-neighbor and next-next-nearest-neighbor distances (indicated by NN, NNN, NNNN, respectively).
    In the right panel, a liquid-like structure is observed beyond the melting transition. The inset shows a zoom on the first peak, which is seen to shift to smaller radii as $\gamma$ increases, reflecting shrinking vortex size.}
    \label{fig:radial_dist_func}
\end{figure}

Figure~\ref{fig:radial_dist_func} shows the radial distribution function observed in our system in the crystalline state and above the melting transition. In panel (a), corresponding to the crystalline state at $\gamma=0.05$, there are pronounced peaks near $r\approx 2\ell_1$ (nearest neighbor distance), $r\approx 2\sqrt{3}\ell_1$ (next nearest neighbor), $r\approx 4\ell_1$ (next next nearest neighbor) and this structure continues to larger radii, modulo increasing fluctuations. Panel (b) shows that a liquid-like structure is observed beyond the melting transition, with a clear peak near the minimum distance between vortex centers, associated with the finite size of the vortices, and successive peaks at larger radii indicating different coordination shells \citep{chandler1987introduction}. A zoom on the first peak highlights that it decreases in radius as $\gamma$ increases, in agreement with \avkrev{direct measurement of the average vorticity profile, as shown in} Fig.~\ref{fig:vorticity_profile_1D}. At large separations $g(r)$ becomes constant, reflecting the random arrangement of vortices in the vortex gas, which might also be called a vortex liquid in view of this result. Indeed, it is known that a dense system of vortices can be treated as a fluid and can itself be described in terms of  anomalous hydrodynamics \citep{wiegmann2014anomalous}, although these authors only consider point vortex flows and do not include the effects of shielding or of the finite size of individual vortices.

\avkrev{The regularity or irregularity of dense vortex configurations in different phases can also be measured using Voronoi diagrams, as described in Appendix~\ref{sec:app_voronoi}. One result of this analysis is that the inter-vortex distance decreases as $\gamma$ increases. However, the Voronoi analysis is based purely on the location of vortex centers and does not include information about the vorticity structure or the vortex size. This analysis thus cannot distinguish between larger gaps between vortices and a change in the vortex size.}
\NEW{
\subsection{Lindemann ratio}
Further insight into the physics of the melting transition of the vortex lattice can be gained by considering the relative displacements of vortices. This can be quantified using an established criterion in terms of the nondimensional Lindemann ratio \citep{goldman2003lattice}, given by
\begin{equation}
    r_L = \frac{\langle |u_m-u_n|^2\rangle }{a^2},
\end{equation}
in terms of the lattice spacing $a$, where $u_m$ is the displacement of vortex $m$ from the perfect lattice and the average is over nearest-neighbor vortex pairs indexed by $m,~n$. In simulations of 2D crystalline atomic lattices at thermal equilibrium, melting has been found to occur when $r_L\approx0.1$ \citep{bedanov1985modified,zheng1998lindemann} and this criterion has been shown to apply also to nonequilibrium systems \citep{goldman2003lattice}. We computed $r_L$ in the vortex crystal phase at the closest available data point to the melting threshold ($\gamma=0.125$), approximating $|u_m-u_n|$ by $|a-|\mathbf{r}_m-\mathbf{r}_n||$ in terms of the actual vortex positions $\mathbf{r}_m$, $\mathbf{r}_n$, to find that $r_L\lesssim 0.002$. This indicates that the displacements of vortices from their lattice sites are small compared to the lattice spacing, even near the onset of melting. This is related to the fact that the lattice spacing in the vortex crystal is nearly identical to the size of individual vortices, implying that displacements of vortices from lattice sites are strongly constrained. The small Lindemann ratio near the onset of melting clearly distinguishes the present nonequilibrium system from the equilibrium examples cited above. 
}

\section{Vorticity profile and vortex numbers}
\label{sec:vorticity_profile}

\begin{figure}
    \centering
    \includegraphics[width=0.49\textwidth]{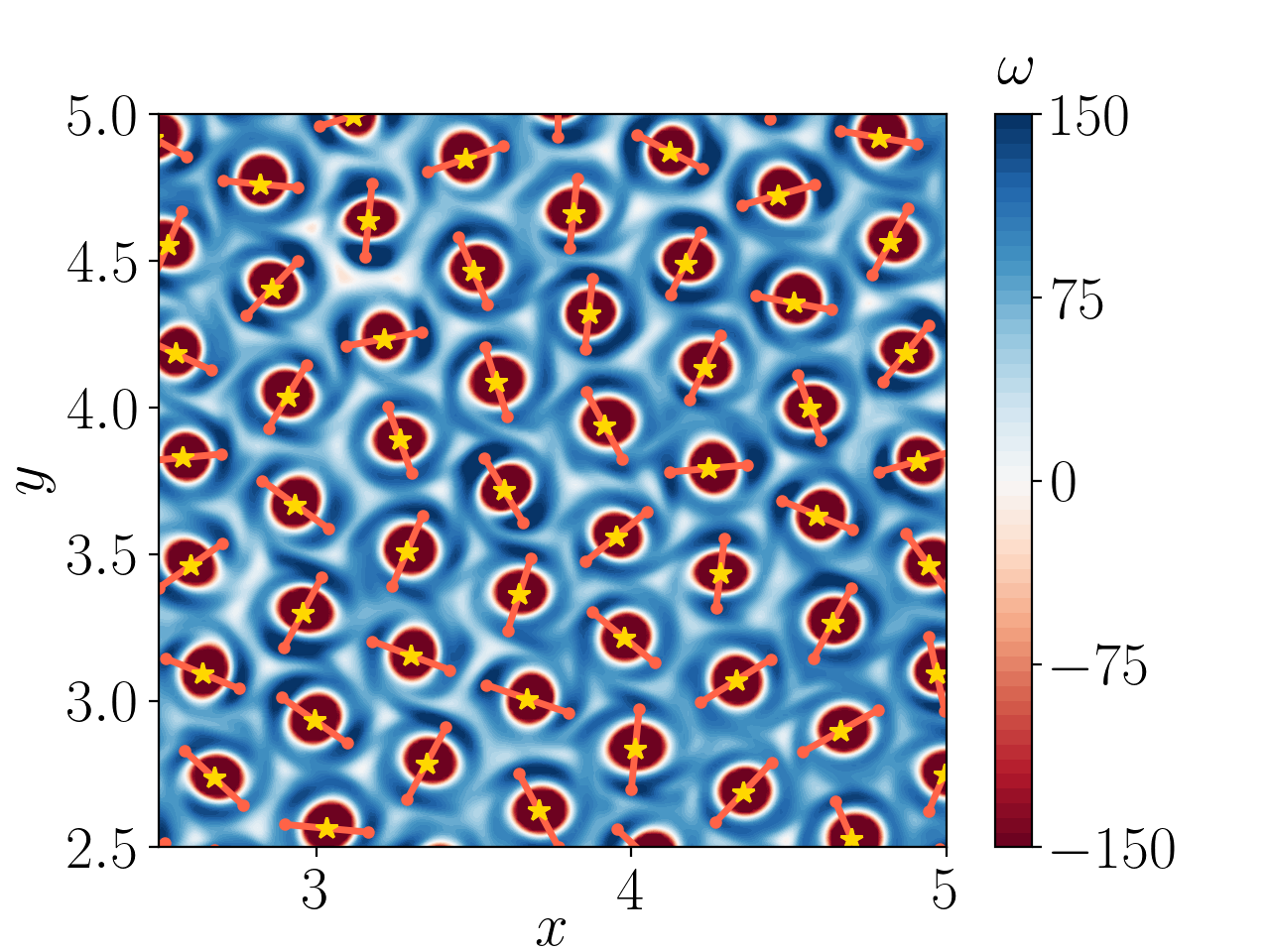}
    \caption{Graphical validation of identified tripole axes in the shielded vortex crystal ($\gamma=0.05)$. \avkrev{Golden stars indicate identified vortex centers while light red lines show the instantaneous tripole axes computed based on the location of the maximum vorticity amplitude in the shield. No polar order is discerned.
    }}
    \label{fig:id_axes}
\end{figure}

The accurate characterisation of the average vorticity profile of tripolar vortices poses a challenge, as one can see in movie SM4: individual tripolar vortices, \NEW{which feature varying degrees of ellipticity}, rotate rapidly about their center. We identify the tripole axis of every vortex in two steps. First, we find the location of the vortex center (position of central extremum). Note that the vortex positions are determined on a grid, implying a spatial resolution of $\Delta x = 2\pi/512 \approx 0.06\ell_1$. Second, we find an extremum of opposite sign in the shield. Then a straight line is drawn through these two points to determine the tripole axis. A graphical validation of the results obtained by this method is shown in Fig.~\ref{fig:id_axes} for the vortex crystal state. The golden crosses shown there indicate the vortex centers, with light red lines indicating the identified tripole axes. \avkrev{Individual vortices in the vortex lattice are found to rotate as approximately rigid bodies. We searched for signs of orientational order, such as local or global phase synchronization among neighboring shielded vortices, but no such effects were detected either in the vortex crystal state or in the vortex gas phase.}

\begin{figure}
 \hspace{0.1\textwidth} (a) \hspace{0.45\textwidth} (b)\\
    \centering
    \includegraphics[width=0.49\textwidth]{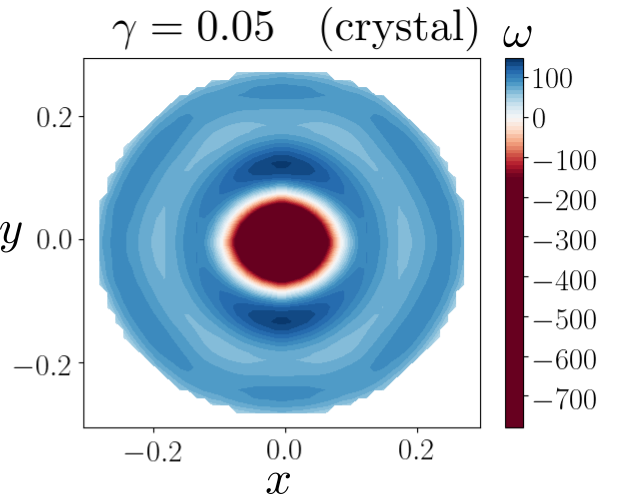}
    \includegraphics[width=0.49\textwidth]{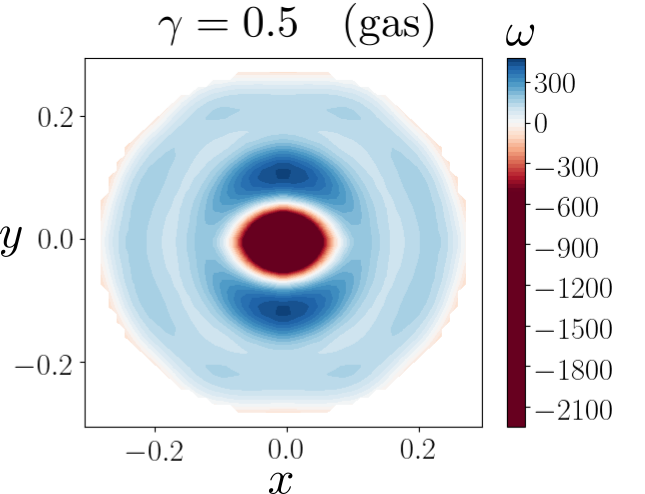}
    \caption{Population-averaged vorticity profiles at $\gamma=0.05$ (panel (a)) and $\gamma=0.5$ (panel (b)) in the crystal and gas phases. The vortex size decreases and the core becomes more elliptical as $\gamma$ increases. Note the different color bars required to accommodate the different vorticity strengths.}
    \label{fig:vorticity_profile}
\end{figure}

\avkrev{To perform a population average, we rotate individual vortices (a linear interpolation is required for this step due to the Cartesian grid) so that their vortex axes are aligned in the $y$ direction and shift all vortices to the origin.} 
The resulting profiles for $\gamma=0.05$ and $\gamma=0.5$ are shown in Fig.~\ref{fig:vorticity_profile}. Three main features can be readily observed: first, increased vorticity amplitude at $\gamma=0.5$ leads to a sharper contrast between the vortex and the background than at $\gamma=0.05$. Second, the vortex core is notably less elliptical in the crystal at $\gamma=0.05$ than at $\gamma=0.5$. Finally, the vortex size is significantly larger at $\gamma=0.05$ (i.e., in the hexagonal vortex crystal) than at $\gamma=0.5$ (the vortex gas).

\begin{figure}
   \hspace{0.2\textwidth} (a) \hspace{0.45\textwidth} (b)\\
    \centering
    \includegraphics[width=0.49\textwidth]{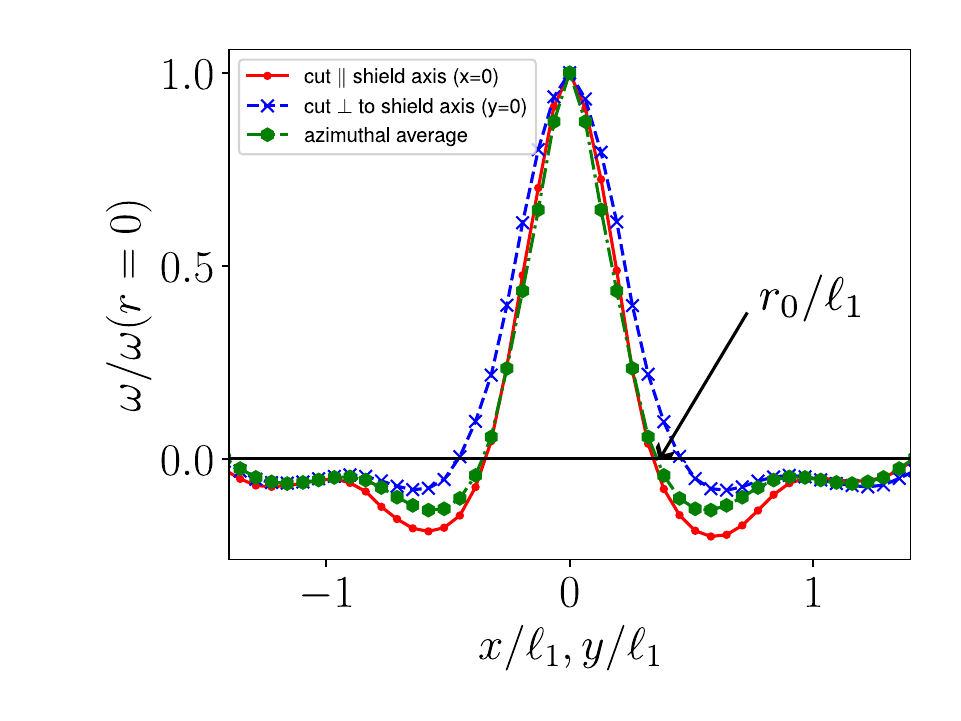}
    \includegraphics[width=0.49\textwidth]{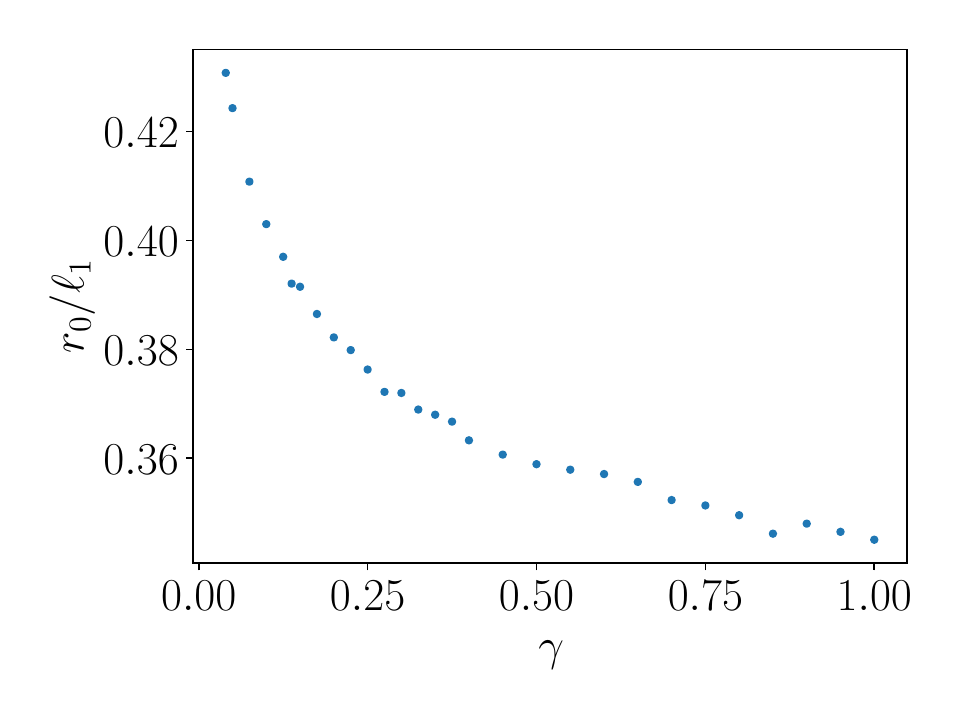}
    \caption{Panel (a): One-dimensional profiles obtained for $\gamma=0.5$ from the two-dimensional population-averaged vorticity profile shown in Fig.~\ref{fig:vorticity_profile}. Panel (b): Radius $r_0$ corresponding to the root (highlighted in the left panel by an arrow) of the azimuthally averaged vorticity profile $\overline{ \omega} (r) = (2\pi)^{-1}\int \omega(r,\phi) d\phi$ (shown in green in the left panel), nondimensionalised by largest scale $\ell_1$ in the forcing range. }
    \label{fig:vorticity_profile_1D}
\end{figure}

Figure~\ref{fig:vorticity_profile_1D}(a) shows one-dimensional radial profiles of the vorticity corresponding to $\gamma=0.5$: cuts through the vortex center along the $x$ and $y$ directions in Fig.~\ref{fig:vorticity_profile}, i.e. parallel and perpendicular to the tripolar shield axis, are shown in blue and red, respectively, together with the azimuthally averaged vorticity profile (in green). We define the radius $r_0$ as the radius at which the azimuthally averaged profile passes through zero. Panel (b) shows that this radius decreases monotonically as $\gamma$ increases, indicating a continuous shrinkage of the tripolar vortices as the contribution of the instability forcing increases. 

\begin{figure}
   \hspace{0.2\textwidth} (a) \hspace{0.45\textwidth} (b)\\
    \centering
    \includegraphics[width=0.49\textwidth]{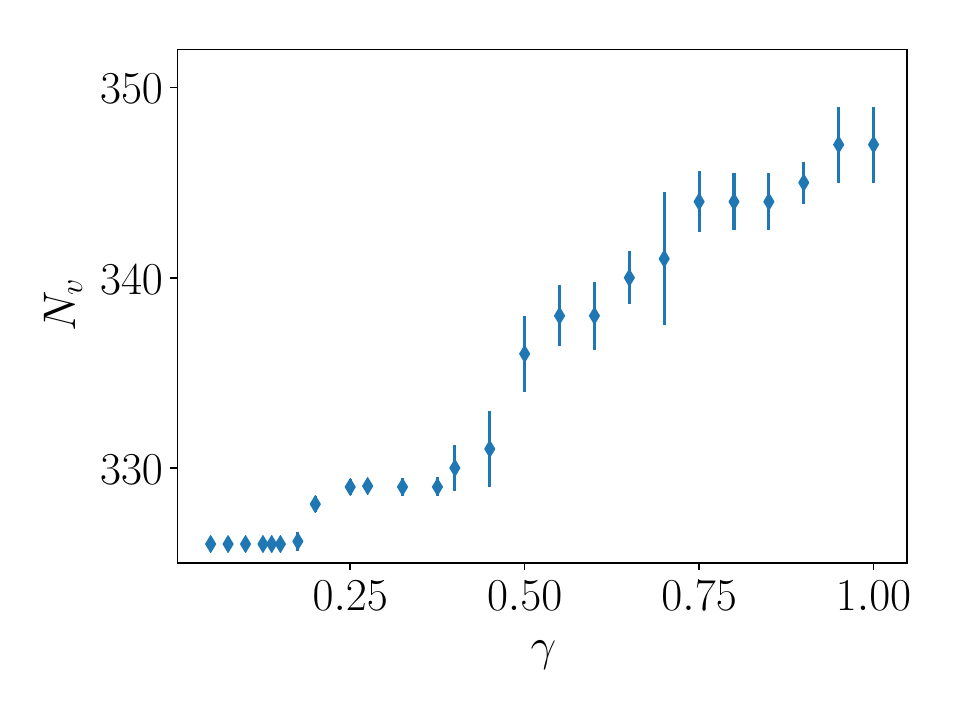}
    \includegraphics[width=0.49\textwidth]{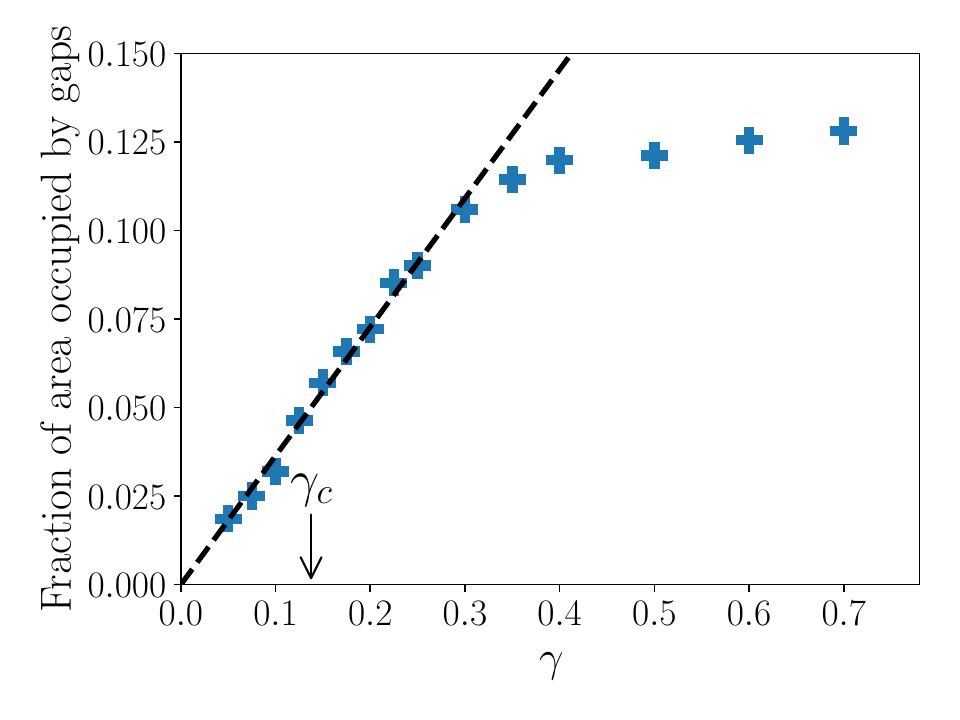}
    \caption{Panel (a): The number $N_v$ of vortices observed in the domain in the statistically stationary state increases with $\gamma$. In the statistically stationary vortex gas state, vortices are occasionally destroyed or created, an effect captured by the error bars that indicate the standard deviation of the number of vortices. The fluctuations in $N_v$ are small compared to the total number of vortices in all cases. \avkrev{Panel (b): Fraction of the domain area occupied by gaps versus $\gamma$. Gaps are identified as regions where the absolute value of vorticity $|\omega|$ is less than or equal to $1\%$ of the maximum value in the vortex core. The gap area increases rapidly and approximately linearly with $\gamma$ at small $\gamma$, an effect important for the melting transition at $\gamma\approx 0.13$, and saturates at larger $\gamma$.}}
    \label{fig:number_vortices}
\end{figure}

As shown in Fig.~\ref{fig:number_vortices}(a), as the vortices shrink with increasing instability growth rate (increasing $\gamma$), the number density of vortices in the stationary state gradually increases, with more and more vortices in the domain. \avkrev{However, the number of vortices increases by less than $10\%$ over the whole range of $\gamma$, while their radial extent decreases more rapidly with $\gamma$, with a reduction by around $20\%$ in vortex radius over the same range (cf.~Fig.~\ref{fig:vorticity_profile_1D}) and faster shrinkage at small values of $\gamma$ compared to larger $\gamma$. Figure~\ref{fig:number_vortices}(b) shows that this results in a rapid and approximately linear growth with $\gamma$ in the fraction of the domain area occupied by gaps between vortices (identified as regions where $|\omega|\leq 0.01\mathrm{max}(|\omega|)$) at small $\gamma$ followed by saturation of this fraction at larger $\gamma$. This increase in the gap area with $\gamma$ is of  great importance in facilitating the observed melting transition. Gaps grow between the lattice sites in the crystal state until, at $\gamma=\gamma_c\approx0.13$, they become sufficiently wide for vortices to begin slipping through and diffuse across the domain.}


\section{Conditions for the maintenance of the vortex crystal}
\label{sec:conditions_suppression_inverse_cascade}
In view of the nontrivial physical properties of the shielded vortex crystal described in the previous sections, one can ask under which conditions this state is stable and whether it can be disrupted in other ways than via the diffusive melting transition discussed in Sec.~\ref{sec:crystallization}. Here, we describe two possibilities for destabilizing the vortex crystal state, both involving the disruption of the vortex shield followed by the subsequent appearance of an inverse energy cascade, which is otherwise suppressed in the crystal when shielding is present. Similar behavior is observed in rapidly rotating Rayleigh-Bénard convection, where the inverse energy cascade is suppressed by the presence of convective Taylor columns \citep{grooms2010model,julien2012statistical}.

\subsection{Crystal decay at very small $\gamma$}
\begin{figure}
    \centering
    \includegraphics[width=0.49\textwidth]{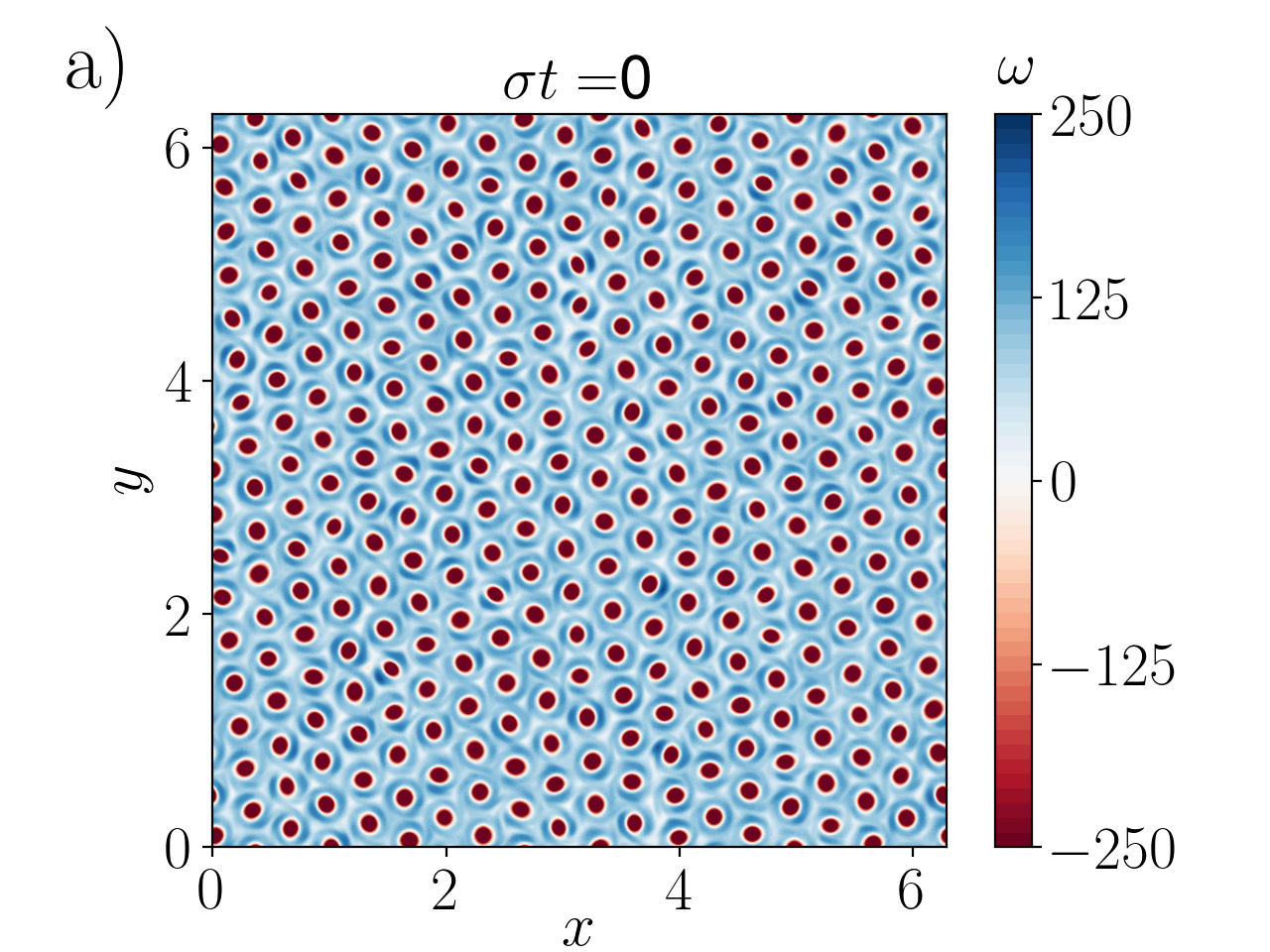}
    \includegraphics[width=0.49\textwidth]{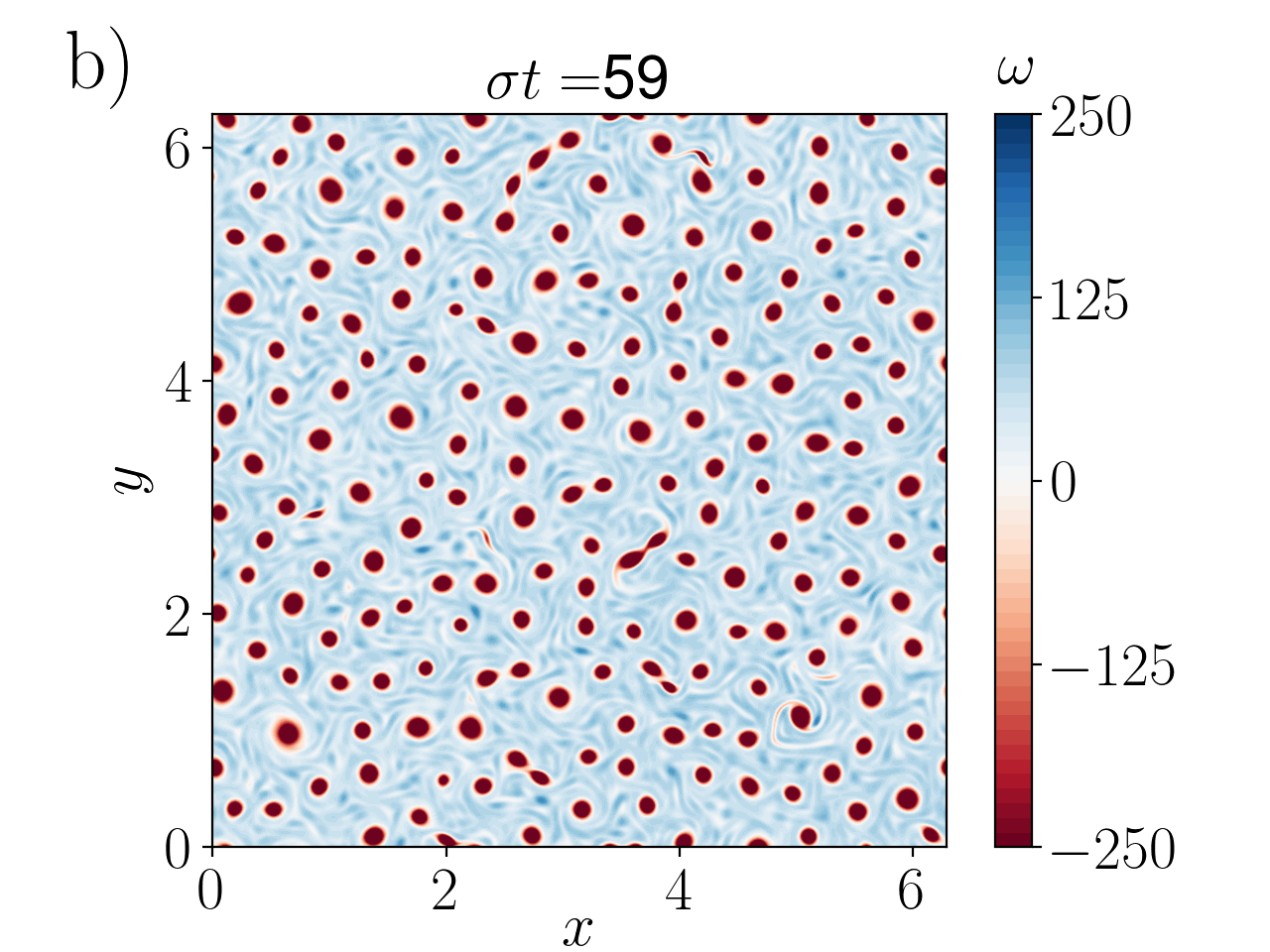}
    \includegraphics[width=0.49\textwidth]{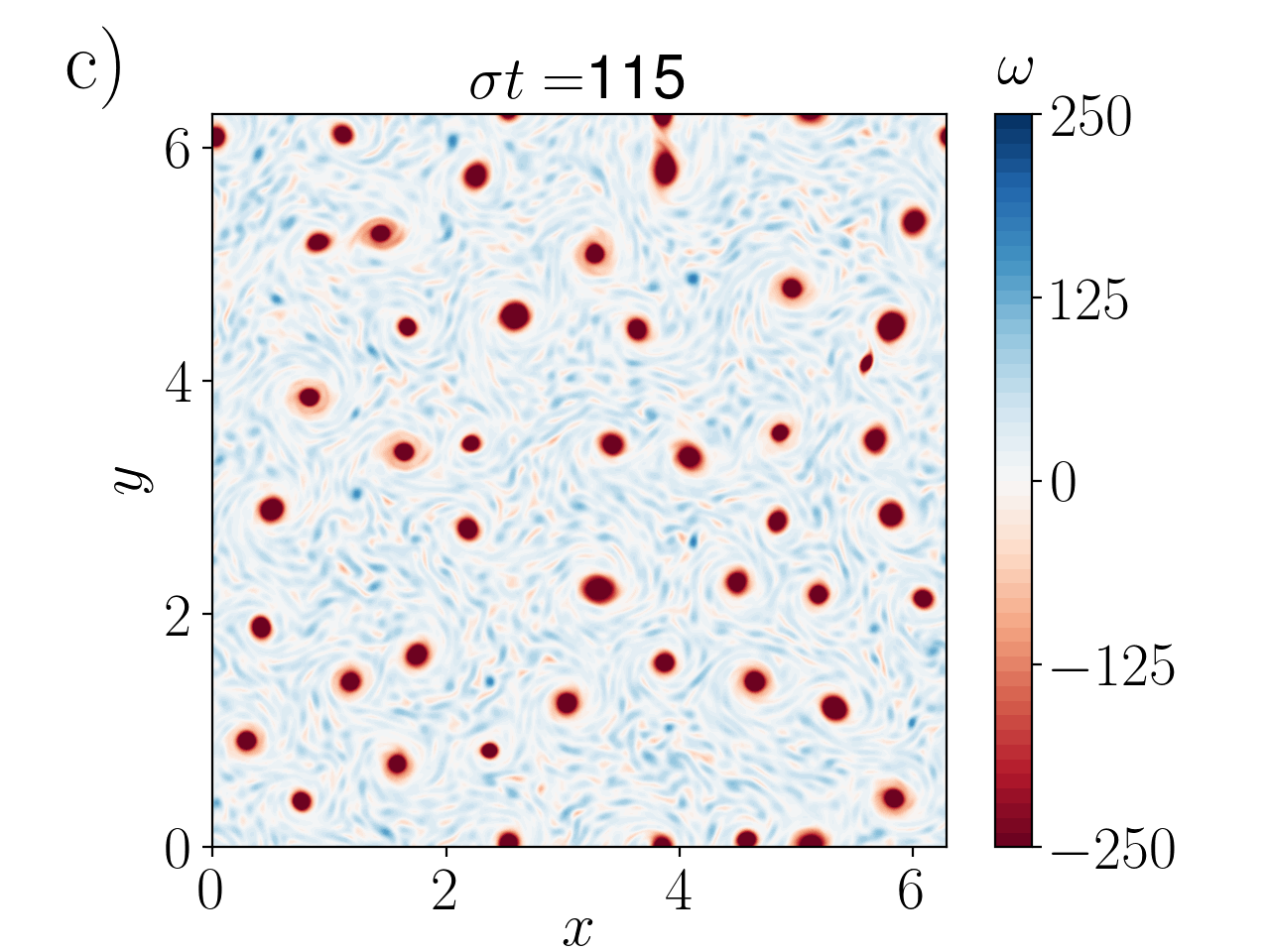}
    \includegraphics[width=0.49\textwidth]{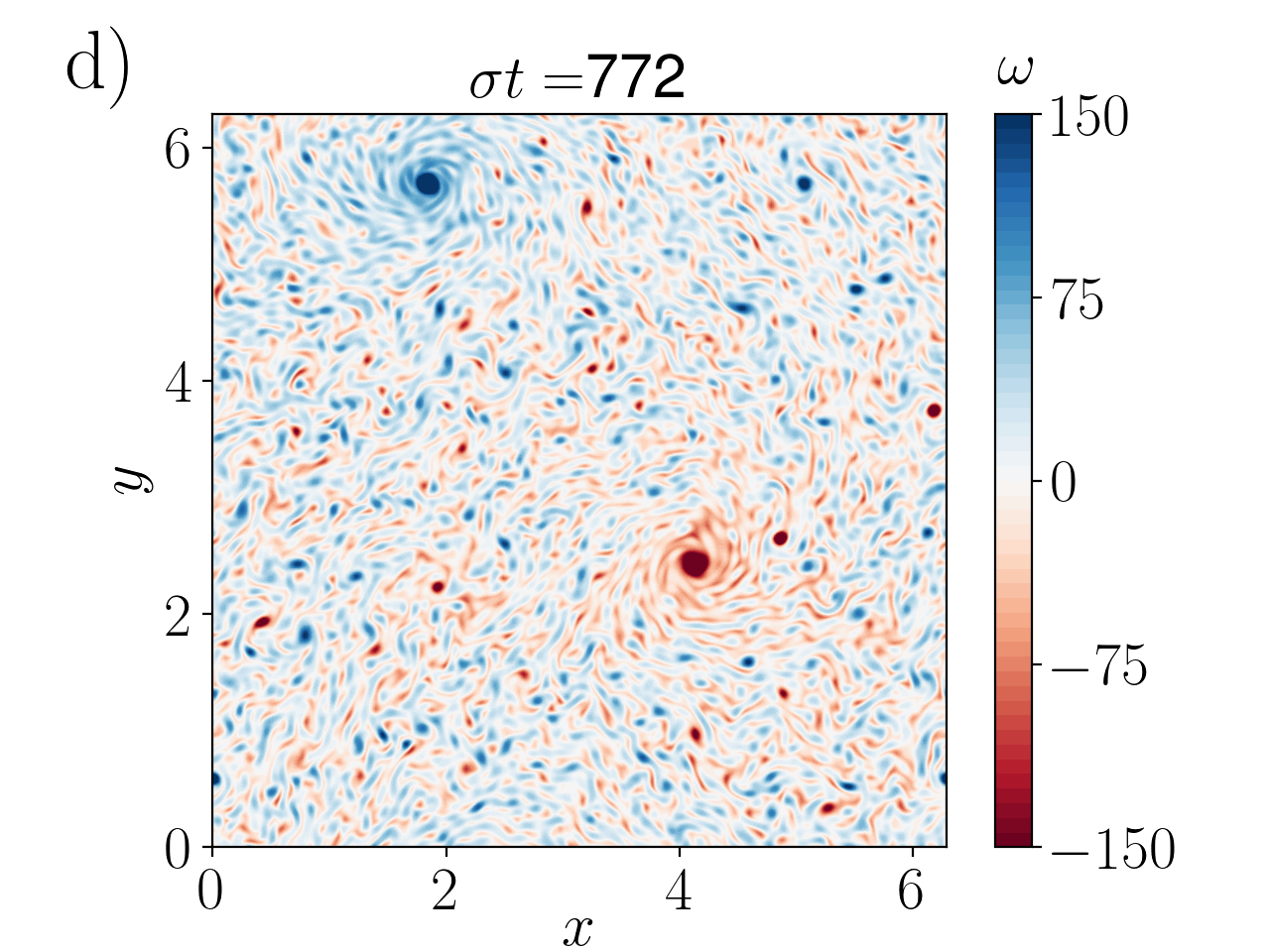}
    \caption{Sequence of states observed at $\gamma=0.03$ with vortex scale damping \NEW{and $\epsilon=1$}, initialised by a vortex crystal state at larger $\gamma$. The dissolution of the shields of tripolar vortices in the vortex crystal is followed by an inverse energy cascade.}
    \label{fig:vortex_crystal_dissolved_shields}
\end{figure}

First, we consider runs from simulation set B, starting in the vortex crystal state (taken from a simulation at $\gamma=0.05$), and continuously reduce $\gamma$ while retaining the full cubic damping term. The vortex crystal is found to remain stable down to $\gamma=0.04$. However, for $\gamma \leq 0.035$, the shields of the tripolar vortices spontaneously dissolve, followed by an inverse energy cascade culminating in the appearance of a large-scale vortex condensate, as illustrated in Fig.~\ref{fig:vortex_crystal_dissolved_shields}. Two possible explanations suggest themselves. Since $\gamma$ is small, one may assume that the random forcing term becomes sufficiently strong to cause the observed dissolution of the shield structures. However, at $\gamma=0.04$ (the smallest $\gamma$ where the crystal is observed to be stable) we observe $\Gamma \approx 11$ (with $\Gamma$ defined in Eq.~(\ref{eq:Gamma})), indicating that the \avkrev{energy injection rate due to the} random forcing  remains subdominant compared with the instability forcing. Instead, we note that the ratio $r(\gamma=0.035)\approx 1.7$, indicating that the time scales of the forcing and the hyperviscous dissipation are comparable. To test whether this is indeed responsible for the observed decay of the crystal at $\gamma\leq 0.035$, we performed additional simulations at $\gamma=0.005,0.01,0.02,0.03$ and $0.035$ with $\epsilon=0$ (no stochastic forcing), leaving all other parameters unchanged (set $F$ in Table \ref{tab:overview_all_runs}). For $\gamma\leq 0.2$, we observe the same temporal evolution from a vortex lattice initial condition: the shields are seen to rapidly dissolve and an inverse cascade ensues. This suggests that the observed decay phenomenology is indeed independent of the stochastic forcing term. \avkrev{Influence of the stochastic forcing was detected only very close to the onset of crystal decay, namely at $\gamma=0.03,~0.035$: the crystal decayed rapidly via an inverse cascade in the presence of noise, while for $\epsilon=0$, the crystal is stable at $\gamma=0.035$ with a random deletion process observed only at $\gamma=0.03$, where single vortices disappear one after another from the lattice, as can be seen in Supplementary Movie SM5.}
Leaving aside these special cases close to the dissolution threshold, we conclude that the observed vortex decay is largely independent of the stochastic forcing except for a small shift in the threshold. We mention that with a different choice of the parameters $\nu_n$, $k_2$ or $\nu_*$ (see Sec. \ref{sec:setup}), one may achieve $\Gamma \approx 1$, while maintaining $r(\gamma) \gg 1$. In this case, the very destabilization of the crystal would likely be facilitated by the dominance of the stochastic force over instability, \avkrev{rather than by (hyper)viscous effects}. 

\subsection{The role of nonlinear damping}
\begin{figure}
 \hspace{0.15\textwidth} (a) \hspace{0.45\textwidth} (b)\vspace{-0.05cm}\\
    \centering
    \includegraphics[width=0.49\textwidth]{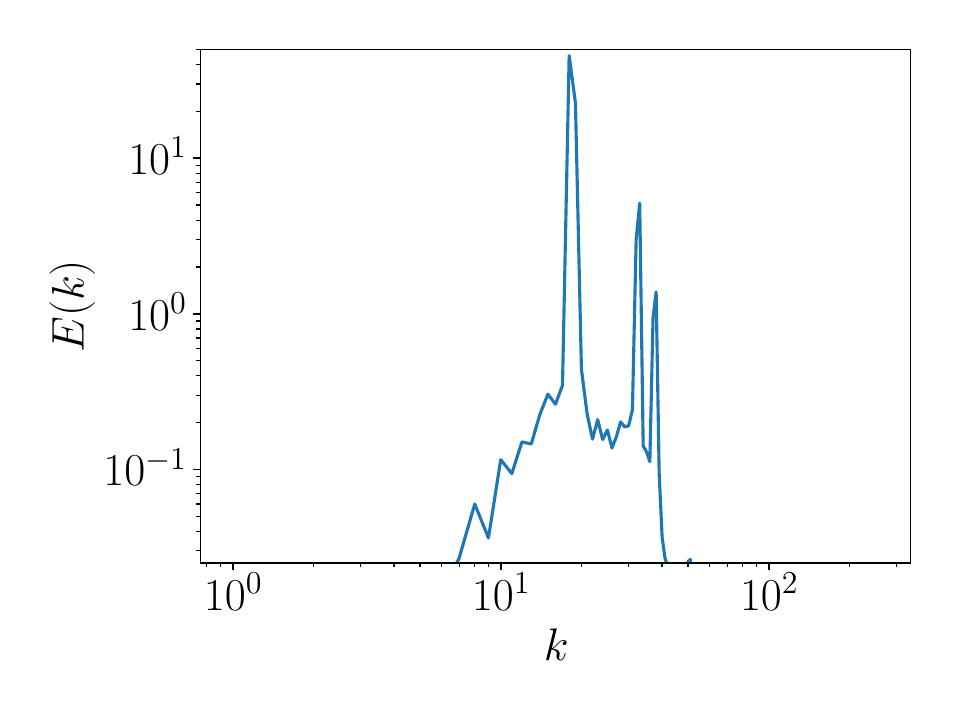}
    \includegraphics[width=0.49\textwidth]{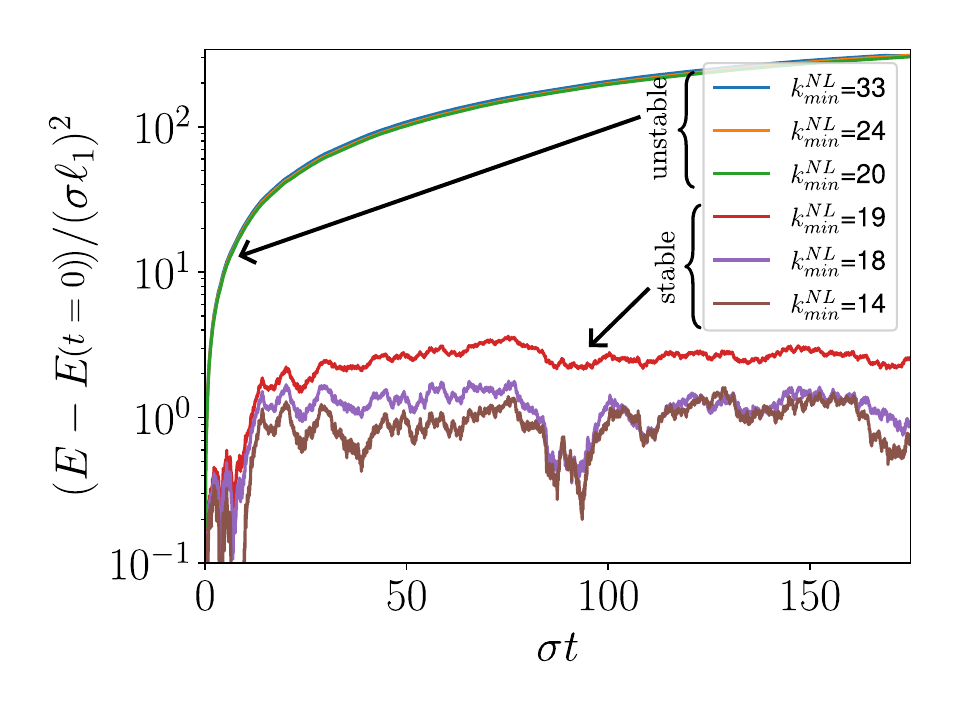}
        \caption{Panel (a): Log-log plot of the energy spectrum in the vortex crystal state ($\gamma=0.05$). The primary peak corresponding to the shielded vortex scale is located at wavenumber $k=k_v\approx 19$. Two secondary peaks are seen at $k\approx \sqrt{3}k_v$ and $k\approx 2 k_v$, corresponding to the next-to-nearest neighbor and next-next-nearest-neighbor distances in the crystal. Panel (b): Lin-log plot of energy deviation from the crystal versus time for different simulations at $\gamma=1$, initialised in the crystal state. In each simulation, a different spectral filter is applied to the nonlinear damping, such that it acts only on the wavenumber window $[k_{min}^{NL},k_2]$, with $k_2=40$ fixed. A sharp transition is observed: when the nonlinear damping acts on the vortex scale $k_v\approx 19$, the vortex crystal is stable, but it breaks down when the damping term does not act on $k_v$. In the latter case, the well-organised shields dissolve and an inverse cascade ensues wherein individual vortex cores merge and one observes a large-scale condensate at late times.}

    \label{fig:varying_nl_window}
\end{figure}

Next, we examine the role of the nonlinear damping. The results presented by \cite{van2022spontaneous} already showed that this term plays a crucial role in fully suppressing the inverse cascade once shielded vortices form. In the absence of nonlinear damping at large scales, a residual inverse cascade persists, albeit much weaker than that observed with stochastic forcing only. However, a more systematic investigation of the role of nonlinear damping in this system is still missing. As shown in Fig.~\ref{fig:varying_nl_window}(a), the energy spectrum in the vortex crystal is sharply peaked at a principal peak corresponding to the vortex scale ($k\approx k_v= 19$), with two minor peaks observed at $k\approx\sqrt{3}k_v$ and $k\approx 2 k_v$. Given this spectral structure, the nonlinear damping in this state will preferentially act on the vortex scale. Hence, it is natural to investigate the situation where the nonlinear damping term is subject to spectral filtering, to include or exclude the vortex scale and study its effect on the stability of the vortex crystal.

Table \ref{tab:overview_set_D} summarises the runs performed in set D, where the cubic nonlinearity is filtered in Fourier space so that it acts on a finite wavenumber interval $[k_{\rm NL,min},k_{\rm NL,max}]$ with $k_{\rm NL,max}=k_2=40$. Table \ref{tab:overview_set_D} indicates that the vortex crystal is only stable when the nonlinear damping acts on the vortex scale. Otherwise the injected energy will inevitably cascade towards larger scales. This is because the nonlinear damping at the shielded vortex scale is strongly amplified, thereby suppressing the inverse cascade.
Figure~\ref{fig:varying_nl_window}(b) shows time series of kinetic energy confirming this expectation, highlighting a sharp transition at $k_{\rm NL,min}=19$: when the filtered damping excludes the vortex scale, the kinetic energy grows and an inverse cascade ensues, while, otherwise, the vortex crystal remains intact. This agrees with the findings of \cite{linkmann2020non}, who observed such an inverse cascade and large-scale condensation in a model of moderate Reynolds number 2D active turbulence with a negative eddy viscosity driving force but no nonlinear damping or random forcing, provided the forcing amplitude was sufficiently large.
\begin{table}
\centering
\begin{tabular}{|c|c|}
\hline
   $k_{\rm NL,min}/k_1$  &  Vortex crystal stable \\ \hline
   $33/33$ & no \\
   $30/33$ & no \\ 
   $27/33$ & no \\
   $24/33$ & no \\
   $21/33$ & no \\
   $20/33$ & no \\
   $19/33$ & yes \\
   $18/33$ & yes \\
   $17/33$ & yes \\
   $14/33$ & yes \\
   $11/33$ & yes  \\ \hline
\end{tabular}
\caption{Stability of the vortex crystal with nonlinear damping applied only to the wavenumber window $[k_{\rm NL,min},k_2]$, for different $k_{\rm NL,min}$. A sharp transition is present at $k_{\rm NL,min}=k_v=19$, i.e., at the energy-containing scale of the vortex crystal (see also Fig.~\ref{fig:varying_nl_window}).}
\label{tab:overview_set_D}
\end{table}

In short, the vortex crystal can be disrupted in \NEW{at least} two ways other than the diffusive melting transition described previously: it breaks down when \NEW{the time scale of (hyper)viscous dissipation at the forcing scale become comparable to the maximum instability growth rate} or when the nonlinear damping is turned off at the vortex scale. A third possible mechanism for the breakdown of the vortex crystal arises when the energy injection rate due to the stochastic forcing term becomes comparable to the energy injected by the instability. However, this is not observed with the parameters considered in our simulations, since here viscous dissipation becomes comparable to the instability forcing already at larger values of $\gamma$. 

\NEW{Although turning off the nonlinear damping at the vortex scale does lead to a breakdown of the crystal state, it should be emphasised that the observed suppression of the inverse energy cascade in that state is not a simple consequence of nonlinear damping.} This is clearly seen from the fact that at sufficiently small $\gamma$, $\gamma\lesssim 0.3$, an inverse cascade ensues \NEW{when the simulations are initialised with small-amplitude initial conditions} despite the presence of nonlinear damping; see also Fig.~\ref{fig:overview_state_space} and \cite{van2022spontaneous}. Instead, it is the presence of coherent, shielded vortices that amplifies the action of the nonlinear damping and leads to the suppression of the inverse energy cascade. Thus both ingredients are needed for the spontaneous suppression of the inverse cascade.

\section{Overview of stationary-state solutions at different $\gamma$}
\label{sec:overview_state_space}
\begin{figure}
    \centering
    \includegraphics[width=0.7\textwidth]{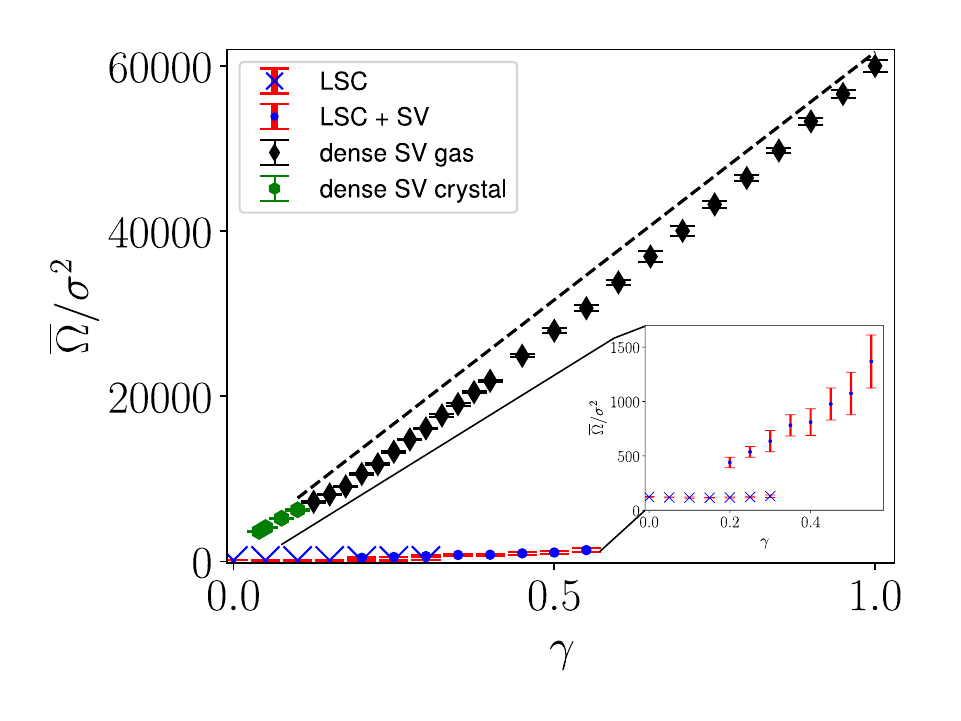}
    \caption{Overview of the stationary states found in nonlinearly damped two-dimensional turbulence driven by the hybrid forcing defined in Eq.~(\ref{eq:forcing}) for different values of $\gamma$. Four qualitatively different states are observed. First, a large-scale condensate (LSC) forms spontaneously from small-amplitude, random initial conditions provided $\gamma<0.35$. Second, a hybrid state with smaller-scale shielded vortices embedded in the LSC (LSV+SV) as reported by \cite{van2022spontaneous}; see also Fig.~\ref{fig:quadratic_drag}. Third, a dense shielded vortex gas and fourth, a shielded vortex crystal, form together the branch of dense states which constitute the primary topic of the present work.}
    \label{fig:overview_state_space}
\end{figure}

Figure~\ref{fig:overview_state_space} shows an overview of the observed stationary states in the model, extending the state diagram previously shown by \cite{van2022spontaneous}. Four qualitatively distinct states are observed: starting in set A from small-amplitude, random initial conditions with predominantly random forcing, i.e. small $\gamma$, a large-scale condensate spontaneously forms (blue crosses). As $\gamma$ increases beyond approximately $\gamma=0.35$, a hybrid state emerges (blue dots) with tripolar, shielded vortices embedded in the remaining large-scale circulation patterns (see \cite{van2022spontaneous} and Fig.~\ref{fig:quadratic_drag}(a,b) for visualisations of such states). Continuing the hybrid states to smaller $\gamma$, a range of bistability is found between the condensate and hybrid states, as shown in the inset. In addition, Fig.~\ref{fig:overview_state_space} shows the branch of high-density vortex states discussed above: the vortex gas (black diamonds) and vortex crystal states (green hexagons), whose amplitude varies approximately linearly with the control parameter $\gamma$, as discussed in Sec.~\ref{sec:late_time_near_gam1}. The dense SV branch formed by the vortex gas and crystal states coexists with the condensate and hybrid states over a wide range of the control parameter $\gamma$, leading to pronounced multistability. The dense SV branch terminates at $\gamma\approx 0.04$ below which hyperviscous dissipation becomes comparable to the instability forcing (rather than by the stochastic force) and the shielded vortices are destabilised. The same four types of qualitatively different states described here for cubic friction are also found with quadratic drag, as shown in Appendix \ref{sec:app_quad_drag}, thereby highlighting the potential relevance of our results to geophysical flows. These are characterised by large-scale, highly anisotropic, quasi-two-dimensional turbulent flows, sustained by various linear instability mechanisms. Indeed, the notion of a vortex gas has recently been used by \cite{gallet2020vortex} to derive a transport closure for turbulence driven by baroclinic instabilities in an oceanic context.

\section{Conclusions and outlook}
\label{sec:conclusions}
In this work we have presented a detailed study, based on extensive DNS, of the statistically stationary states observed in 2D turbulence driven by a hybrid forcing that interpolates between stochastic stirring and linear instability in the presence of nonlinear damping. Our simulations reveal an approximately self-similar evolution from small-amplitude random initial conditions to a high-density vortex gas state when the forcing is dominated by the instability term. The observed evolution begins with a short-lived initial inverse cascade that gives way to the formation of shielded vortices of positive and negative polarities. This brief phase is in turn followed by spontaneous symmetry breaking resulting in a state with vortices of one sign only. This broken-symmetry state further evolves via a slow, random nucleation process through which ever more coherent vortices are generated, up to a critical threshold where the number of shielded vortices (equivalently, the enstrophy) is about two fifths of its maximum value. Once this threshold is reached, \NEW{the interstitial space between vortices has shrunk sufficiently for the coherent vortices to efficiently impart vorticity to these interstices, leading to the formation of vortex seeds and} a rapid increase in vortex number, resulting in convergence towards a high-density statistically stationary state. The interplay between turbulence and the coherent vortices remains incompletely understood. Specifically, the dependence of the nucleation time scale on the forcing strength and the detailed mechanism of vortex nucleation and turbulence suppression, as well as the exact threshold density at which it occurs all require further investigation, \avkrev{as does the impact of the domain size}, all of which are left for a future study. 

The second key result reported here is that the high-density vortex gas state obtained from the observed self-similar, explosive evolution that sets in when the above threshold is exceeded exists over a wide range of forcing parameters, and undergoes a \avkrev{continuous} or supercritical phase transition to a hexagonal vortex state at a threshold $\gamma=\gamma_c\approx0.13$, below which individual vortices are trapped in a lattice and no longer diffuse across the domain. \avkrev{We showed that this transition is facilitated by rapid growth of inter-vortex gaps with increasing $\gamma$ when $\gamma$ is small.} This transition was found to be sharp in the presence of noise, and becomes imperfect in its absence. This somewhat counterintuitive behaviour appears to be associated with defects which anneal in the presence of noise but persist in its absence. The loss of positional order in the vortex gas was quantified using methods from statistical physics and crystallography \avkrev{focusing on the radial distribution function, with brief remarks on Voronoi diagrams and the Lindeman parameter.} 
The results revealed that the vortex "gas" is in fact liquid-like in terms of the distribution of relative vortex positions. It is important to stress here that the type of diffusive melting transition described in this work differs qualitatively from that described by \cite{james2021emergence}, since no vortices are annihilated here, the spontaneous symmetry breaking is maintained during the transition and no hexatic phase is observed, although with suitable initial conditions a transient hexatic phase exists in the present problem as well but was not observed to arise spontaneously. Somewhat surprisingly, we do not observe any spontaneous transitions between the vortex crystal and turbulence, in contrast with the results of \cite{james2021emergence}, although we cannot exclude the possibility that such behaviour might occur in the present model with different parameter values.

Another important result obtained in this paper is the population-averaged vorticity profile of the tripolar shielded vortices in this system. This profile indicates that the structure of the coherent tripolar vortices changes with the control parameter $\gamma$: the vortex size decreases monotonically with increasing $\gamma$ and vortices in the crystal state feature a less elliptical vortex core than at larger $\gamma$. No theoretical explanation is available for these observations and any future progress on these questions will be a significant contribution to our understanding of this prominent building block of instability-driven 2D turbulence. \avkrev{In fact, the gradual decrease in vortex size was shown to be offset, in part, by a correspondingly gradual increase in the number of vortices in the stationary state, resulting in an approximately linear growth in the gap area (as measured by the area where the vorticity magnitude is below $1\%$ of the maximum) with the control parameter $\gamma$.} The vortex intensity was likewise found to scale approximately linearly with $\gamma$ over a wide range of $\gamma$. 

The conditions for the suppression of the inverse cascade in the vortex crystal were analysed, revealing that \NEW{three key ingredients are required. First, the dominant energy injection must stem from the instability forcing. Second, the nonlinear damping must act at the scale of individual vortices, and third, the time scale of viscous dissipation must be large compared to the inverse instability growth rate at the most linearly unstable scales.} Violation of any of these conditions results in the dissolution of the shields of the coherent vortices, triggering a standard inverse energy cascade and the formation of a large-scale condensate. Finally, all known statistically stationary states of this problem were discussed, including earlier results from \cite{van2022spontaneous}, namely the large-scale condensate, hybrid states consisting of large-scale circulation patterns with embedded shielded vortices, and the dense vortex gas and vortex crystal states.

Many questions remain open. Since transient evolution towards an active vortex crystal, reminiscent of the transient dynamics observed here, has been observed in a model of active turbulence \citep{PhysRevFluids.3.061101}, it is natural to ask if the self-similarity of this evolution and the nontrivial scaling of its time scale with the driving strength is found in the moderate-Reynolds number flows of active turbulence models as well. Moreover, while it is well known that large-scale self-organization occurs in highly anisotropic 3D turbulent flows, such as flows in thin layers or rapidly rotating flows, it is as yet unclear what will be the fate of the vortex crystal described here in quasi-2D flows, i.e., when the flow is no longer required to be strictly 2D. Three-dimensionality is typically not taken into account in numerical studies of vortex crystals (see e.g. \cite{james2021emergence}), although numerically metastable 2D lattices of 3D vortices are observed in rapidly rotating turbulence \citep{di2020phase}. It is therefore important to clarify the stability of such vortex crystals in a quasi-2D setting, both from a fundamental standpoint, and because laboratory realisations of such states will necessarily be 3D and hence prone to 3D instabilities (e.g. \cite{kerswell2002elliptical} and \cite{seshasayanan2020onset}). 

It will also be of considerable interest to develop a hydrodynamic description of the vortex gas (or liquid!) extending the approach of \cite{wiegmann2014anomalous} to include the effects of vortex shielding and possibly the tripolar vortex structure. More generally, since the statistical mechanics of point vortex systems is well developed \citep{onsager1949statistical,eyink2006onsager}, including the aforementioned effects in this context would yield significant insights into the properties of the shielded vortex gas identified here.

\avkrev{We anticipate that the predictions of the present model may be verified in highly anisotropic flows driven by spectrally localised instabilities, where ambient fluctuations might play the role of the stochastic forcing term. It is important to stress that, for the dense vortex states, the ratio $\Gamma$ is always much greater than one, implying that the impact of the presence or absence of a stochastic forcing is minor. This observation suggests that states of this type are likely to be robust, provided only that the turbulent state is forced via an instability.}

\section*{Acknowledgements}
This work was supported by the National Science Foundation (Grants DMS-2009319, DMS-2009563, DMS-2308337 and DMS-2308338) and by the German Research Foundation (DFG Projektnummer: 522026592). A. v. K. thanks Ian Grooms for fruitful discussions and Paul Holst for pointing out useful references on backscatter parameterisations. \NEW{This research used the Savio computational cluster resource provided by the Berkeley Research Computing program at the University of California, Berkeley (supported by the UC Berkeley Chancellor, Vice Chancellor for Research, and Chief Information Officer). In addition, this work also utilised the Alpine high performance computing resource at the University of Colorado, Boulder. Alpine is jointly funded by the University of Colorado Boulder, the University of Colorado Anschutz, and Colorado State University. Data storage for this project was supported by the University of Colorado Boulder PetaLibrary.}  

\section*{Declaration of interests}
The authors report no conflict of interest.
\appendix

\section{Quadratic bottom drag}
\label{sec:app_quad_drag}
Many studies of nearly 2D turbulence include a linear Rayleigh drag law to model the effect of bottom friction \citep{boffetta2012two}. However, in the context of geophysical turbulence numerous works consider a quadratic bottom drag law, e.g. \cite{jansen2015parameterization} and \cite{gallet2020vortex}. Such a quadratic drag law can be obtained from dimensional considerations and is widely used in numerical ocean models \citep{gill1982atmosphere,willebrand2001circulation,egbert2004numerical,couto2020mixing}.

To test the robustness of the results presented here with respect to the choice of damping, we summarise here the results from runs in set C, i.e., DNS of Eqs.~(\ref{eq:nse})-(\ref{eq:forcing}) with quadratic damping ($m=1$ in Eq.~(\ref{eq:nse})) and $\beta=0.1$.
\begin{figure}
    \centering
    \includegraphics[width=0.49\textwidth]{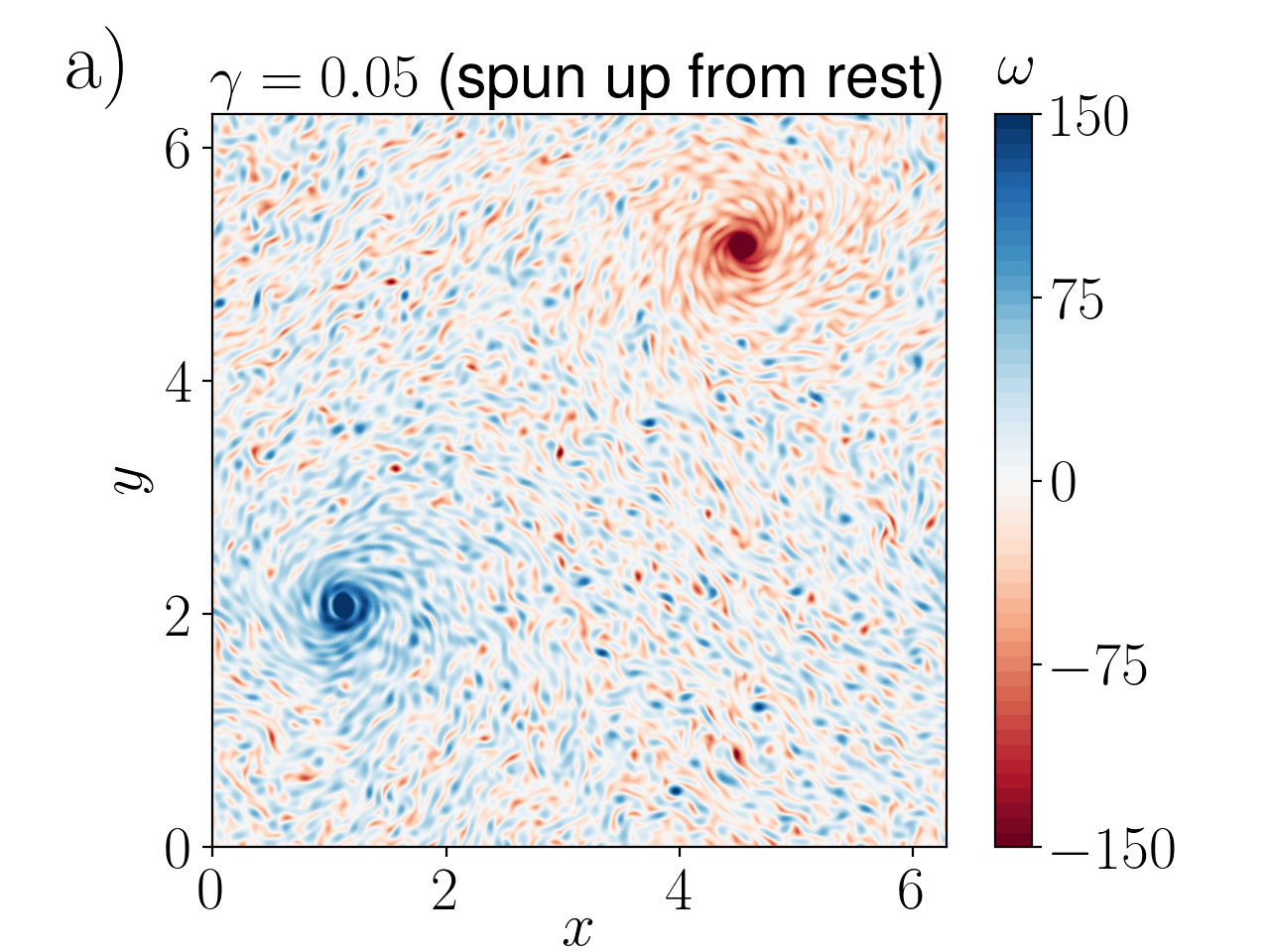}
     \includegraphics[width=0.49\textwidth]{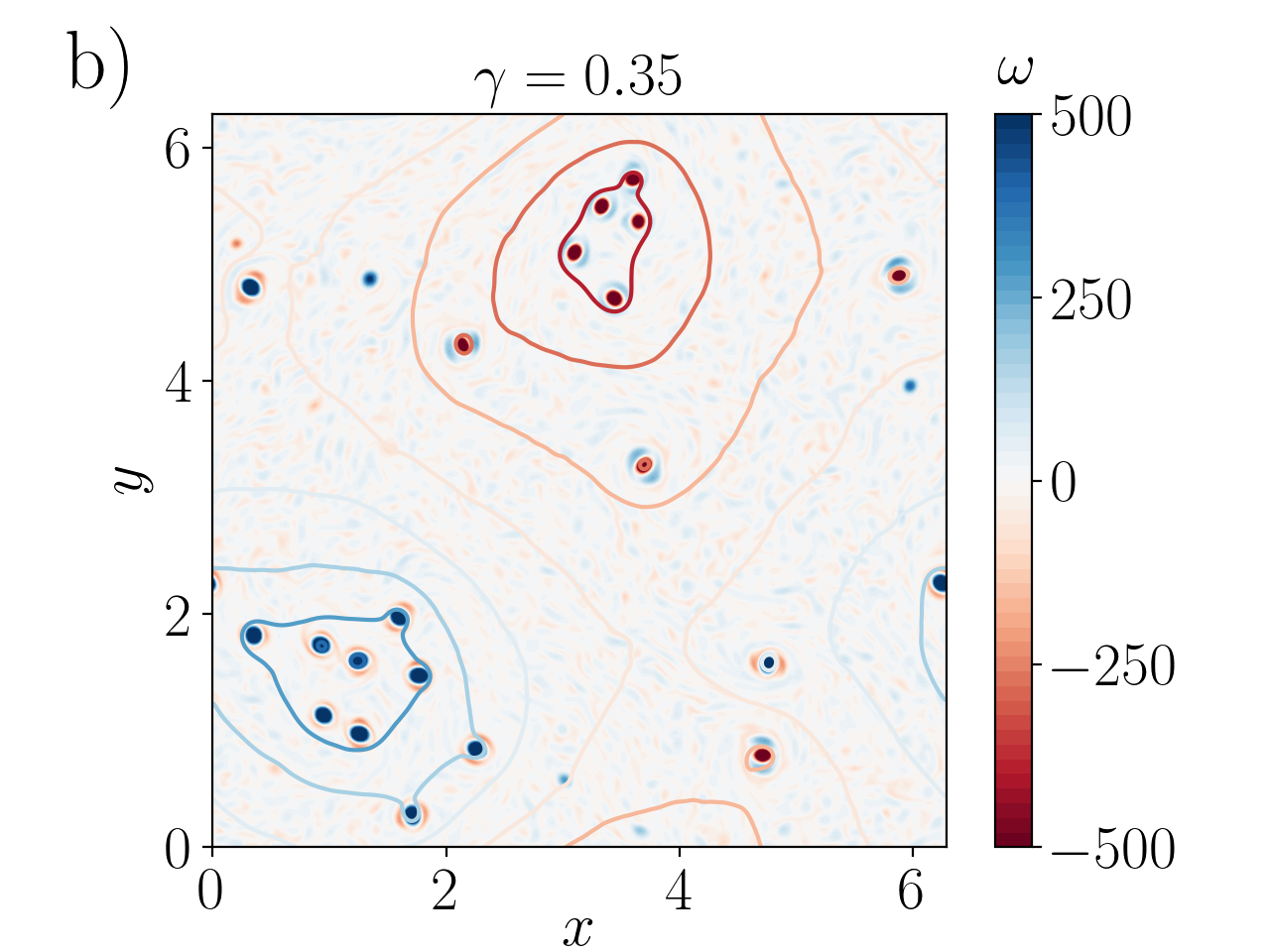}
    \includegraphics[width=0.49\textwidth]{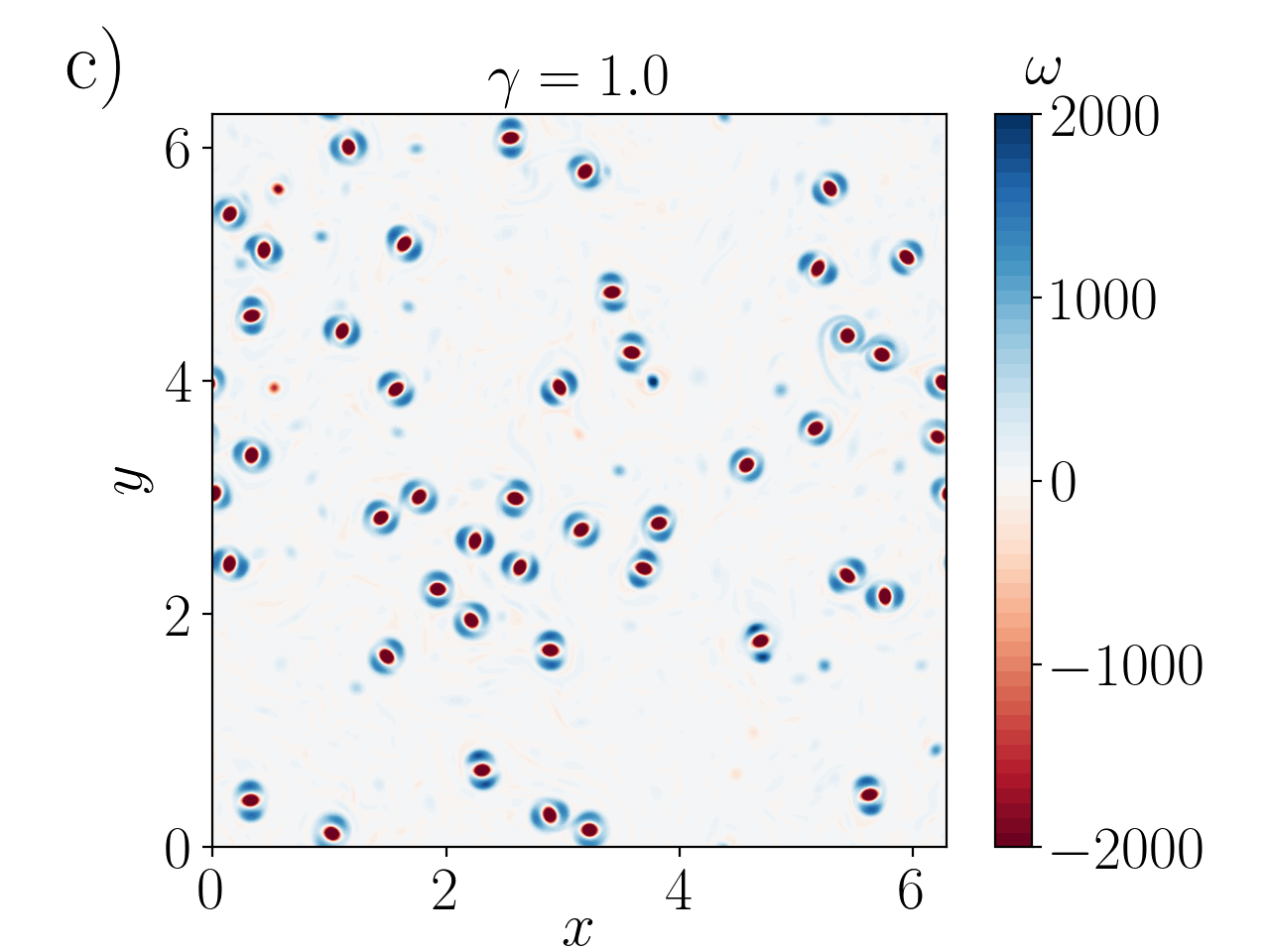}
    \includegraphics[width=0.49\textwidth]{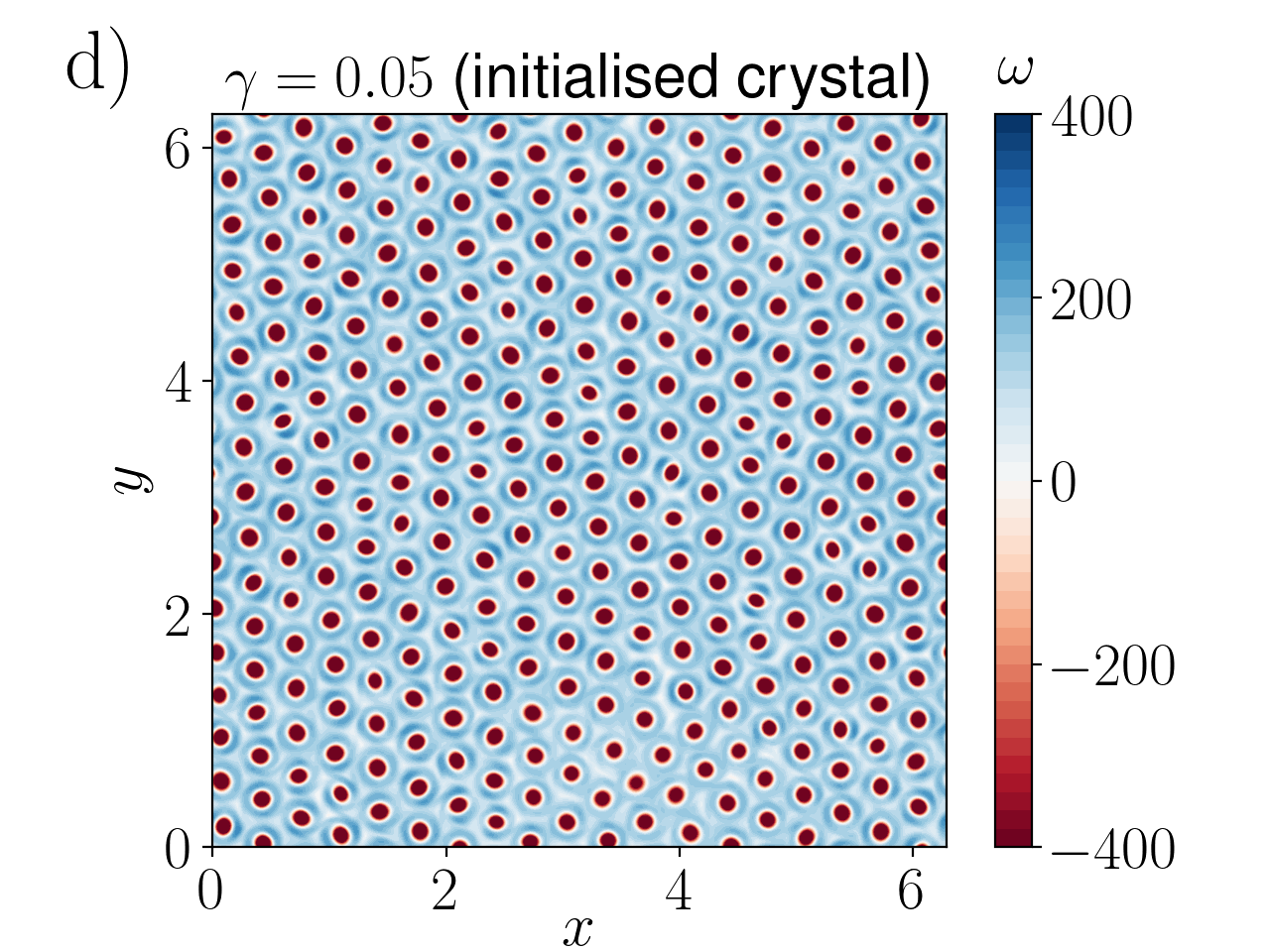}
    \caption{Overview of the stationary states observed with quadratic damping at different values of $\gamma$. These states are qualitatively similar to those observed with cubic damping. Filled contours show vorticity, while the contour lines in panel (b) show the large scale streamfunction.}
    \label{fig:quadratic_drag}
\end{figure}
Figure~\ref{fig:quadratic_drag} shows snapshots of solutions obtained with this quadratic friction and confirms that the four stationary states reported for cubic friction by \cite{van2022spontaneous} also exist in the geophysically relevant case of quadratic friction. Panel (a) shows the classical large-scale condensate obtained from small-amplitude initial conditions for $\gamma=0.05$. Panel (b) shows a hybrid state, again obtained from small-amplitude initial conditions, with two large-scale counter-rotating patches with embedded tripolar vortices whose sign is aligned with the background circulation. Panel (c) shows a dilute vortex gas, also obtained from small-amplitude initial conditions for $\gamma=1.0$. Finally, panel (d) shows a stable vortex lattice obtained by initialising with a similar lattice state at $\gamma=0.05$.

\section{\avkrev{Voronoi analysis of spatial order in dense vortex states}\label{sec:app_voronoi}}

\avkrev{An alternative tool for quantifying changes in regularity in the spatial distribution of the vortices is the Voronoi tesselation, a classical tool in the study of crystals \citep{blatov2004voronoi}. The Voronoi tesselation associated with a given set $M$ of points in the plane is given by a set of polygons, each corresponding to a unique point in the set $M$, defined as the set of points whose Euclidean distance from the point is smaller than from any other point in $M$. In our application, we take $M$ to be the set of vortex centers. Since we are considering a finite periodic domain, while Voronoi tesselations are constructed for the entire plane, it is necessary to exclude polygons at the edges of the domain to avoid spurious boundary effects. {We make use of the \textit{spatial.Voronoi} module provided by the scipy package \citep{virtanen2020scipy} to compute the Voronoi diagrams.}}

\begin{figure}
    \centering
    \includegraphics[width=0.328\textwidth]{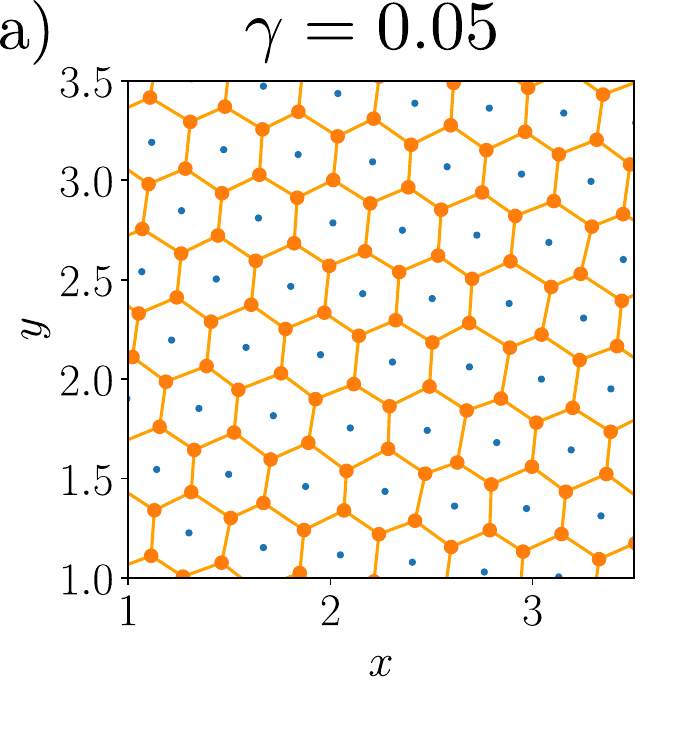}
     \includegraphics[width=0.328\textwidth]{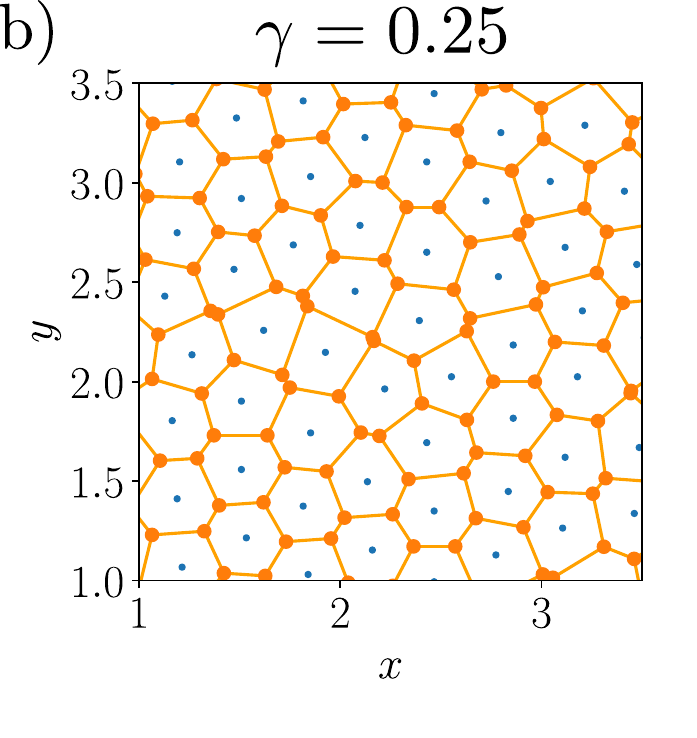}
     \includegraphics[width=0.328\textwidth]{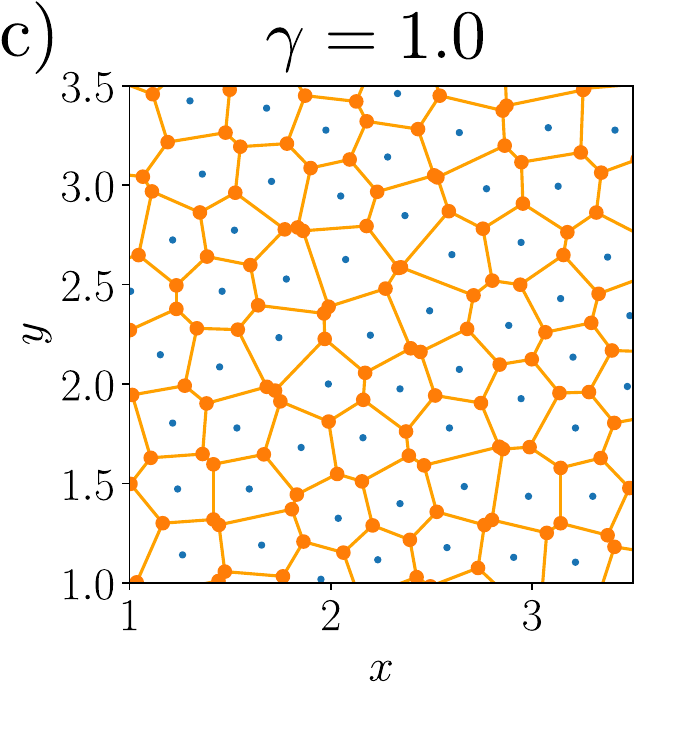}
    \includegraphics[width=0.32\textwidth,height=0.25\textwidth]{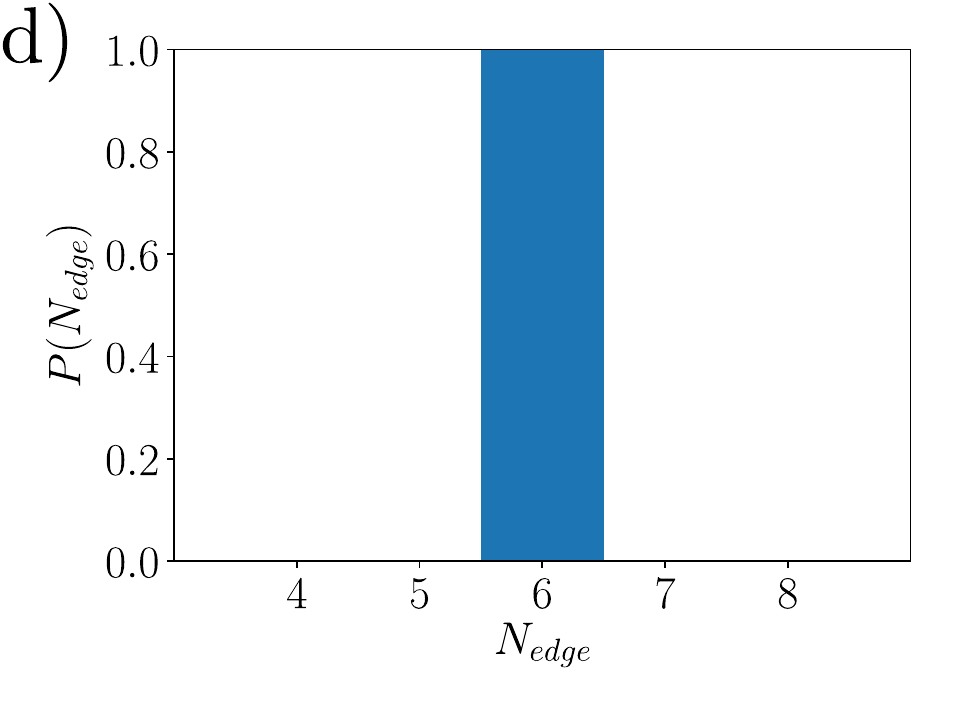}
    \includegraphics[width=0.32\textwidth,height=0.25\textwidth]{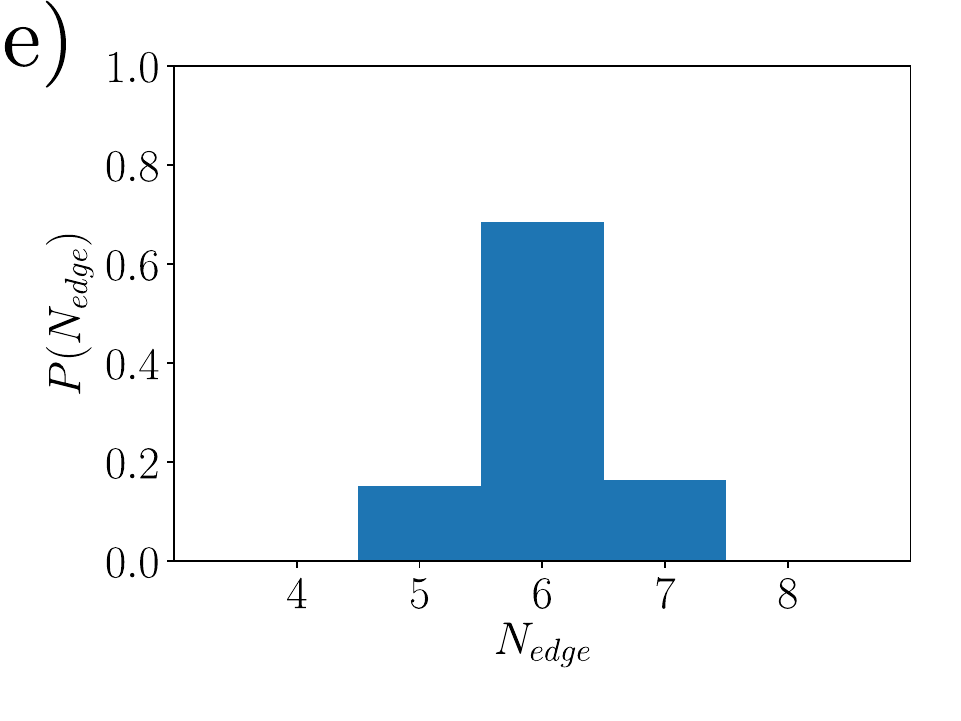}
    \includegraphics[width=0.32\textwidth,height=0.25\textwidth]{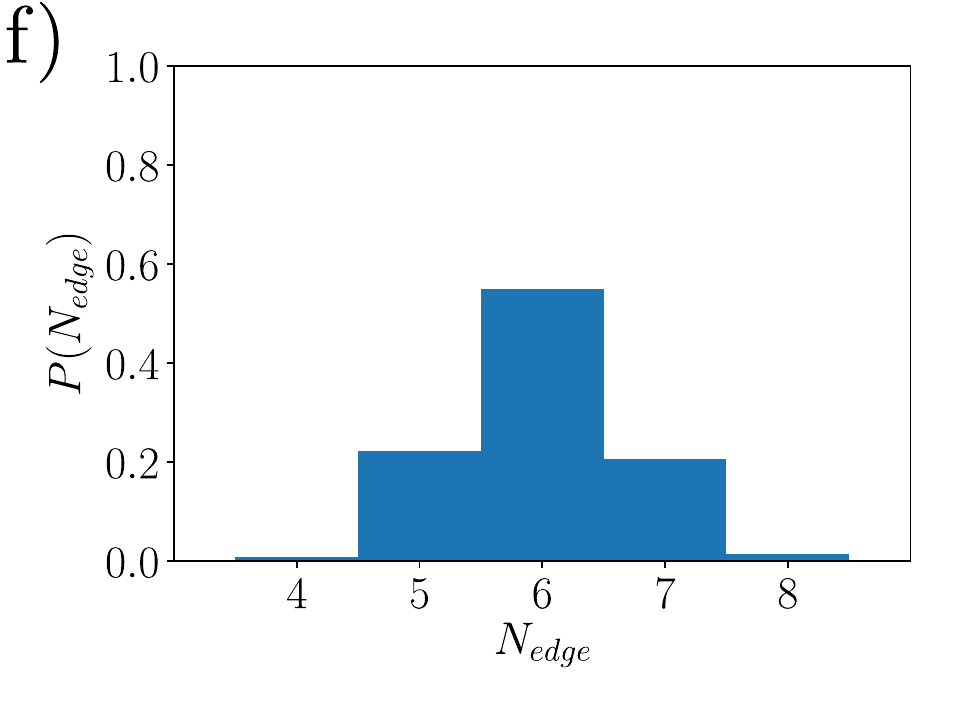}

    \caption{\avkrev{Panels (a)-(c): Voronoi tesselations based on the vortex center locations show a change in regularity with increasing $\gamma$. Note that only a fraction of the domain is shown. Panels (d)-(f): {Time-averaged} histograms of the number of edges in the above Voronoi diagrams at the same values of $\gamma$. The variance of the number of edges increases with increasing $\gamma$: the lattice state at $\gamma=0.05$ is perfectly hexagonal, while pentagons and hexagons appear at $\gamma=0.25$, and a small number of octagons and tetragons at $\gamma=1$.}}
    \label{fig:voronoi_diagrams}
\end{figure}

\avkrev{Figures~\ref{fig:voronoi_diagrams}(a)-(c) show that the Voronoi tesselation associated with the dense vortex states varies strongly with $\gamma$. At $\gamma=0.05$, where the hexagonal vortex crystal is observed, the corresponding Voronoi tesselation also consists of hexagons. As $\gamma$ increases above the melting threshold, the polygons in the Voronoi tessellation vary more and more in shape. Figures~\ref{fig:voronoi_diagrams}(d)-(f) quantify this visual insight in terms of histograms of the number of edges at $\gamma=0.05$ (in the crystalline state) and at $\gamma=0.25$ and $\gamma=1$ (above the melting threshold).
At $\gamma=0.05$, all polygons are indeed hexagonal, while at $\gamma=0.25$, there are also pentagons and heptagons. At $\gamma=1$, the fraction of hexagons has decreased further, while pentagons and heptagons are more frequent, and a small number of octagons and tetragons also appear. In short, the variance of the number of edges increases as $\gamma$ increases. }

\begin{figure}
    \centering
    \includegraphics[width=0.5\textwidth]{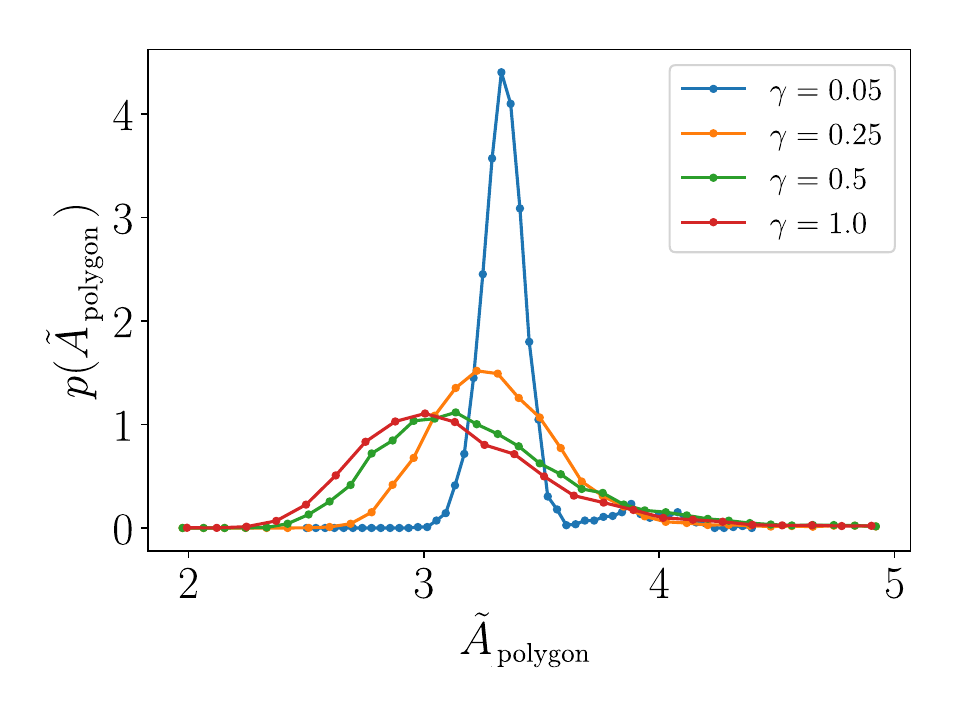}
    \caption{\avkrev{Probability density function (PDF) of the area $A_{\mathrm{polygon}}$ of the polygons in the Voronoi tesselation for different values of $\gamma$, nondimensionalised according to $\tilde{A}_{\mathrm{polygon}} = A_{\mathrm{polygon}}/\ell_1^2$. The PDF is sharply peaked at $\gamma=0.05$ in the crystalline state, and broadens significantly as $\gamma$ increases above the melting threshold. The mean area decreases as $\gamma$ increases, indicating that the distance between vortex cores also decreases. See also Fig.~\ref{fig:vorticity_profile_1D}.}}
    \label{fig:voronoi_area_pdfs}
\end{figure}

\avkrev{In addition to the number of edges, another quantifiable characteristic of the Voronoi polygons is their area. Figure~\ref{fig:voronoi_area_pdfs} shows histograms of the polygon areas (nondimensionalised by $\ell_1^2$ where $\ell_1=2\pi/k_1$ is the largest forcing scale). The histogram is sharply peaked in the crystalline state, and its variance increases monotonically with $\gamma$ in the vortex gas state. In addition, both the average polygon area and its most probable value shrink as $\gamma$ increases, observations that are indicative of a reduced average distance between vortex centers. }

\section{Impact of noise on the crystallisation transition}
\label{app:phase_transition_with_different_forcings}
\begin{figure}
   \hspace{0.2\textwidth} (a) \hspace{0.45\textwidth} (b)\vspace{-0.05cm}\\
    \centering
    \includegraphics[width=0.49\textwidth]{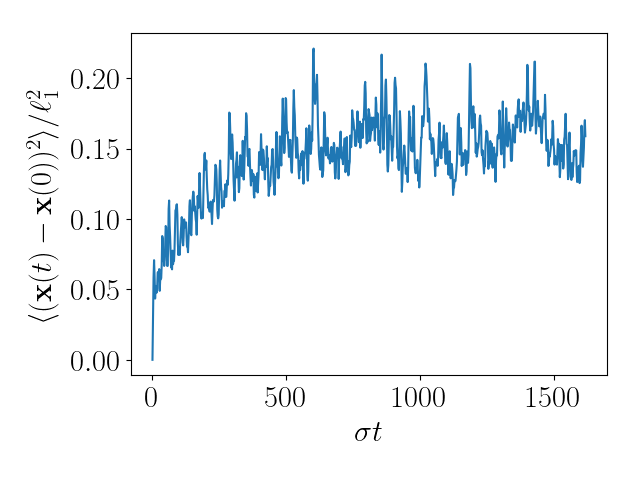}
    \includegraphics[width=0.49\textwidth]{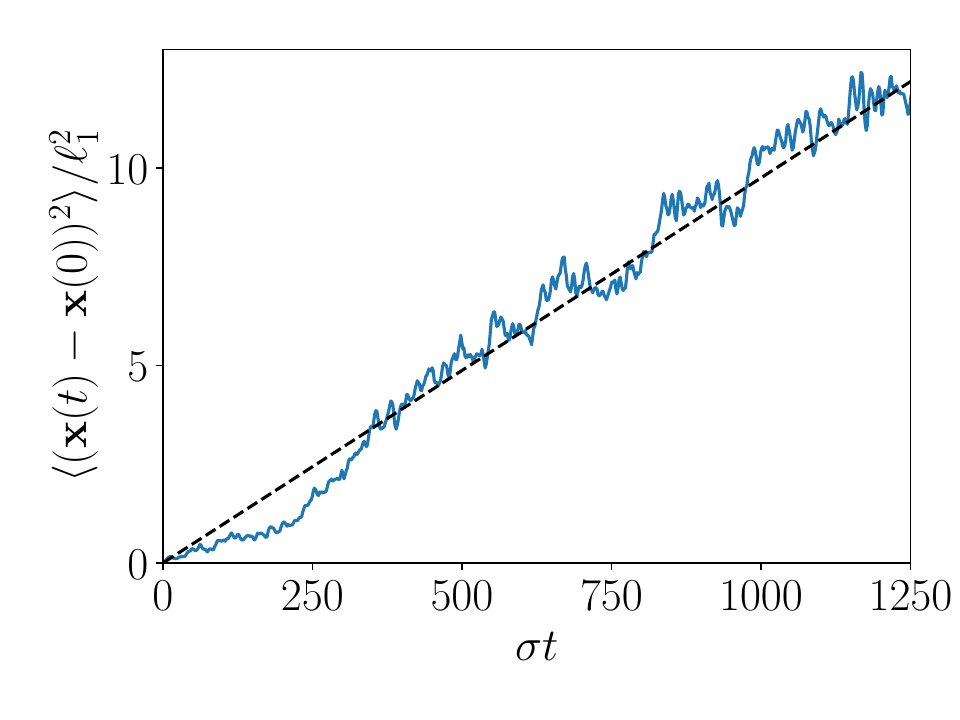}
    \caption{Mean squared displacement versus time at $\gamma=0.125$ (panel (a)) and $\gamma=0.15$ (panel (b)) with the stochastic forcing acting on $[k_1=33,k_2=40]$, instead of a thin shell near $k_2=40$. Individual vortices in the system are trapped at $\gamma=0.125$, but diffuse across the domain at $\gamma=0.15$, indicating that the melting transition near $\gamma=\gamma_c\approx 0.13$ is robust to changes in the width of the forcing window. Dashed line in panel (b) indicates a linear fit.}
    \label{fig:trapped_vs_non_trapped_xms_at_eps1_thin} 
\end{figure}

One can ask to what extent the melting/crystallisation transition observed here depends on the forcing characteristics. Figure~\ref{fig:trapped_vs_non_trapped_xms_at_eps1_thin} shows the mean squared displacement of shielded vortices observed with a modified random forcing with the same energy injection rate $\epsilon$ but acting over the wider range $k\in[33,40]$ instead of a thin shell near $k_2=40$. The instability forcing term is unchanged. A sharp transition is observed from a trapped state at $\gamma=0.125$ to a diffusing state at $\gamma=0.15$, consistent with the results described in the main text. The transition behavior and the location of the transition is thus robust to changes in the scales subject to random forcing.

\begin{figure} 
    \centering
    \includegraphics[width=0.6\textwidth]{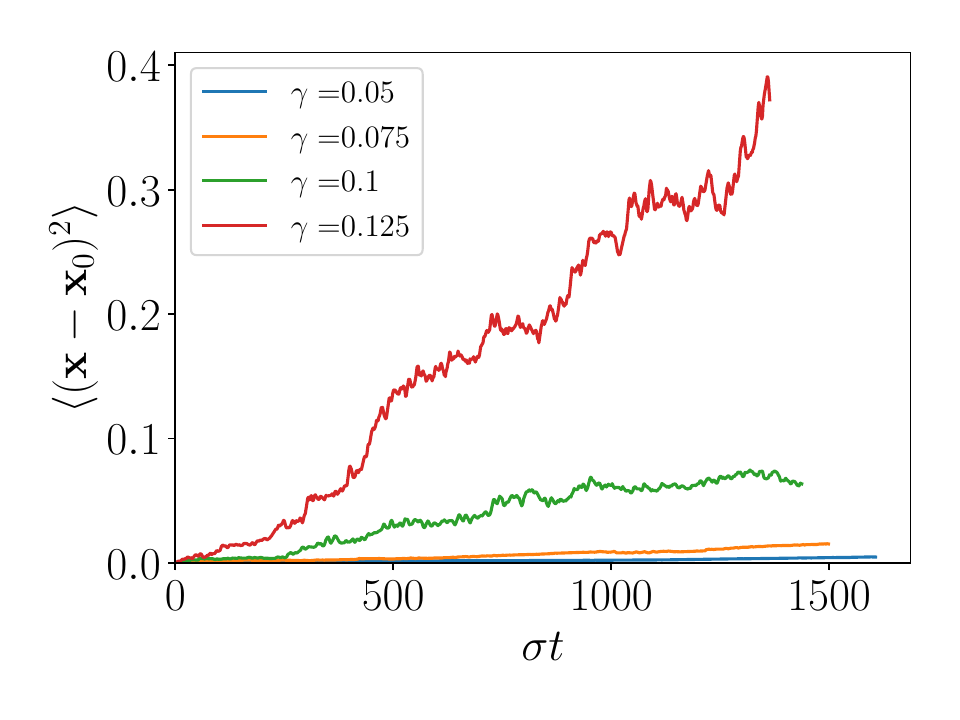}
    \caption{Mean squared displacement of vortices at $\epsilon=0$ versus nondimensional time. At the same values of $\gamma$, vortices are trapped when the random forcing is active ($\epsilon=1$). Here, in contrast, a small residual diffusion is observed. }
    \label{fig:xms_at_eps0_below_threshold}
\end{figure}

\begin{figure}
  (a) \hspace{0.45\textwidth} (b)\vspace{-0.05cm}\\
    \centering
    \includegraphics[width=0.49\textwidth]{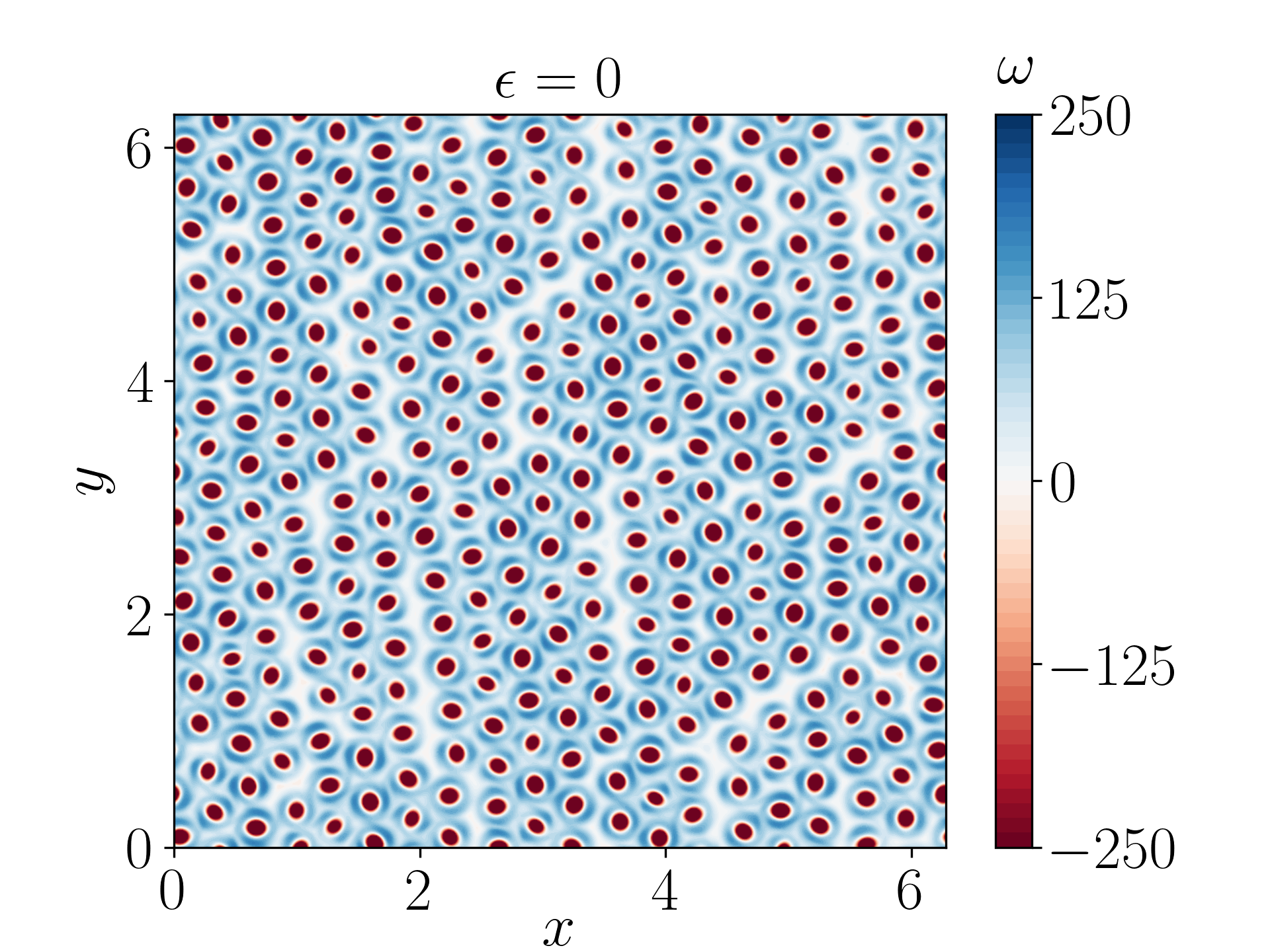}
    \includegraphics[width=0.49\textwidth]{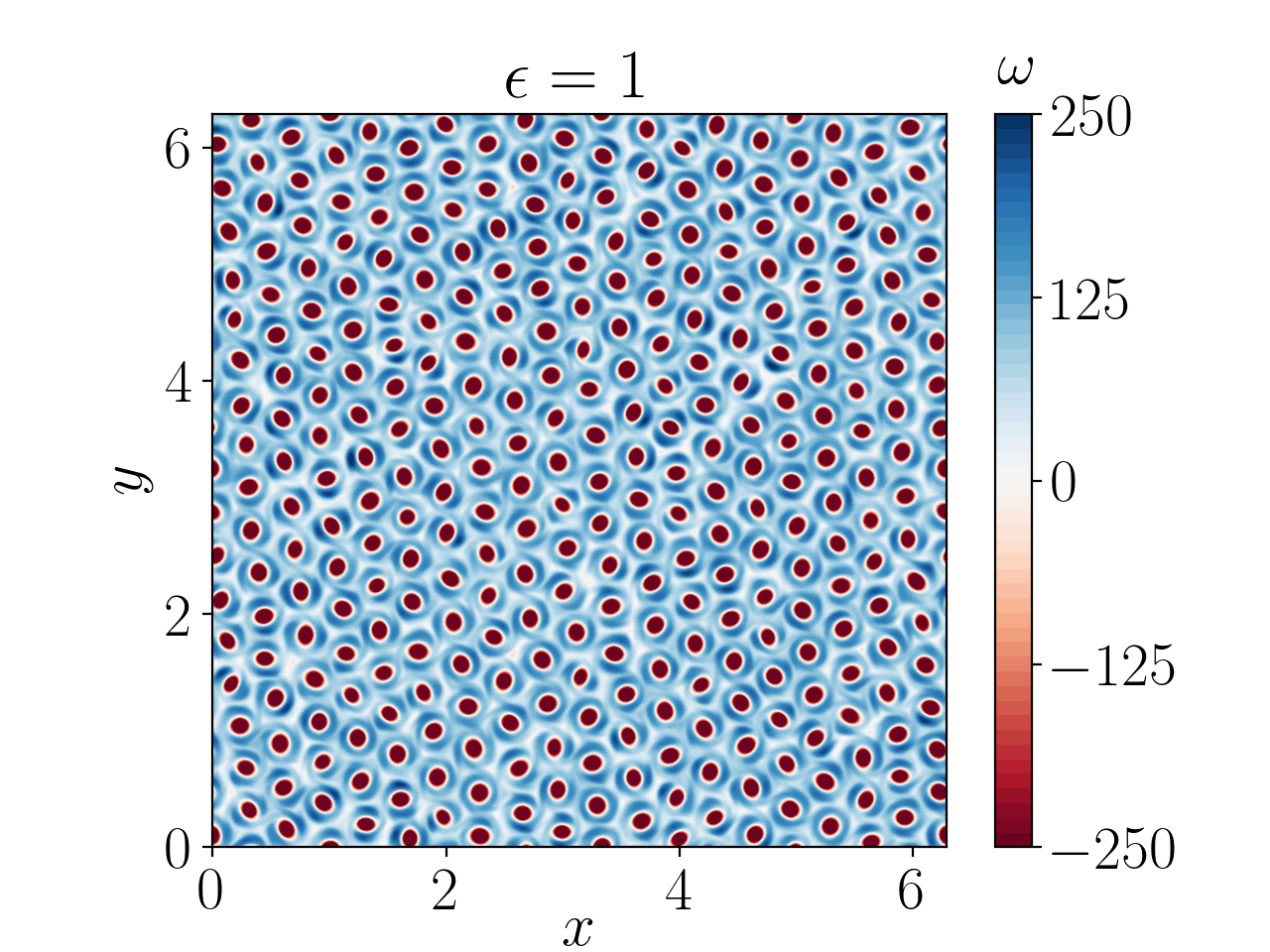}
    \caption{Snapshots of the vorticity field from set B at $\gamma=0.075$ with random forcing turned off ($\epsilon=0$, panel (a)) and turned on ($\epsilon=1$, panel (b)). }
    \label{fig:defects}
\end{figure}

In contrast, when the energy injection rate $\epsilon$ by the random forcing is set to zero, a qualitative change in the system behaviour is observed, with finite diffusion rates below the threshold in the presence of noise, as shown in Fig.~\ref{fig:xms_at_eps0_below_threshold} in terms of the mean squared displacement (cf. Fig.~\ref{fig:diffusivity_measurements}). In the absence of random forcing ($\epsilon=0$), long-lived line defects spontaneously form as shown in Fig.~\ref{fig:defects}(a)), which lead to residual vortex diffusion below the threshold at $\gamma_c\approx0.13$ observed in the presence of the stochastic forcing. \NEW{Supplementary movie SM6 shows that these defects are long-lived and thus significantly affect the crystal structure: contrast this with the regular lattice structure seen in Fig.~\ref{fig:defects}(b) and supplementary movie SM7, with identical parameters as in SM6 except for $\epsilon=1$ (random forcing turned on), where no defects are present.} The difference between these two situations, with and without stochastic forcing, may be interpreted in terms of \textit{annealing} of the crystal structure by the random stirring force. Other phenomena where a transition is sharpened in the presence of noise include noise-induced synchronisation \citep{kurths,kori}.

\bibliography{jfm_bib}
\bibliographystyle{abbrvnat}
\end{document}